\documentclass[prb,longbibliography,twocolumn,reprint]{revtex4}

\usepackage{amssymb,amsmath}

\newcommand\beq{\begin{equation}}
\newcommand\eeq{\end{equation}}
\newcommand\beqa{\begin{eqnarray}}
\newcommand\eeqa{\end{eqnarray}}

\newcommand{\nn}{\nonumber\\}

\newcommand{\al}{\alpha}

\usepackage{color}
\definecolor{darkgreen}{rgb}{0,0.6,0.0}
\newcommand{\vicente}[1]{{ #1}}

\usepackage{graphicx}
\begin{document}
\title{Navier--Stokes transport coefficients for a model of a confined quasi-two-dimensional granular binary mixture}
\author{Vicente Garz\'o}
\email{vicenteg@unex.es} \homepage{http://www.unex.es/eweb/fisteor/vicente/}
\affiliation{Departamento de
F\'{\i}sica and Instituto de Computaci\'on Cient\'{\i}fica Avanzada (ICCAEx), Universidad de Extremadura, E-06071 Badajoz, Spain}
\author{Ricardo Brito}
\affiliation{Departamento de Estructura de la Materia, F\'{\i}sica T\'ermica y Electr\'onica and GISC, Universidad Complutense de Madrid, E-28040 Madrid, Spain}
\author{Rodrigo Soto}
\affiliation{Departamento de
F\'{\i}sica, Facultad de Ciencias F\'{\i}sicas y Matem\'aticas, Universidad de Chile, 8370449 Santiago, Chile}

\begin{abstract}

The Navier--Stokes transport coefficients for a model of a confined quasi-two-dimensional granular binary mixture of inelastic hard spheres are determined from the Boltzmann kinetic equation. A \emph{normal} or hydrodynamic solution to the Boltzmann equation is obtained via the Chapman--Enskog method for states near the local version of the homogeneous time-dependent state. The mass, momentum, and heat fluxes are determined to first order in the spatial gradients of the hydrodynamic fields, and the associated transport coefficients are identified. They are given in terms of the solutions of a set of coupled linear integral equations. In addition, in contrast to previous results obtained for low-density granular mixtures, there are also nonzero contributions to the first-order approximations to the partial temperatures $T_i^{(1)}$ and the cooling rate $\zeta^{(1)}$. Explicit forms for the diffusion transport coefficients, the shear viscosity coefficient, and the quantities $T_i^{(1)}$ and $\zeta^{(1)}$ are obtained by assuming steady-state conditions and by considering
the leading terms in a Sonine polynomial expansion. The above transport  coefficients are given in terms of the coefficients of restitution, concentration, and  the masses and diameters of the components of the mixture. The results apply in principle for arbitrary degree of inelasticity and are not limited to specific values of concentration, mass and/or size ratios. As a simple application of these results, the violation of the Onsager reciprocal relations for a confined granular mixture is quantified in terms of the parameter space of the problem.

\end{abstract}



\date{\today}
\maketitle

\section{Introduction}
\label{sec1}

Granular gases can be considered as a collection of discrete macroscopic particles (typically of the order of micrometers or larger). Normally, grains differ in size, mass or in their mechanical properties and, as a consequence, granular gases require a multicomponent description.
Due to their macroscopic dimensions, in contrast to molecular or ordinary gases, all collisions between grains are inelastic and so the total kinetic energy of the particles decreases with time. \cite{BP04,G19} Thus, on order to maintain the system in the so-called rapid flow regime, an external energy input is needed to inject energy into the system and compensate for the energy dissipated by collisions. When both mechanisms cancel each other the system achieves a \emph{steady} nonequilibrium state. The injection of energy can be done for instance by vibrating walls \cite{YHCMW02,HYCMW04} or by bulk driving, as in air-fluidized beds. \cite{AD06,SGS05} However, this way of providing energy develops in most cases strong spatial gradients and hence, the theoretical description of the system is quite complicated. To avoid the above problem, it is  common in theoretical and computational studies to feed energy into the system by means of external driving forces or
\emph{thermostats}~\cite{PLMV99,CLH00,PEU02,PTNE02,PBL02,FAZ09,KSZ10,VAZ11,GSVP11,PGGSV12}.
A remarkable observation is that the transport properties of granular systems not only depend on the mechanical properties of the grains,
but also on the thermostating method~(see, for example, the comparison between
the Navier--Stokes shear viscosity obtained from the Chapman--Enskog expansion around the homogeneous cooling state~\cite{BDKS98} and the non-Newtonian shear viscosity of the uniform shear flow state;~\cite{SGD04,CRS05} a detailed discussion on this issue can be found in Ref.\ \onlinecite{G19}).

An alternative to the use of external forces has been proposed in the past few years. \cite{OU98,PMEU04,MVPRKEU05,CCDHMRV08,PCR09,RPGRSCM11,CMS12,mujica2016dynamics} The idea is to employ a particular geometry where the granular gas is confined in a box  \vicente{whose $z$-direction is slightly larger than one particle diameter, so particles are confined in the vertical direction. We refer to this geometry as a quasi-two-dimensional geometry. The box is} vertically vibrated so that energy is injected into the vertical degrees of freedom of particles via the collisions of grains with the top and bottom plates. The
energy gained by collisions with the walls is then transferred to the horizontal degrees of freedom \vicente{when a collision between particles takes place.}  Under certain conditions, the system presents a liquid-solid like phase separation. Complementarily, it remains in a homogeneous fluidized state (when it is observed from above) for a wide range of parameters \vicente{(see Ref.~\onlinecite{mujica2016dynamics} for a review of this geometry).}

\vicente{The full collisional dynamics in this geometry is highly complex, particularly due to the severe restrictions to the set of possible impact parameters imposed by the confinement.~\cite{MGB19a} To advance in the understanding of the  quasi-two-dimensional geometry, a collisional model for the transfer of energy from the vertical to horizontal degrees of freedom  was proposed years ago by Brito \textit{et al.}\ \cite{BRS13} As mentioned earlier, the vertical vibration accumulates energy into the $z$-component of the velocity.  In this model, particles move only in two-dimensions but, when a collision between particles occurs, part of the accumulated energy is released into the horizontal components of the velocity}. \vicente{In practice, an extra velocity $\Delta$ is added to the relative motion of colliding spheres and hence, the magnitude of the normal component of the relative velocity is increased by a given factor in the collision.}
The magnitude of the factor $\Delta$ can be related with the intensity of the vertical vibrations in the experiments.~\cite{MGB19a} For simplicity, $\Delta$ is assumed to be constant; this choice has the advantage of adding only a single parameter to the conventional inelastic hard sphere (IHS) model. The parameter $\Delta$ fixes the energy scale of the steady state. Moreover, it has been shown for constant $\Delta$ that the system remains homogeneous for all values of the global density. \cite{BRS13} A more realistic version of the above collisional model has been recently proposed \cite{RSG18}, where $\Delta$ is assumed to be a function of the local density. Such model gives rise to a van der Waals loop and a phase separation, in agreement with experiments. \cite{OU98,PMEU04,MVPRKEU05,CCDHMRV08} However, the derivation of the hydrodynamic equations from this model is much more involved than that of the model where $\Delta$ is constant since an additional hydrodynamic field is needed.

The collisional model with constant $\Delta$ (referred here to as the $\Delta$-model) has been widely employed by several groups in the past few years \vicente{to describe the properties of the quasi-two-dimensional geometry}. In particular, Brey and coworkers have considered this model in the dilute regime (i) to analyze the homogeneous state, \cite{BGMB13,BMGB14} (ii) to derive the Navier--Stokes hydrodynamic equations with explicit forms for the corresponding transport coefficients, \cite{BBMG15} and (iii) to perform a linear stability analysis of the homogeneous time-dependent state. \cite{BBGM16} The shear viscosity of a dilute granular gas has been also independently determined \cite{SRB14} and theoretical predictions compare quite well with computer simulations. The above previous works \cite{BBMG15,SRB14} of the $\Delta$-model have been recently extended to moderate densities by considering the Enskog kinetic equation and explicit forms of the Navier--Stokes transport coefficients have been explicitly obtained in terms of the coefficient of restitution and the density. \cite{GBS18} In addition, very recently, the $\Delta$-model has been extended to binary mixtures where the lack of equipartition has been analyzed in the stationary state.~\cite{BSG20} Besides this work,  we are not aware of any previous study on granular hydrodynamics in the relevant case of multicomponent systems in the context of the $\Delta$-model.

The goal of this paper is to provide a description of hydrodynamics in binary granular mixtures at low density with a comparable accuracy to that for the monocomponent case, namely, valid over the broadest parameter range including strong inelasticity. \cite{BBMG15,SRB14,GBS18} As a previous step, the reference homogeneous time-dependent state for a binary mixture has been discussed in detail recently. \cite{BSG20} The characterization of this reference state is crucial to provide the proper basis of transport due to spatial inhomogeneities.

As in previous works on granular mixtures, \cite{GD02,GMD06,SGNT06,GM07,SNTG09,GMV13a} the Chapman--Enskog method \cite{CC70} conveniently adapted to account for dissipative collisions is used to solve the coupled set of the Boltzmann equations for the two components. \cite{G19} In the first order of spatial gradients, the constitutive equations for the mass, momentum, and heat fluxes are derived and the transport coefficients of the mixture identified: three coefficients ($D$, $D_p$, and $D_T$) associated with the mass flux, the shear viscosity coefficient $\eta$ associated with the pressure tensor, and three coefficients ($D''$, $L$, and $\lambda$) associated with the heat flux. In addition, there are also contributions to the partial temperatures $T_i^{(1)}$ and the cooling rate $\zeta_U$ proportional to the divergence of the flow velocity field. While these two latter quantities vanish in the conventional IHS model for dilute gases \cite{GD02,GM07,GMV13a} (but not for dense mixtures, \cite{KS79b,GGG19b}), they are different from zero in the $\Delta$-model. The \emph{seven} relevant Navier--Stokes transport coefficients of the mixture as well as $T_i^{(1)}$ and $\zeta_U$ are given in terms of the solutions of a set of 9 coupled linear integral equations. This is, of course, a cumbersome task. For this reason, in this work we will address the determination of the set of transport coefficients $\left\{D,D_p, D_T,\eta\right\}$ and the quantities $T_i^{(1)}$ and $\zeta_U$. The thermal heat flux transport coefficients ($D''$, $L$, and $\lambda$) will be obtained only to the lowest order, which give trivial vanishing values when the two components are mechanically equivalent.

As usual, approximate forms of the above transport coefficients will be obtained by solving the integral equations by considering the leading terms in a Sonine polynomial expansion of the first-order distribution function. However, given the technical difficulties for obtaining explicit forms of the transport coefficients in the time-dependent problem, here the relevant state of a confined granular mixtures with \emph{steady} temperature is considered. This simplification offers the possibility of providing analytical expressions of transport properties in terms of the parameter space of the system.

As a simple application of the present results, the violation of Onsager's reciprocity relations is studied. To accomplish it, as said before, the heat flux transport coefficients ($D''$, $L$, and $\lambda$) can be expressed in terms of the diffusion coefficients when only the first Sonine approximation is retained. The study of Onsager's relations in the $\Delta$-model complements a previous analysis carried out years ago in the conventional IHS model. \cite{GMD06} As expected, since time reversal invariance does not fulfill in granular systems, Onsager's relations do not apply for finite inelasticity. However, it is interesting to gauge the deviations of the above relations as inelasticity increases.

The plan of the paper is as follows. In Sec.\ \ref{sec2}, the $\Delta$-model for granular mixtures is introduced and the coupled set of Boltzmann equations and the hydrodynamic equations are recalled. The Chapman--Enskog method adapted to inelastic binary mixtures is described in Sec.\ \ref{sec3} while the determination of the Navier--Stokes transport coefficients and the first-order contributions to the partial temperatures and the cooling rate is worked out in Sec.\ \ref{sec4}. Technical details on this derivation are relegated to three Appendices. Explicit expressions for all the above quantities are obtained at steady state conditions in Sec.\ \ref{sec5}. In dimensionless forms, they are given in terms of the coefficients of restitution, the concentration, and the mass and diameter ratios. The results for the above 6 quantities are illustrated in a two-dimensional system for a common coefficient of restitution and several values of the remaining parameters. The deviations from ordinary gases are in general significant but smaller than those previously found in the conventional IHS model. \cite{GMD06} The usual Onsager relations among the mass and heat flux transport coefficients for ordinary gases are then
noted and tested for the granular gas in Sec.\ \ref{sec6}. The expected violation is demonstrated as a function of the coefficient of restitution. Finally, the results are discussed in Sec.\ \ref{sec7}.

\section{Boltzmann kinetic equation for a model of a confined quasi-two-dimensional granular binary mixture}
\label{sec2}

\subsection{Collision rules for the $\Delta$-model}

Let us consider a granular binary mixture modeled as a gas of smooth inelastic hard spheres of masses $m_i$ and diameters $\sigma_i$ ($i=1, 2$). Let $(\mathbf{v}_1, \mathbf{v}_2)$ denote the pre-collisional velocities of two spherical particles of species $i$ and $j$, respectively, while $(\mathbf{v}_1',\mathbf{v}_2')$ denote their corresponding post-collisional velocities. The collision rules in the so-called $\Delta$-model read
\beq
\label{1.1}
\mathbf{v}_1'=\mathbf{v}_1-\mu_{ji}\left(1+\alpha_{ij}\right)(\widehat{{\boldsymbol {\sigma }}}\cdot \mathbf{g})\widehat{{\boldsymbol {\sigma }}}-2\mu_{ji}\Delta_{ij} \widehat{{\boldsymbol {\sigma }}},
\eeq
\beq
\label{1.2}
{\bf v}_{2}'=\mathbf{v}_{2}+\mu_{ij}\left(1+\alpha_{ij}\right)(\widehat{{\boldsymbol {\sigma }}}\cdot \mathbf{g})\widehat{{\boldsymbol {\sigma }}}+2\mu_{ij}\Delta_{ij} \widehat{{\boldsymbol {\sigma }}},
\eeq
where $\mu_{ij}=m_i/(m_i+m_j)$, $\mathbf{g}=\mathbf{v}_1-\mathbf{v}_2$ is the relative velocity, $\widehat{{\boldsymbol {\sigma}}}$ is the unit collision vector joining the centers of the two colliding spheres and pointing from particle 1 to particle 2. Particles are approaching if $\widehat{{\boldsymbol {\sigma}}}\cdot \mathbf{g}>0$. In Eqs.\ \eqref{1.1} and \eqref{1.2}, $0<\al_{ij}\leq 1$ is the (constant) coefficient of normal restitution for collisions $i$-$j$, and $\Delta_{ij}$ is an extra velocity added to the relative motion. This extra velocity points outward in the normal direction $\widehat{\boldsymbol {\sigma}}$, as required by the conservation of angular momentum. \cite{L04bis} The relative velocity after collision is
\beq
\label{1.3}
\mathbf{g}'=\mathbf{v}_1'-\mathbf{v}_2'=\mathbf{g}-(1+\al_{ij})(\widehat{{\boldsymbol {\sigma}}}\cdot \mathbf{g})
\widehat{\boldsymbol {\sigma}}-2\Delta_{ij} \widehat{{\boldsymbol {\sigma }}},
\eeq
so that
\beq
\label{1.4}
(\widehat{{\boldsymbol {\sigma}}}\cdot \mathbf{g}')=-\al_{ij} (\widehat{{\boldsymbol {\sigma}}}\cdot \mathbf{g})-2\Delta_{ij}.
\eeq
Similarly, the collision rules for the so-called restituting collisions $\left(\mathbf{v}_1'',\mathbf{v}_2''\right)\to \left(\mathbf{v}_1,\mathbf{v}_2\right)$ with the same collision vector $\widehat{{\boldsymbol {\sigma }}}$ are defined as
\beq
\label{1.7a}
\mathbf{v}_1''=\mathbf{v}_1-\mu_{ji}\left(1+\alpha_{ij}^{-1}\right)(\widehat{{\boldsymbol {\sigma }}}\cdot \mathbf{g})\widehat{{\boldsymbol {\sigma }}}-2\mu_{ji}\Delta_{ij}\al_{ij}^{-1} \widehat{{\boldsymbol {\sigma }}},
\eeq
\beq
\label{1.8a}
\mathbf{v}_2''=\mathbf{v}_2+\mu_{ij}\left(1+\alpha_{ij}^{-1}\right)(\widehat{{\boldsymbol {\sigma }}}\cdot \mathbf{g})\widehat{{\boldsymbol {\sigma }}}+2\mu_{ij}\Delta_{ij}\al_{ij}^{-1} \widehat{{\boldsymbol {\sigma }}}.
\eeq
Equations \eqref{1.7a}--\eqref{1.8a} yield the relationship
\beq
\label{1.8.1}
(\widehat{{\boldsymbol {\sigma}}}\cdot \mathbf{g}'')=-\al_{ij}^{-1}(\widehat{{\boldsymbol {\sigma}}}\cdot \mathbf{g})-2\Delta_{ij}\al_{ij}^{-1}.
\eeq

\subsection{Boltzmann kinetic equation}

In the low-density regime and neglecting the effect of the gravity field, the one-particle distribution function $f_i(\mathbf{r}, \mathbf{v};t)$ of the species or component $i$ obeys the Boltzmann kinetic equation
\beq
\label{1.5}
\frac{\partial}{\partial t}f_i+\mathbf{v}\cdot \nabla f_i=\sum_{j=1}^2\; J_{ij}[\mathbf{r},\mathbf{v}|f_i,f_j], \quad (i=1,2)
\eeq
where the Boltzmann collision operators $J_{ij}$ of the $\Delta$-model read
\beqa
\label{1.6}
& & J_{ij}[\mathbf{v}_1|f_i,f_j]\equiv \sigma_{ij}^{d-1} \int d{\bf v}_{2}\int d \widehat{\boldsymbol{\sigma}}\;
\Theta (-\widehat{{\boldsymbol {\sigma }}}\cdot {\bf g}-2\Delta_{ij})\nonumber\\
& & \times
(-\widehat{\boldsymbol {\sigma }}\cdot {\bf g}-2\Delta_{ij})
\al_{ij}^{-2}f_i(\mathbf{r},\mathbf{v}_1'';t)f_j(\mathbf{r},\mathbf{v}_2'';t)\nonumber\\
& & -\sigma_{ij}^{d-1}\int d {\bf v}_{2}\int d\widehat{\boldsymbol{\sigma}}\;
\Theta (\widehat{{\boldsymbol {\sigma }}}\cdot {\bf g})
(\widehat{\boldsymbol {\sigma }}\cdot {\bf g})f_i(\mathbf{r},\mathbf{v}_1;t)\nonumber\\
& &\times
f_j(\mathbf{r},\mathbf{v}_2;t),
\eeqa
where $\Theta(x)$ is the Heaviside step function, $\boldsymbol{\sigma}_{ij}=\sigma_{ij}\widehat{\boldsymbol{\sigma}}$ and $\sigma_{ij}=(\sigma_i+\sigma_j)/2$.
Note that although the $\Delta$-model was built to describe quasi-two dimensional systems, the calculations worked out
here will be performed for an arbitrary number of dimensions $d$.

An important property of the \vicente{Boltzmann} collision operators is \cite{BGMB13,SRB14}
\beqa
\label{1.6.1}
I_{\psi_i}&\equiv& \int\; d \mathbf{v}_1\; \psi_i(\mathbf{v}_1) J_{ij}[\mathbf{v}_1|f_i,f_j]\nonumber\\
&=&\sigma_{ij}^{d-1}\int d \, \mathbf{v}_1\int\ d {\bf v}_{2}\int d \widehat{\boldsymbol{\sigma}}\,
\Theta (\widehat{{\boldsymbol {\sigma }}}\cdot {\bf g})(\widehat{\boldsymbol {\sigma }}\cdot {\bf g})\nonumber\\
& & \times  f_i(\mathbf{r},\mathbf{v}_1;t)
f_j(\mathbf{r},\mathbf{v}_2;t)\left[\psi_i(\mathbf{v}_1')-\psi_i(\mathbf{v}_1)\right],\nonumber\\
\eeqa
where $\mathbf{v}_1'$ is defined by Eq.\ \eqref{1.1}. The property \eqref{1.6.1} is identical to the one obtained in the conventional IHS model ($\Delta_{ij}=0$). \cite{BP04,G19}

The relevant hydrodynamic fields in a granular mixture are the number densities $n_{i}$, the
flow velocity $\mathbf{U}$, and the granular temperature $T$. They are
defined in terms of velocity moments of the velocity distributions $f_{i}$ as
\begin{equation}
n_{i}=\int d{\bf v}f_{i}({\bf v})\;,\quad \rho {\bf U}=\sum_{i=1}^2m_{i}
\int d {\bf v}{\bf v}f_{i}({\bf v})\;,
\label{1.7b}
\end{equation}
\begin{equation}
nT=p=\sum_{i=1}^2 n_i T_i=\sum_{i=1}^2\frac{m_{i}}{d}\int d{\bf
v}V^{2}f_{i}({\bf v})\;, \label{1.8b}
\end{equation}
where ${\bf V}={\bf v}-{\bf U}$ is the peculiar velocity, $
n=n_{1}+n_{2}$ is the total number density, $\rho
=m_{1}n_{1}+m_{2}n_{2}$ is the total mass density, and $p$ is the
hydrostatic pressure. The third equality of Eq.\ (\ref{1.8b})
defines the kinetic temperatures $T_i$ for each component, which
measure their mean kinetic energies.
It is convenient also to work with the local mole fractions (or concentrations) $x_i=n_i/n$.

The collision operators conserve the particle number of each component and the
total momentum but the total energy is not conserved. This yields the conditions
\begin{equation}
\int d{\bf v}_{1}J_{ij}[{\bf v}_{1}|f_{i},f_{j}]=0,  \label{1.9}
\end{equation}
\begin{equation}
\sum_{i=1}^2\sum_{j=1}^2\int d{\bf v}_{1}m_{i}{\bf v}_{1}J_{ij}[{\bf v}_{1}|f_{i},f_{j}]=0,  \label{1.10}
\end{equation}
\begin{equation}
\sum_{i=1}^2\sum_{j=1}^2\int d{\bf v}_{1}\frac{1}{2}m_{i}v_{1}^{2}J_{ij}[{\bf v}
_{1}|f_{i},f_{j}]=-\frac{d}{2}nT\zeta,  \label{1.11}
\end{equation}
where $\zeta $ is identified as the ``cooling rate'' due to inelastic
collisions among all components. From Eqs.\ (\ref{1.9})--(\ref{1.11}) and using the property \eqref{1.6.1} for $\psi_i\equiv \left\{1,m_i\mathbf{v},\frac{1}{2}m_iv^2\right\}$, the macroscopic balance equations for the mixture can be easily obtained. In fact, their structure is similar to that of the conventional IHS model \cite{GD02,GM07,GMV13a} and they are given by
\begin{equation}
D_{t}n_{i}+n_{i}\nabla \cdot {\bf U}+\frac{\nabla \cdot {\bf j}_{i}}{m_{i}}
=0,  \label{1.12}
\end{equation}
\begin{equation}
D_{t}{\bf U}+\rho ^{-1}\nabla \cdot \mathsf{P}=0,  \label{1.13}
\end{equation}
\begin{equation}
D_{t}T-\frac{T}{n}\sum_{i=1}^2\frac{\nabla \cdot {\bf j}_{i}}{m_{i}}+\frac{2}{dn}
\left( \nabla \cdot {\bf q}+\mathsf{P}:\nabla {\bf U}\right) =-\zeta \,T.
\label{1.14}
\end{equation}
In the above equations, $D_{t}=\partial_{t}+{\bf U}\cdot \nabla $ is the
material derivative,
\begin{equation}
{\bf j}_{i}=m_{i}\int d{\bf v}_{1}\,{\bf V}_{1}\,f_{i}({\bf v}_{1}),
\label{1.15}
\end{equation}
is the mass flux for the component $i$ relative to the local flow
\begin{equation}
\mathsf{P}=\sum_{i=1}^2\,\int d{\bf v}_{1}\,m_{i}{\bf V}_{1}{\bf V}_{1}\,f_{i}({\bf
v}_{1}),  \label{1.16}
\end{equation}
is the total pressure tensor, and
\begin{equation}
{\bf q}=\sum_{i=1}^2\,\int d{\bf v}_{1}\,\frac{1}{2}m_{i}V_{1}^{2}{\bf V}
_{1}\,f_{i}({\bf v}_{1}),  \label{1.17}
\end{equation}
is the total heat flux.

It is quite apparent that the balance equations \eqref{1.12}--\eqref{1.14} do not constitute a closed set of hydrodynamic equations for the
fields $n_{i}$, ${\bf U}$ and $T$. This can be achieved when the fluxes (\ref{1.15})--(\ref{1.17})
and the cooling rate $\zeta$ are expressed in terms of the above hydrodynamic fields and their spatial gradients. To get this functional dependence one has to solve the Boltzmann equation by means of the Chapman--Enskog method \cite{CC70} conveniently modified to account for the inelasticity of collisions.

\section{Chapman--Enskog expansion}
\label{sec3}

The Chapman--Enskog method \cite{CC70} is applied in this section to solve the set of Boltzmann equations \eqref{1.1} for the binary mixture up to first order in spatial gradients. The first-order solution will be used later to determine the Navier--Stokes transport coefficients in terms of the coefficients of restitution $\al_{ij}$, the masses $m_i$ and diameters $\sigma_i$ of grains, the parameters $\Delta_{ij}$, the local mole fraction $x_1$, and the temperature $T$.

\subsection{Sketch of the Chapman--Enskog method}

\vicente{As discussed in different textbooks, \cite{CC70,FK72,GS03,S16} two separate stages are identified in the relaxation of an \emph{ordinary} gas towards equilibrium. A kinetic stage (for times of the order of the mean free time) is first identified where the main effect of collisions is to  relax quickly the gas towards a local equilibrium state. This stage depends on the initial conditions of the system. Then, a slow stage is identified where the gas has completely forgotten its initial preparation and so, its microscopic state  is governed by the hydrodynamic fields. The second stage is usually referred to as the \emph{hydrodynamic} regime. In the case of \emph{granular} gases, the above two-stage regimes are also expected to be identified. However, in the kinetic stage the distribution function will relax to a time-dependent nonequilibrium distribution (the so-called local homogeneous cooling state in the conventional IHS model)\cite{GD99b} instead of the local equilibrium distribution. This time-dependent distribution must be consistently obtained as the solution to the Boltzmann kinetic equation in the absence of spatial gradients. Moreover, in the hydrodynamic stage, although the granular temperature $T$ is not a conserved field due to  \emph{inelastic} collisions, it is still considered as a \emph{slow} hydrodynamic variable (i.e., its time evolution is much slower than other velocity moments of $f_i$ such as those related with the irreversible fluxes). This assumption has been clearly confirmed by the good agreement found between granular hydrodynamics and computer simulations in different nonequilibrium situations. \cite{BRC99,BRCG00,BRM01,DHGD02,LBD02,MG02,MG03,GM03,BRM05,LLC07,MDCPH11,BR13,MGH14}

This way, according to the above scenario, in the hydrodynamic regime the velocity distribution functions $f_i(\mathbf{r}, \mathbf{v};t)$ of the mixture are expected to depend on space and time through a \emph{functional} dependence on the hydrodynamic fields}. In this paper, similarly as in previous works on dilute granular mixtures, \cite{GD02,GMD06,GM07,GMV13a} we will chose the set $\xi\equiv \left\{x_1,p,T,\mathbf{U}\right\}$ as the hydrodynamic fields of the binary mixture instead of $\left\{ n_1,n_2,T,\mathbf{U}\right\}$:
\begin{equation}
f_{i}({\bf r},{\bf v}_{1},t)=f_{i}\left[ {\bf v}_{1}|x_{1}
(t), p(t), T(t), {\bf U}(t) \right] \;.
\label{2.1}
\end{equation}
The solution \eqref{2.1} is called a ``normal'' solution. Notice that functional dependence in \eqref{2.1} means that in order to determine $f_i$ at the point $\mathbf{r}$ one needs to know the hydrodynamic fields and all their spatial gradients at $\mathbf{r}$.

\vicente{It is quite apparent that the determination of the normal solution \eqref{2.1} from the set of coupled Boltzmann kinetic equations \eqref{1.5} is in general a very complex problem. This task becomes more accessible when the spatial gradients are small. In this case, the Chapman--Enskog method \cite{CC70} converts the functional dependence \eqref{2.1} into a local space dependence through an expansion of $f_i$ in powers of the Knudsen number Kn (Kn$=\ell/h$, where $\ell$ is the mean free path and $h$ is a characteristic hydrodynamic length). Since $h$ is the typical distance over which the distributions $f_i$ change substantially, then $h^{-1}\sim |\nabla \ln f_i|$ and hence, the expansion in Kn is actually equivalent to an expansion in powers of the spatial gradients of the hydrodynamic fields. In addition, as usual in the perturbation expansions, a bookkeeping parameter $\epsilon$ is introduced to label the relative orders of magnitude of the different terms appearing in the expansion. Thus, for small spatial variations, $f_i$ is written as a series expansion in powers of $\epsilon$:
\begin{equation}
f_{i}=f_{i}^{(0)}+\epsilon \,f_{i}^{(1)}+\epsilon^2 \,f_{i}^{(2)}+\cdots \;,
\label{2.2}
\end{equation}
where, for instance, a term of order $\epsilon$ is of first order in gradients ($\epsilon\sim \nabla \xi$) while a term of order $\epsilon^2$ is either a product of two first-order gradients [$(\partial_i \xi)(\partial_j \xi)$] or one second degree gradient ($\partial_i^2 \xi$). The formal parameter $\epsilon$ is taken to be equal to 1 at the end of the calculations}.

The expansions \eqref{2.2} and \eqref{2.2.1} yield similar expansions for the fluxes and the cooling rate when substituted into Eqs.\ \eqref{1.11}, and \eqref{1.15}--\eqref{1.17}:
\beq
\label{2.2.2}
\mathbf{j}_i=\mathbf{j}_i^{(0)}+\epsilon\, \mathbf{j}_i^{(1)}+\cdots, \quad \mathsf{P}=\mathsf{P}^{(0)}+\epsilon\, \mathsf{P}^{(1)}+\cdots,
\eeq
\beq
\label{2.2.3}
\mathbf{q}=\mathbf{q}^{(0)}+\epsilon\, \mathbf{q}^{(1)}+\cdots, \quad \zeta=\zeta^{(0)}+\epsilon\, \zeta^{(1)}+\cdots.
\eeq
Although the partial temperatures $T_i$ are not hydrodynamic quantities, they are also involved in the evaluation of the cooling rate. \cite{GGG19b} Its expansion is
\beq
\label{2.2.4}
T_i=T_i^{(0)}+\epsilon\,T_i^{(1)}+\cdots
\eeq

\vicente{To obtain the kinetic equations verifying the approximations $f_i^{(k)}$, since the term $\nabla f_i^{(k)}$ is of order $k+1$ in spatial gradients, then one formally replaces $\nabla \to \epsilon \nabla$ in the Boltzmann equation \eqref{1.1} and expands the time derivative $\partial_t$ as} \cite{note1}
\begin{equation}
\label{2.2.1}
\partial_t=\partial_t^{(0)}+\epsilon\partial_t^{(1)}+\cdots
\end{equation}
The action of the time derivatives $\partial_t^{(k)}$ on $x_1$, $\mathbf{U}$, $p$, and $T$ can be obtained from the balance equations \eqref{1.12}--\eqref{1.14} after taking into account the expansions \eqref{2.2.2}--\eqref{2.2.4}, \vicente{setting $\nabla \to \epsilon \nabla$, and collecting terms of the same order of $\epsilon$. In particular, the action of the operators $\partial_t^{(0)}$ and $\partial_t^{(1)}$ on $n_i$, $\mathbf{U}$, and $T$ can be easily obtained from Eqs.\ \eqref{1.12}--\eqref{1.14} as
\beq
\label{n1}
\partial _{t}^{(0)}n_{i}=\partial_t^{(0)}U_\lambda=0,\quad
T^{-1}\partial_{t}^{(0)}T=-\zeta ^{(0)},
\eeq
\beq
\label{n2}
\partial_t^{(1)}n_i=-n_i\nabla\cdot \mathbf{U},
\eeq
\beq
\label{n3}
\partial_t^{(1)}U_\lambda=-\mathbf{U}\cdot \nabla U_\lambda-\rho^{-1}\nabla p,
\eeq
\beq
\label{n4}
\partial_t^{(1)}T=-\mathbf{U}\cdot \nabla T+\frac{2}{d}T \nabla\cdot \mathbf{U}-T\zeta^{(1)}T.
\eeq
Here, to be consistently verified later, use has been made of the results $\mathbf{j}_i^{(0)}=\mathbf{q}^{(0)}=\mathbf{0}$ and $P_{\lambda\beta}^{(0)}=p\delta_{\lambda\beta}$}.

As usual in the Chapman--Enskog method, \cite{CC70} the hydrodynamic fields $n_i$, $p$, $T$, and $\mathbf U$ are defined in terms of the zeroth-order distributions
\beq
\label{2.2.5}
\int d\mathbf{v}\left(f_i-f_i^{(0)}\right)=0,
\eeq
\beq
\label{2.2.6}
\sum_{i=1}^2\int d\mathbf{v}\; \left\{m_i\mathbf{v}, \frac{m_i}{2}V^2\right\}\left(f_i-f_i^{(0)}\right)=\left\{\mathbf{0},0\right\}.
\eeq
Since the constraints \eqref{2.2.5} and \eqref{2.2.6} must hold at any order in $\epsilon$, the remainder of the expansion must obey the orthogonality conditions
\beq
\label{2.2.7}
\int d\mathbf{v}f_i^{(k)}=0,
\eeq
and
\beq
\label{2.2.8}
\sum_{i=1}^2\int d\mathbf{v}\; \left\{m_i\mathbf{v}, \frac{m_i}{2}V^2\right\}f_i^{(k)}=\left\{\mathbf{0},0\right\},
\eeq
for $k\geq 1$. The constraints \eqref{2.2.8} yield the relations
\begin{equation}
\label{2.2.9}
\mathbf{j}_1^{(k)}=-\mathbf{j}_2^{(k)}, \quad n_1 T_1^{(k)}=-n_2 T_2^{(k)},
\end{equation}
for $k\geq 1$. As expected, the second condition in Eq.~\eqref{2.2.9} prevents that the total temperature $T$ is affected by the spatial gradients

\vicente{The kinetic equations obeying the successive approximations $f_i^{(k)}$ can be easily obtained by inserting the expansions in power of $\epsilon$ in the Boltzmann equation \eqref{1.1} and equating terms of the same order in $\epsilon$. In this paper, only terms up to first-order in $\epsilon$ (Navier--Stokes hydrodynamic order) will be accounted for to compute the irreversible fluxes and the cooling rate.}

\subsection{Zeroth-order approximation}

To zeroth-order in $\epsilon$ [which is equivalent to neglect all gradients in the normal solution \eqref{2.1}], the Boltzmann kinetic equation \eqref{1.5} becomes
\beq
\label{2.3}
\partial_t^{(0)} f_i^{(0)}=\sum_{j=1}^2\; J_{ij}[\mathbf{v}|f_i^{(0)},f_j^{(0)}].
\eeq
\vicente{Since $f_i^{(0)}$ is a normal solution, its time dependence only occurs through its dependence on the hydrodynamic fields $\xi$:
\beqa
\label{2.3.1}
\partial_t^{(0)} f_i^{(0)}&=&\frac{\partial f_i^{(0)}}{\partial x_1}\partial_t^{(0)}x_1+\frac{\partial f_i^{(0)}}{\partial U_\lambda}\partial_t^{(0)}U_\lambda+\frac{\partial f_i^{(0)}}{\partial p}\partial_t^{(0)}p\nonumber\\
& &+\frac{\partial f_i^{(0)}}{\partial T}\partial_t^{(0)}T =-\zeta^{(0)}\Big(T\frac{\partial f_i^{(0)}}{\partial T}+p\frac{\partial f_i^{(0)}}{\partial p}\Big),\nn
\eeqa
}where the cooling rate $\zeta^{(0)}$ is determined by Eq.\ (\ref{1.11}) to
zeroth order, namely,
\begin{equation}
\zeta^{(0)}=-\frac{2}{d n T}\sum_{i,j}\int d{\bf v}_{1}\,\frac{1}{2}
m_{i}\,V_{1}^{2}J_{ij}[{\bf v}_{1}|f_{i}^{(0)},f_{j}^{(0)}]\;.  \label{2.5}
\end{equation}
To obtain Eq.\ \eqref{2.3.1} use has been made of the results $\partial _{t}^{(0)}x_{1}=0$ and
\beq
\label{2.5.0}
p^{-1}\partial _{t}^{(0)}p=-\zeta ^{(0)}.
\eeq
According to Eq.\ \eqref{2.3.1}, the Boltzmann equation \eqref{2.3} can be rewritten as
\beq
\label{2.5.1}
-\zeta^{(0)}\Big(T\frac{\partial f_i^{(0)}}{\partial T}+p\frac{\partial f_i^{(0)}}{\partial p}\Big)
=\sum_{j=1}^2\; J_{ij}[\mathbf{v}|f_i^{(0)},f_j^{(0)}].
\eeq

\vicente{For elastic collisions and $\Delta_{ij}=0$, the cooling rate vanishes ($\zeta^{(0)}=0$) and hence Eq.\ \eqref{2.5.1} becomes
\beq
\label{2.5.2}
0=\sum_{j=1}^2\; J_{ij}^{(\text{el})}[\mathbf{v}|f_i^{(0)},f_j^{(0)}],
\eeq
where $J_{ij}^{(\text{el})}$ is defined by Eq.\ \eqref{1.6} with $\al_{ij}=1$ and $\Delta_{ij}=0$. The solution to Eq.\ \eqref{2.5.2} is simply the \emph{local} equilibrium distribution function $f_{\text{LE}}^{(i)}(\mathbf{V})$ given by
\beq
\label{2.5.3}
f_{\text{LE}}^{(i)}(\mathbf{V})=n_i\Big(\frac{m_i}{2\pi T}\Big)^{d/2}e^{-m_iV^2/2T}.
\eeq
Note that in Eq.\ \eqref{2.5.3} the fields are evaluated at the point $\mathbf{r}$ and time $t$.

In the case of inelastic collisions ($\al_{ij}\neq 1$), $\zeta^{(0)}\neq 0$ and to date an exact solution to Eq.\ \eqref{2.5.1} has not been found, even for monocomponent granular gases. On the other hand, in the hydrodynamic regime, dimensional analysis and symmetry considerations suggest that the solution to Eq.\ \eqref{2.3} must be of the form \cite{BMGB14,BSG20}}
\beq
\label{2.7}
f_i^{(0)}=x_i \frac{p}{T}v_\text{th}^{-d}\varphi_i(\mathbf{c},\Delta_{11}^*, \Delta_{22}^*, \Delta_{12}^*),
\eeq
where $\mathbf{c}=\mathbf{V}/v_\text{th}$ and $\Delta_{ij}^*=\Delta_{ij}/v_\text{th}$, $v_\text{th}=\sqrt{2T/\overline{m}}$ being the thermal velocity. Here, $\overline{m}=(m_1+m_2)/2$. \vicente{The consistency of the assumption \eqref{2.7} has been confirmed by computer simulations carried out for monocomponent \cite{BGMB13,BMGB14} and multicomponent \cite{BSG20} granular gases. Apart from its dependence on $\mathbf{c}$ and $\Delta_{ij}^*$, the scaled distributions $\varphi_i$ are expected to be also functions of the coefficients of restitution $\al_{ij}$ and the parameters of the mixture (mole fraction, masses and diameters).}

Since the dependence of $f_i^{(0)}$ on $T$ and $p$ is of the form $pT^{-(1+\frac{d}{2})}\varphi_i(V/\sqrt{T},\Delta_{ij}/\sqrt{T})$, one has the relations
\beq
\label{2.9}
T\frac{\partial f_i^{(0)}}{\partial T}=-f_i^{(0)}-\frac{1}{2}\frac{\partial}{\partial \mathbf{V}}\cdot \left(\mathbf{V}f_i^{(0)}\right)
-\frac{1}{2}f_i^{(0)}\Delta^*\frac{\partial \ln \varphi_i}{\partial \Delta^*},
\eeq
\beq
\label{2.10}
p\frac{\partial f_i^{(0)}}{\partial p}=f_i^{(0)},
\eeq
where we have introduced the shorthand notation
\beq
\label{2.11.1}
\Delta^*\frac{\partial}{\partial \Delta^*}\equiv \left(\Delta_{11}^*\frac{\partial}{\partial \Delta_{11}^*}+\Delta_{22}^*\frac{\partial}{\partial \Delta_{22}^*}+\Delta_{12}^*\frac{\partial}{\partial \Delta_{12}^*}\right).
\eeq
Note that in the particular case $\Delta_{11}^*=\Delta_{22}^*=\Delta_{12}^*=\Delta^*$, only one of the three terms of the identity \eqref{2.11.1} must be considered.

The Boltzmann equation \eqref{2.5.1} can be more usefully written by employing the relations \eqref{2.9} and \eqref{2.10} as
\begin{multline}
\label{2.6.1}
\frac{1}{2}\zeta^{(0)}\frac{\partial}{\partial \mathbf{V}}\cdot \left(\mathbf{V}f_i^{(0)}\right)
+\frac{1}{2}\zeta^{(0)}f_i^{(0)}\Delta^*\frac{\partial \ln \varphi_i}{\partial \Delta^*}\\=\sum_{j=1}^2\; J_{ij}[\mathbf{v}|f_i^{(0)},f_j^{(0)}].
\end{multline}
As expected, Eq.\ \eqref{2.6.1} has the same form as the Boltzmann equation for the $\Delta$-model in time-dependent homogeneous states. \cite{BSG20} \vicente{Thus, as is standard for ordinary gases, the gradient expansion in the Chapman--Enskog method is taken with respect to the reference \emph{local} time-dependent state, i.e., that resulting from the neglect of all gradients but evaluated at the value of the hydrodynamic fields at the chosen point and time. In other words, the election of the reference state in the Chapman--Enskog method for granular gases comes from the solution to the Boltzmann equation to zeroth-order in gradients and cannot not be chosen a priori.}

In dimensionless form, the Boltzmann equation  \eqref{2.6.1} for $\varphi_i$ can be rewritten as
\beq
\label{2.12}
\frac{1}{2}\zeta_0^*\frac{\partial}{\partial \mathbf{c}}\cdot \left(\mathbf{c}\varphi_i\right)
+\frac{1}{2}\zeta_0^*\Delta^*\frac{\partial \varphi_i}{\partial \Delta^*}
=\sum_{j=1}^2\; J_{ij}^*[\varphi_{i},\varphi_{j}],
\eeq
where $\zeta_0^*=\zeta^{(0)}/\nu$, $J_{ij}^*=(v_\text{th}/n_i \nu)J_{ij}$, and
\beq
\label{2.12.0}
\nu=n \sigma_{12}^{d-1}v_\text{th}=\sqrt{\frac{2}{\overline{m}}}\sigma_{12}^{d-1}p T^{-1/2}
\eeq
is an effective collision frequency. The equation for the temperature ratio
\beq
\label{2.12.1}
\gamma_i\equiv \frac{T_i^{(0)}}{T}=\frac{2}{d}\frac{m_i}{\overline{m}}\int\; d \mathbf{c}\; c^2 \varphi_i(\mathbf{c})
\eeq
can be easily derived by multiplying both sides of Eq.\ \eqref{2.12} by $c^2$ and integrating over velocity. The result is
\beq
\label{2.12.2}
\frac{1}{2}\zeta_0^*\Delta^*\frac{\partial \gamma_i}{\partial \Delta^*}=\gamma_i\left(\zeta_0^*-\zeta_i^*\right),
\eeq
where the (reduced) partial cooling rates $\zeta_i^*$ are defined as
\beq
\label{2.12.3}
\zeta_i^*=-\frac{2}{d}\theta_i \sum_{j=1}^2 \int\;d {\bf c}\; c^{2}J_{ij}^*[\varphi_i,\varphi_j],
\eeq
and $\theta_i= m_i/(\overline{m}\gamma_i)$.

As expected, since the distribution functions $f_i^{(0)}$ are isotropic in velocity space, then
\beq
\label{2.8}
\mathbf{j}_1^{(0)}=\mathbf{q}^{(0)}=\mathbf{0}, \quad P_{\lambda\beta}^{(0)}=p\delta_{\lambda\beta}.
\eeq

\vicente{Although the exact form of the distributions $f_i^{(0)}$ is not known, an indirect information on them is provided by the kurtosis (or fourth-cumulants) $a_2^{(i)}$ ($i=1,2$). These quantities measure the departure of $f_i^{(0)}$ from its Maxwellian form. In the context of the $\Delta$-model, the kurtosis has been evaluated for monocomponent gases and the results show that its magnitude is in general  small. \cite{BGMB13,BMGB14} The asymptotic steady  homogenous state of a granular binary mixture has been recently studied \cite{BSG20} by assuming Maxwellian distributions at $T_i^{(0)}$ for $f_i^{(0)}$.
Despite this approximation, theory compares in general quite well with Monte Carlo and molecular dynamics simulations when computing the temperature ratio $T_1^{(0)}/T_2^{(0)}$ and the global temperature, specially for low-density systems.
Thus, non-Gaussian corrections to $f_i^{(0)}$ can be neglected for practical purposes.}

To determine the Navier--Stokes transport coefficients under steady state conditions, one needs to evaluate derivatives such as $(\partial \gamma_i/\partial \Delta^*)_s$ and $(\partial \zeta_0^*/\partial \Delta^*)_s$. Here, the subscript $s$ means that the derivatives are evaluated in the steady state (i.e., when $\zeta_1^*=\zeta_2^*=0$). The above derivatives measure the departure of the steady state from the perturbed time-dependent state. The calculation of these derivatives is performed in the Appendix~\ref{appA}.

\section{First-order approximation: Navier--Stokes transport coefficients}
\label{sec4}

The first-order contribution $f_i^{(1)}$ to the distribution functions are considered in this section. Given that the mathematical steps involved in this calculation are quite similar to those carried out years ago \cite{GD02,GM07} in the conventional IHS model, some parts of this derivation are omitted here. We refer to the interested reader to Ref.~\onlinecite{G19} or Refs.\ \onlinecite{GD02,GM07} for specific details. The only subtle point (which is absent in the conventional IHS model) is that there are non vanishing contributions to the partial temperatures $T_i^{(1)}$ and the cooling rate $\zeta^{(1)}$ in the first-order solution. Some technical details are provided in the Appendix \ref{appB}.

The first-order distribution function $f_i^{(1)}(\mathbf{V})$ is given by
\begin{multline}
f_{i}^{(1)}={\boldsymbol {\cal A}}_{i}\cdot \nabla x_{1}+{\boldsymbol {\cal B}}_{i}\cdot
\nabla p+{\boldsymbol {\cal C}}_{i}\cdot \nabla T\\+\mathcal{D}_{i,\beta\lambda
}\frac{1}{2}\left(\frac{\partial U_\beta}{\partial r_\lambda}+
\frac{\partial U_\lambda}{\partial r_\beta}-\frac{2}{d}\delta_{\beta\lambda}\nabla\cdot \mathbf{U}\right)
+\mathcal{E}_i \nabla\cdot \mathbf{U}\;,
\label{4.0}
\end{multline}
where $\beta$ and $\lambda$ refer to Cartesian components and an implicit summation over  repeated indices is used. The quantities ${\boldsymbol {\cal A}}_{i}(\mathbf{V})$, ${\boldsymbol {\cal B}}_{i}(\mathbf{V})$, ${\boldsymbol {\cal C}}_{i}(\mathbf{V})$, $\mathcal{D}_{i,\beta\lambda}(\mathbf{V})$, and $\mathcal{E}_i(\mathbf{V})$ obey the linear coupled integral equations \eqref{2.30}--\eqref{2.34}, respectively. Use of Eq.\ \eqref{4.0} in the definitions \eqref{1.15}--\eqref{1.17} allows us to obtain the forms of the irreversible fluxes. As expected, they are given by
\beq
\label{3.2}
{\bf j}_{1}^{(1)}=-\frac{m_{1}m_{2}n}{\rho} D\nabla x_{1}-\frac{\rho}{p}D_{p}\nabla p-\frac{\rho}{T}D_T\nabla T,
\eeq
\beq
\label{4.1}
P_{\lambda\beta}^{(1)}=-\eta\left(\frac{\partial U_\lambda}{\partial r_\beta}+\frac{\partial U_\beta}{\partial r_\lambda}-\frac{2}{d}\delta_{\lambda\beta}\nabla \cdot \mathbf{U}\right),
\eeq
\beq
\label{4.2}
\mathbf{q}^{(1)}=-T^2 D''\nabla x_1-L \nabla p-\lambda \nabla T.
\eeq
The Navier--Stokes transport coefficients in Eqs.\ \eqref{3.2}--\eqref{4.2} are the diffusion coefficient $D$, the pressure diffusion coefficient $D_p$, the thermal diffusion coefficient $D_T$, the shear viscosity coefficient $\eta$, the Dufour coefficient $D''$, the pressure energy coefficient $L$, and the thermal conductivity coefficient $\lambda$. \cite{note2}

The transport coefficients associated with the mass flux are identified as
\begin{equation}
D=-\frac{1}{d}\frac{\rho}{m_{2}n}\int d{\bf v}\,{\bf V}\cdot {\boldsymbol {\cal A}}_{1},  \label{3.3}
\end{equation}
\begin{equation}
D_{p}=-\frac{1}{d}\frac{m_{1}p}{\rho}\int d{\bf v}\,{\bf V}\cdot {\boldsymbol {\cal B}}_{1},  \label{3.4}
\end{equation}
\begin{equation}
D_T=-\frac{1}{d}\frac{m_{1}T}{\rho}\int d{\bf v}\,{\bf V}\cdot {\boldsymbol {\cal C}}_{1}.  \label{3.5}
\end{equation}
The shear viscosity $\eta$ is
\beq
\label{4.3}
\eta=-\frac{1}{(d-1)(d+2)} \sum_{i=1}^2\; \int d{\bf v}\, V_\lambda  V_\beta \mathcal{D}_{i,\lambda\beta}(\mathbf{V}).
\eeq
Finally, the transport coefficients for the heat flux are
\beq
\label{4.4}
D''=-\frac{1}{d T^2}\sum_{i=1}^2\; \int d{\bf v}\, \frac{m_i}{2} V^2 \mathbf{V}\cdot {\boldsymbol {\cal A}}_{i},
\eeq
\beq
\label{4.5}
L=-\frac{1}{d}\sum_{i=1}^2\; \int d{\bf v}\, \frac{m_i}{2} V^2 \mathbf{V}\cdot {\boldsymbol {\cal B}}_{i},
\eeq
\beq
\label{4.6}
\lambda=-\frac{1}{d}\sum_{i=1}^2\; \int d{\bf v}\, \frac{m_i}{2} V^2 \mathbf{V}\cdot {\boldsymbol {\cal C}}_{i}.
\eeq

As mentioned before, apart from the transport coefficients, an interesting quantity in the $\Delta$-model is the first-order contribution $T_i^{(1)}$ to the partial temperature $T_i$. Since this quantity is a scalar, it is coupled to the divergence of the flow velocity $\nabla \cdot \mathbf{U}$:
\beq
\label{4.7}
T_i^{(1)}=\varpi_i \nabla \cdot \mathbf{U},
\eeq
where
\beq
\label{4.8}
\varpi_i=\frac{m_i}{d n_i}\int  d\mathbf{v}\; V^2 \; \mathcal{E}_i(\mathbf{V}).
\eeq
Note that $T_i^{(1)}=0$ at low-density in the conventional IHS model, \cite{GD02,GMD06,GM07,GMV13a} although it is different from zero at finite densities. \cite{KS79b,GGG19b}

As usual, to obtain the explicit forms of the transport coefficients and the quantities $\varpi_i$, one has to resort to the leading terms in a Sonine polynomial expansion of the unknowns $\left\{ {\boldsymbol {\cal A}}_{i},{\sf {\cal B}}_{i},{\sf {\cal
C}}_{i}, {\sf {\cal D}}_{i,\beta\lambda}, \mathcal{E}_i\right\}$. This task is carried out below.

\subsection{Diffusion transport coefficients}

The lowest order Sonine polynomial approximations for ${\boldsymbol {\cal A}}_{i}$ , ${\boldsymbol {\cal B}}_{i}$, and ${\boldsymbol {\cal C}}_{i}$ are
\beq
\label{4.8.1}
\left(
\begin{array}{c}
{\boldsymbol {\cal A}}_{i}\\
{\boldsymbol {\cal B}}_{i}\\
{\boldsymbol {\cal C}}_{i}
\end{array}
\right)\longrightarrow f_{i,\text{M}}\mathbf{V}\left(
\begin{array}{c}
a_{i}\\
b_{i}\\
c_{i}
\end{array}
\right),
\eeq
where
\beq
\label{4.13}
f_{i,\text{M}}({\bf V})=n_i\left(\frac{m_i}{2\pi T_i^{(0)}}\right)^{d/2}\exp\left(-
\frac{m_i V^2}{2T_i^{(0)}}\right)
\eeq
is the Maxwellian distribution characterized by the partial temperature $T_i^{(0)}$. The coefficients $a_i$, $b_i$, and $c_i$ are related in this approximation to the transports coefficients $D$, $D_p$, and $D_T$ through Eqs.\ \eqref{3.3}--\eqref{3.5} as
\begin{equation}
\label{4.10}
a_1=-\frac{n_2T_2^{(0)}}{n_1T_1^{(0)}}a_2=-\frac{m_1m_2n}{\rho n_1T_1^{(0)}}D,
\end{equation}
\begin{equation}
\label{4.11}
b_1=-\frac{n_2T_2^{(0)}}{n_1T_1^{(0)}}b_2=-\frac{\rho}{p n_1T_1^{(0)}}D_p,
\end{equation}
\begin{equation}
\label{4.12}
c_1=-\frac{n_2T_2^{(0)}}{n_1T_1^{(0)}}c_2=-\frac{\rho}{T n_1T_1^{(0)}}D_T.
\end{equation}
In Eqs.\ \eqref{4.10}--\eqref{4.11}, $n_1T_1^{(0)}+n_2T_2^{(0)}=nT=p$.

The transport coefficients $D$, $D_p$, and $D_T$ are determined by substitution of Eq.\ \eqref{4.8.1} into the integral equations \eqref{2.30}--\eqref{2.32}. Next, multiplication of the above equations by $m_1 {\bf V}$ and integration over velocity yields
\begin{widetext}
\beq
\label{4.14}
\left[-\frac{1}{2}\zeta^{(0)}\left(1-\Delta^*\frac{\partial \ln D^*}{\partial \Delta^*}\right)+\nu_D\right]D=\frac{\rho}{m_1m_2 n}\left[\left(\frac{\partial}{\partial x_1}n_1 T_1^{(0)}\right)_{p,T}+\rho
\left(\frac{\partial \zeta ^{(0)}}{\partial x_{1}}\right)_{p,T}\left(D_p+D_T\right)\right],
\eeq
\beq
\label{4.15}
\left[\frac{1}{2}\zeta^{(0)}\left(1+\Delta^*\frac{\partial \ln D_p^*}{\partial \Delta^*}\right)-2\zeta^{(0)}+\nu_D\right]D_p=\frac{n_1T_1^{(0)}}{\rho }\left(1-\frac{m_1 nT}{\rho T_1^{(0)}}\right)+\zeta^{(0)} D_T,
\eeq
\beq
\label{4.16}
\Bigg[\frac{1}{2}\zeta^{(0)}\Delta^*\left(\frac{\partial \ln D_T^*}{\partial \Delta^*}+\frac{\partial \ln \zeta_0^*}{\partial \Delta^*}\right)+\nu_D\Bigg]D_T=-\frac{n_1T}{2\rho}\Delta^*\frac{\partial \gamma_1}{\partial \Delta^*}-\frac{\zeta^{(0)}}{2}\left(1+\Delta^*\frac{\partial \ln \zeta_0^*}{\partial \Delta^*}\right)D_p,
\eeq
where
\beq
\label{3.13}
\nu_D=-\frac{1}{dn_1T_1^{(0)}}\int d\mathbf{v}_1 m_1 \mathbf{V}_1 \cdot \left(J_{12}[f_{1,M}\mathbf{V}_1,f_{2}^{(0)}]-\frac{n_1 T_1^{(0)}}{n_2T_2^{(0)}}J_{12}[f_1^{(0)},f_{2,M}\mathbf{V}_2]\right).
\eeq
\end{widetext}
\vicente{In Eq.\ \eqref{4.14}, the derivatives with respect to $x_1$ at constant pressure and temperature are given by
\beq
\label{3.13.1}
\left(\frac{\partial}{\partial x_1}n_1 T_1^{(0)}\right)_{p,T}=p\Big(\gamma_1+x_1\frac{\partial \gamma_1}{\partial x_1}\Big),
\eeq
\beq
\label{3.13.2}
\left(\frac{\partial \zeta^{(0)}}{\partial x_1}\right)_{p,T}=\nu\Bigg[\left(\frac{\partial \zeta_0^*}{\partial x_1}\right)_{\gamma_1}+
\frac{\partial \zeta_0^*}{\partial \gamma_1}\frac{\partial \gamma_1}{\partial x_1}\Bigg].
\eeq
Note that $\gamma_2=(1-x_1\gamma_1)/(1-x_1)$ and hence, $\partial_{x_1}\gamma_2$ can be easily expressed in terms of $\partial_{x_1}\gamma_1$. The derivative $\partial_{x_1}\gamma_1$ is evaluated in the Appendix \ref{appA} in the steady state}. Moreover, in the definition of $\nu_D$ the self-collision terms arising from $J_{11}$ do not contribute since they conserve momentum of species $1$. In addition, upon obtaining Eqs.\ \eqref{4.14}--\eqref{4.16}, we have introduced the dimensionless transport coefficients
\beq
\label{4.16Dstar}
D^*=\frac{m_{1}m_{2}\nu}{\rho T}D,\quad
D_{p}^*=\frac{\rho \nu}{nT}D_{p},\quad D_T^*=\frac{\rho \nu}{nT}D_T,
\eeq
and use has been made of the relations
\beq
\label{4.17}
\left(T\frac{\partial}{\partial T}+p\frac{\partial}{\partial p}\right)D=\frac{D}{2}\left(1-\Delta^*\frac{\partial \ln D^*}{\partial \Delta^*}\right),
\eeq
\beq
\label{4.16.1}
\left(T\frac{\partial}{\partial T}+p\frac{\partial}{\partial p}\right)\frac{\rho}{p}D_p=-\frac{\rho}{2p}D_p\left(1+\Delta^*\frac{\partial \ln D_p^*}{\partial \Delta^*}\right),
\eeq
\beq
\label{4.18}
\left(T\frac{\partial}{\partial T}+p\frac{\partial}{\partial p}\right)\frac{\rho}{T}D_T=-\frac{\rho}{2T}D_T\left(1+\Delta^*\frac{\partial \ln D_T^*}{\partial \Delta^*}\right).
\eeq
An explicit expression for $\nu_D$ can be obtained when $f_i^{(0)}$ is replaced by its Maxwellian approximation $f_{i,M}$. The result is (see Appendix \ref{appC} for more details)
\beqa
\label{4.18.1}
\nu_D&=&\frac{2\pi^{(d-1)/2}}{d\Gamma\left(\frac{d}{2}\right)}\left(x_1\mu_{12}+x_2\mu_{21}\right)n\sigma_{12}^{d-1}v_\text{th}\nonumber\\
& & \times
\left[\left(\frac{\theta_1+\theta_2}{\theta_1\theta_2}\right)^{1/2}(1+\al_{12})+\sqrt{\pi}\Delta_{12}^*\right].\nonumber\\
\eeqa
In contrast to the conventional IHS model, the diffusion transport coefficients are obtained as the solutions of the set of coupled nonlinear differential equations \eqref{4.14}--\eqref{4.16}. When $\Delta_{11}^*=\Delta_{22}^*=\Delta_{12}^*=0$, Eqs.\ \eqref{4.14}--\eqref{4.16} are consistent with those obtained in the IHS model. \cite{GD02,GM07}

\subsection{Shear viscosity coefficient}

According to Eq.\ \eqref{4.3}, $\eta=\eta_1+\eta_2$ where
\beq
\label{4.19}
\eta_i=-\frac{1}{(d-1)(d+2)} \int d{\bf v}\, V_\lambda  V_\beta \mathcal{D}_{i,\lambda\beta}(\mathbf{V}).
\eeq
To get the coefficients $\eta_i$, one considers now the leading Sonine approximation for the function $\mathcal{D}_{i,\lambda\beta}(\mathbf{V})$:
\beq
\label{4.20}
\mathcal{D}_{i,\lambda\beta}(\mathbf{V}) \rightarrow -f_{i,\text{M}}(\mathbf{V}) R_{i,\lambda\beta}(\mathbf{V})\frac{\eta_i}{n_i T_i^{(0)2}},
\eeq
where
\beq
\label{4.21}
R_{i,\lambda\beta}(\mathbf{V})=m_i\left(V_\lambda V_\beta-\frac{1}{d}\delta_{\lambda\beta}V^2\right).
\eeq
As in the case of the diffusion coefficients, the partial contributions $\eta_i$ are obtained by substituting Eq.\ \eqref{4.20} into the integral equation \eqref{2.33}, multiplying it by $R_{i,\lambda\beta}$ and integrating over the velocity. After some algebra, one achieves the result
\beq
\label{4.22}
\sum_{j=1}^2\left[\tau_{ij}-\frac{1}{2}\zeta^{(0)}\left(1-\Delta^*\frac{\partial \ln \eta_i^*}{\partial \Delta^*}\right)\right]\eta_j=n_i T_i^{(0)},
\eeq
where $\eta_i^*=(\nu/n_i T)\eta_i$ and the collision frequencies $\tau_{ij}$ are defined as
\begin{widetext}
\beq
\label{4.23}
\tau_{ii}=-\frac{1}{(d-1)(d+2)}\frac{1}{n_i T_i^{(0)2}}\Bigg(\int d\mathbf{v}R_{i,\lambda\beta}
J_{ii}\left[f_i^{(0)},f_{i\text{M}}
R_{i,\lambda\beta}\right]+\sum_{j=1}^2\int d\mathbf{v}R_{i,\lambda\beta}J_{ij}
\left[f_{i,\text{M}}R_{i,\lambda\beta},f_j^{(0)}\right]\Bigg),
\eeq
\beq
\label{4.24}
\tau_{ij}=-\frac{1}{(d-1)(d+2)}\frac{1}{n_j T_j^{(0)2}}\int d\mathbf{v}R_{i,\lambda\beta}J_{ij}
\left[f_i^{(0)},f_{j,\text{M}}R_{j,\lambda\beta}\right], \quad (i\neq j).
\eeq
\end{widetext}
The expressions of the collision frequencies $\tau_{ii}$ and $\tau_{ij}$ in the Maxwellian approximation are given in the Appendix~\ref{appC}. Upon deriving Eq.\ \eqref{4.22} we have accounted for that $\eta_i \propto (p/\sqrt{T}) \eta_i^*(\Delta^*)$ and hence,
\beq
\label{4.24.1}
\left(T\frac{\partial}{\partial T}+p\frac{\partial}{\partial p}\right) \eta_i=\frac{1}{2}\eta_i\left(1-\Delta^*\frac{\partial \ln \eta_i^*}{\partial \Delta^*}\right).
\eeq
As in the case of the diffusion coefficients, when $\Delta_{11}^*=\Delta_{22}^*=\Delta_{12}^*=0$ Eq.\ \eqref{4.22} is consistent with the one derived for the shear viscosity in the IHS model. \cite{GD02,GM07}

\subsection{First-order contributions to the partial temperatures}

The first-order contributions to the partial temperatures are defined by Eqs.\ \eqref{4.7} and \eqref{4.8}. To determine $\varpi_i$, we consider the leading Sonine approximation to $\mathcal{E}_i(\mathbf{V})$ given by
\beq
\label{4.25}
\mathcal{E}_i(\mathbf{V})\rightarrow f_{i\text{M}}(\mathbf{V})W_i(\mathbf{V})\frac{\varpi_i}{T_i^{(0)}},\quad
W_i(\mathbf{V})=\frac{m_iV^2}{2T_i^{(0)}}-\frac{d}{2}.
\eeq
The coefficients $\varpi_i$ are coupled with the first-order contribution to the cooling rate $\zeta^{(1)}=\zeta_U \nabla \cdot \mathbf{U}$. The relationship between  $\varpi_i$ and  $\zeta_U$ can be made more explicit when one substitutes Eq.\ \eqref{4.25} into Eqs.\ \eqref{2.35} and \eqref{2.35.1}. After some algebra, one gets the result
\beq
\label{4.26a}
\zeta_U=\sum_{i=1}^2\; \xi_i \varpi_i,
\eeq
where $\xi_i=\xi_i^{(0)}+\xi_i^{(1)}$,
\begin{widetext}
\beq
\label{4.27}
\xi_i^{(0)}=\frac{3\pi^{(d-1)/2}}{2d\Gamma\left(\frac{d}{2}\right)}\frac{m_i v_\text{th}^3}{n T T_i^{(0)}}\sum_{j=1}^2 n_i n_j \sigma_{ij}^{d-1}
\mu_{ji}(1-\al_{ij}^2)\left(\theta_i+\theta_j\right)^{1/2}\theta_i^{-3/2}\theta_j^{-1/2}\Big],
\eeq
\beq
\label{4.27.1}
\xi_i^{(1)}=-\frac{4\pi^{(d-1)/2}}{d\Gamma\left(\frac{d}{2}\right)}\frac{v_\text{th}}{n T}\sum_{j=1}^2 n_i n_j \sigma_{ij}^{d-1}
\mu_{ji}\Delta_{ij}^*\Bigg\{\sqrt{\pi}\al_{ij}+\left(\theta_i+\theta_j\right)^{-1/2}\theta_i^{3/2}\theta_j^{-1/2}\Delta_{ij}^*
\Big[d-d\left(\theta_i+\theta_j\right)\theta_i^{-1}+(d+1)\theta_i \theta_j^{-1}\Big]\Bigg\}.
\eeq
The set of differential equations obeying the coefficients $\varpi_i$ are obtained by multiplying both sides of Eq.\ \eqref{2.34} by $m_i V^2$ and integrating over $\mathbf{V}$. After some algebra, one gets
\beq
\label{4.26b}
\sum_{j=1}^2\left[\omega_{ij}+\frac{1}{2}\zeta^{(0)}\left(1-\Delta^*\frac{\partial \ln \varpi_i^*}{\partial \Delta^*}\right)\delta_{ij}-\frac{1}{2}T\Delta^*\frac{\partial \gamma_i}{\partial \Delta^*}\xi_j+T_i^{(0)}\xi_j\right]\varpi_j=\frac{T}{d}\Delta^*\frac{\partial \gamma_i}{\partial \Delta^*},
\eeq
where $\varpi_i^*=(n\sigma_{12}^{d-1} v_\text{th}/T)\varpi_i$, and
\beq
\label{4.27.2}
\omega_{ii}=\frac{1}{dn_iT_i^{(0)}}\Bigg(\sum_{j=1}^2\int d \mathbf{v}\; m_iV^2J_{ij}\left[f_{i,\text{M}}W_i,f_j^{(0)}\right]
+\int d \mathbf{v}\; m_iV^2 J_{ii}\left[f_i^{(0)},f_{i,\text{M}}W_i\right]\Bigg) ,
\eeq
\beq
\label{4.28}
\omega_{ij}=\frac{1}{dn_iT_j^{(0)}}\int d \mathbf{v}\; m_iV^2 J_{ij}\left[f_i^{(0)},f_{j,\text{M}}W_j\right], \quad (i\neq j).
\eeq
\end{widetext}
The expressions of $\omega_{ii}$ and $\omega_{ij}$ in the Maxwellian approximation are given in the Appendix \ref{appC}. To achieve Eq.\ \eqref{4.26b}, we have taken into account the relation
\beq
\label{4.28.1}
\left(T\frac{\partial}{\partial T}+p\frac{\partial}{\partial p}\right) \varpi_i=\frac{1}{2}\varpi_i\left(1-\Delta^*\frac{\partial \ln \varpi_i^*}{\partial \Delta^*}\right).
\eeq
In the case that $\Delta_{11}^*=\Delta_{22}^*=\Delta_{12}^*=0$, Eq.\ \eqref{4.26b} yields $\varpi_i=0$ as expected.

\subsection{Heat flux transport coefficients}

The evaluation of the heat flux transport coefficients $D''$, $L$, and $\lambda$ is more involved since it requires going up to the second Sonine approximation. This calculation lies beyond the scope of the present paper. However, it is still  possible to obtain expressions for these coefficients when the first Sonine approximations \eqref{4.10}--\eqref{4.12} are considered for ${\boldsymbol {\cal A}}_i$, ${\boldsymbol {\cal B}}_i$, and ${\boldsymbol {\cal C}}_i$, respectively. In this approximation, one gets
\beq
\label{4.29}
D''=\frac{d+2}{2}\frac{n m_1m_2}{\rho T}\left(\frac{\gamma_1}{m_1}-\frac{\gamma_2}{m_2}\right)D,
\eeq
\beq
L=\frac{d+2}{2}\frac{\rho}{n}\left(\frac{\gamma_1}{m_1}-\frac{\gamma_2}{m_2}\right)D_p,
\eeq
\beq
\label{4.30}
\lambda=\frac{d+2}{2}\rho\left(\frac{\gamma_1}{m_1}-\frac{\gamma_2}{m_2}\right)D_T.
\eeq
According to Eqs.\ \eqref{4.29}--\eqref{4.30}, \vicente{for mechanically equivalent components (i.e., when $\sigma_1=\sigma_2$, $m_1=m_2$, $\al_{11}=\al_{22}=\al_{12}$, and $\Delta_{11}^*=\Delta_{22}^*=\Delta_{12}^*$), energy equipartition holds ($\gamma_1=\gamma_2$) \cite{BSG20} and so the first Sonine approximation to the heat transport coefficients vanishes ($D''=L=\lambda=0$)}. Hence, the forms \eqref{4.29}--\eqref{4.30} are not able to reproduce the monocomponent limit. Nevertheless, these expressions are consistent in the order of approximation used to obtain the mass flux transport coefficients and, consequently, can be employed to study the violation of Onsager's relations in Sec.~\ref{sec6}.

\begin{figure}[h!]
\centering
\includegraphics[width=0.4\textwidth]{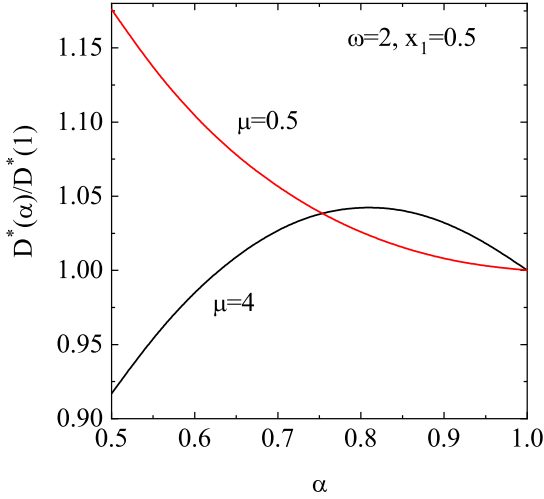}
\caption{Plot of the dimensionless diffusion transport coefficient $D^*(\alpha)/D^*(1)$ as a function of the (common) coefficient of restitution $\al_{ij}\equiv \alpha$ for $d=2$, $\omega= \sigma_1/\sigma_2=2$, $x_1=0.5$, and two different values of the mass ratio $\mu= m_1/m_2$: $\mu=0.5$ and $\mu=4$. Here, $D^*(1)$ refers to the value of the diffusion coefficient $D^*$ for elastic collisions ($\al=1$). }
\label{fig1}
\end{figure}
\begin{figure}[h!]
\centering
\includegraphics[width=0.4\textwidth]{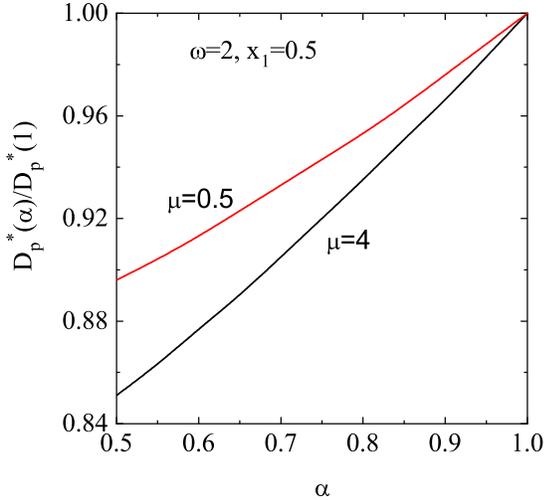}
\caption{The same as in Fig.\ \ref{fig1} but for the dimensionless pressure diffusion coefficient $D_p^*(\alpha)/D_p^*(1)$. Here, $D_p^*(1)$ refers to the value of the pressure diffusion coefficient $D_p^*$ for elastic collisions ($\al=1$). }
\label{fig2}
\end{figure}
\begin{figure}[h!]
\centering
\includegraphics[width=0.4\textwidth]{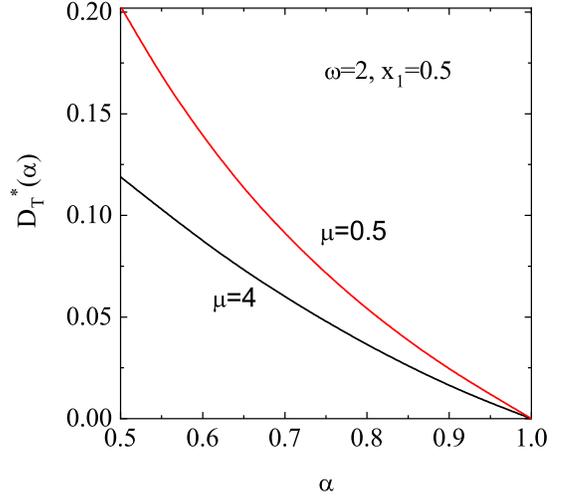}
\caption{The same as in Fig.\ \ref{fig1} but for the dimensionless thermal diffusion coefficient $D_T^*(\alpha)$.}
\label{fig3}
\end{figure}

\section{Transport coefficients at the stationary temperature}
\label{sec5}

As mentioned before, the determination of the Navier--Stokes transport coefficients as well as the first-order contributions to the partial temperatures requires to numerically solve intricate first-order differential equations in \vicente{the (dimensionless) $\Delta$-parameters $\Delta_{11}^*$, $\Delta_{22}^*$, and $\Delta_{12}^*$}. A detailed study of the dependence of the transport coefficients on the inelasticity was made in Ref.\ \onlinecite{BBMG15} for a monocomponent low-density granular gas \vicente{(i.e., when $\al_{ij}\equiv \al$ and $\Delta_{ij}^*\equiv \Delta^*$)}. Here, since we want to get analytical forms for those coefficients, the relevant state of a two-dimensional confined granular mixture with stationary temperature is considered. In this case ($\partial_t^{(0)}T=\partial_t^{(0)}p=0$), according to Eqs.\ \eqref{n1} and \eqref{2.5.0} $\zeta_1^{(0)}=\zeta_2^{(0)}=\zeta^{(0)}=0$ and hence, Eqs.\ \eqref{4.14}--\eqref{4.16}, \eqref{4.22}, and \eqref{4.26b} become linear algebraic equations.

In dimensionless forms, in the steady state, the diffusion transport coefficients $D_p^*$, $D_T^*$, and $D^*$ are given by
\beq
\label{5.2}
D_p^*=\frac{x_1}{\nu_D^*}\Bigg(\gamma_1-\frac{\mu}{x_2+\mu x_1}\Bigg),
\eeq
\beq
\label{5.2.1}
D_T^*=-\frac{x_1 \Delta_s^*\left(\frac{\partial \gamma_1}{\partial \Delta^*}\right)_s+\Delta_s^*\left(\frac{\partial \zeta_0^*}{\partial \Delta^*}\right)_s D_p^*}{2\nu_D^*+\Delta_s^*\left(\frac{\partial \zeta_0^*}{\partial \Delta^*}\right)_s},
\eeq
\beq
\label{5.1}
D^*=\frac{\gamma_1+x_1\left(\frac{\partial \gamma_1}{\partial x_1}\right)_s+\left(D_p^*+D_T^*\right)\left(\frac{\partial \zeta_0^*}{\partial x_1}\right)_s}{\nu_D^*},
\eeq
where $\nu_D^*=\nu_D/\nu$, $\mu=m_1/m_2$ is the mass ratio, and the subindex $s$ means that the quantities must be evaluated in the steady state. In addition, we recall that the operator $\Delta_s^* \partial_{\Delta^*}$ is defined by Eq.\ \eqref{2.11.1} and the derivatives $\partial_{x_1}\gamma_1$, $\partial_{x_1}\zeta_0^*$, $\partial_{\Delta^*}\gamma_1$, and $\partial_{\Delta^*}\zeta_0^*$ are determined in the Appendix \ref{appA}. Since $\mathbf{j}_1^{(1)}=-\mathbf{j}_2^{(1)}$ and $\nabla x_1=-\nabla x_2$, $D$ must be symmetric while $D_p$ and $D_T$ must be antisymmetric with respect to the change $1\leftrightarrow 2$. This can be easily verified from Eqs.\ \eqref{5.1}--\eqref{5.2.1} by noting that $x_1\gamma_1+x_2\gamma_2=1$ and $x_1 \partial \gamma_1/\partial \Delta^*=-x_2 \partial \gamma_2/\partial \Delta^*$.

The (reduced) shear viscosity coefficient $\eta^*=(\nu/p) \eta$ can be obtained from Eq.\ \eqref{4.22} as
\beq
\label{5.3}
\eta^*=\frac{\left(\tau_{22}^*-\tau_{21}^*\right)x_1\gamma_1+\left(\tau_{11}^*-\tau_{12}^*\right)x_2\gamma_2}
{\tau_{11}^*\tau_{22}^*-\tau_{12}^*\tau_{21}^*},
\eeq
where $\tau_{ij}^*=\tau_{ij}/\nu$. Finally, the dimensionless coefficient $\varpi_1^*=(n\sigma_{12}^{d-1}v_\text{th}/T)\varpi_1$
can be determined from Eq.~\eqref{4.26b} as
\beq
\label{5.4}
\varpi_1^*=\frac{1}{d}\frac{\Delta_s^*\left(\frac{\partial \gamma_1}
{\partial \Delta^*}\right)_s}{\Lambda_1^*},
\eeq
where
\beqa
\label{4.5.1}
\Lambda_1^*&=&\omega_{11}^*-\frac{x_1}{x_2} \omega_{12}^*-\Bigg[\frac{1}{2}T\Delta_s^*
\left(\frac{\partial \gamma_1}{\partial \Delta^*}\right)_s-\gamma_1\Bigg]\nonumber\\
& & \times\Bigg(\xi_1^*-\frac{x_1}{x_2} \xi_2^*\Bigg).
\eeqa
Here, $\omega_{ij}^*=\omega_{ij}/n\sigma_{12}^{d-1}v_\text{th}$ and $\xi_{i}^*=\xi_{i}/n\sigma_{12}^{d-1}Tv_\text{th}$.
The expression for $\varpi_2^*$ can be easily obtained from Eq.~\eqref{5.4} by making the changes $1\leftrightarrow 2$. The solution \eqref{5.4} must be indeed consistent with the requirement $x_1\varpi_1^*+x_2\varpi_2^*=0$. This is because $x_1\gamma_1+x_2\gamma_2=1$, $\partial \gamma_1/\partial \Delta^*=-(x_2/x_1)\partial \gamma_2/\partial \Delta^*$, and $\omega_{11}^*-(x_1/x_2)\omega_{12}^*+(\xi_1^*/x_1)=\omega_{22}^*-(x_2/x_1)\omega_{21}^*+(\xi_2^*/x_2)$.

\subsection{Some illustrative systems}

Now, we want to assess the dependence of the diffusion coefficients, the shear viscosity, and the first-order contributions to the partial temperature and the cooling rate on the parameter space of the system. For the sake of illustration, it is more convenient to plot the transport coefficients in their dimensionless forms. In the case of the diffusion coefficients, $D_p^*$, $D_T^*$, and  $D^*$ are given by
Eqs.\ \eqref{5.2}--\eqref{5.1}, respectively, the shear viscosity $\eta^*$ is defined by Eq. \ \eqref{5.3}, the coefficient $\zeta_U$ is given by Eqs.\ \eqref{4.26a}--\eqref{4.27.1}, and the expression of $\varpi_1^*$ is provided by Eq.\ \eqref{5.4}.  Note that the coefficient $\zeta_U$ is already a dimensionless quantity.

It is quite apparent that the above transport coefficients depend on the mass ratio $\mu=m_1/m_2$, the ratio of diameters $\omega= \sigma_1/\sigma_2$, the mole fraction $x_1$, the dimensionless parameters $\Delta_{ij}^*$, and the coefficients of restitution $\al_{ij}$. For the sake of simplicity, we will assume the case $\Delta_{11}^*=\Delta_{22}^*=\Delta_{12}^*\equiv \Delta^*$ and will take a common coefficient of restitution $\al_{11}=\al_{22}=\al_{12}\equiv \al$. Moreover, a two-dimensional system ($d=2$) will be considered. Since in the steady state, $\Delta^*$ is a function of $\al$, $x_1$, and the mechanical parameters of the mixture, then the parameter space is reduced to three quantities: $\left\{\mu, \omega, x_1\right\}$.

\begin{figure}[h!]
\centering
\includegraphics[width=0.4\textwidth]{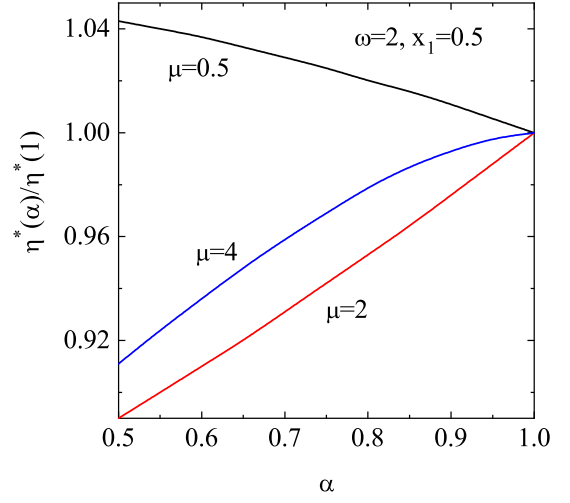}
\caption{Plot of the dimensionless shear viscosity coefficient $\eta^*(\alpha)/\eta^*(1)$ as a function of the (common) coefficient of restitution $\al_{ij}\equiv \alpha$ for $d=2$, $\omega=\sigma_1/\sigma_2=2$, $x_1=0.5$, and three different values of the mass ratio $\mu=m_1/m_2$: $\mu=0.5$, $\mu=2$, and $\mu=4$.  Here, $\eta^*(1)$ refers to the value of the shear viscosity coefficient $\eta^*$ for elastic collisions ($\al=1$).}
\label{fig4}
\end{figure}
\begin{figure}[h!]
\centering
\includegraphics[width=0.4\textwidth]{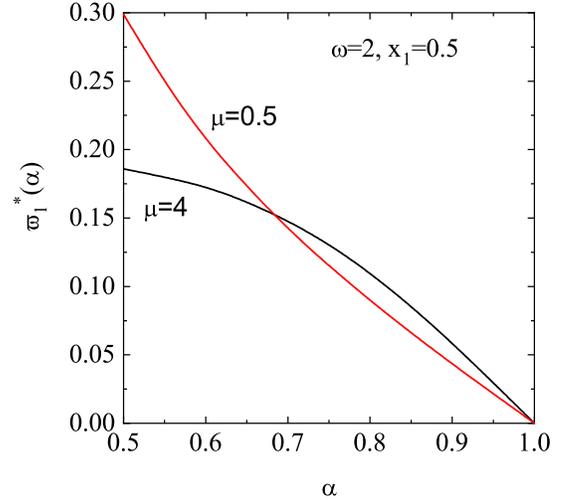}
\caption{Plot of the dimensionless coefficient $\varpi_1^*(\alpha)$ as a function of the (common) coefficient of restitution $\al_{ij}\equiv \alpha$ for $d=2$, $\omega=\sigma_1/\sigma_2=2$, $x_1=0.5$, and two different values of the mass ratio $\mu=m_1/m_2$: $\mu=0.5$ and $\mu=4$.}
\label{fig5}
\end{figure}
\begin{figure}[h!]
\centering
\includegraphics[width=0.4\textwidth]{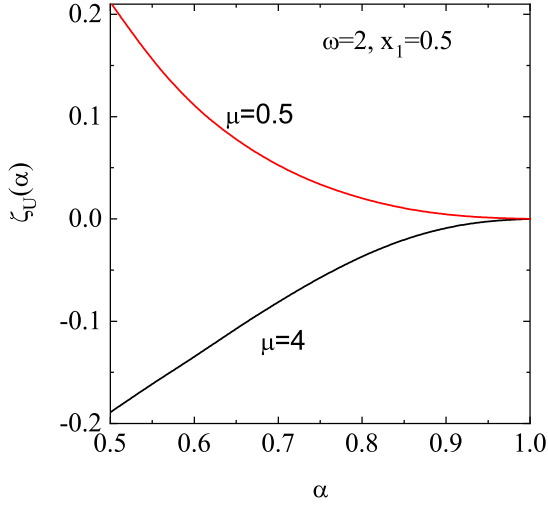}
\caption{The same as in Fig.\ \ref{fig5} for the dimensionless coefficient $\zeta_U(\alpha)$.}
\label{fig6}
\end{figure}

Given that the most interesting feature in a granular mixture is the dependence of the transport coefficients on inelasticity, we will normalize the values of the (dimensionless) transport coefficients with respect to their values for elastic collisions. Figures \ref{fig1}--\ref{fig3} show $D^*(\al)/D^*(1)$, $D_p^*(\al)/D_p^*(1)$, and $D_T^*(\al)$ as a function of $\al$ for $x_1=0.5$, $\omega=2$, and two different values of the mass ratio $\mu$: $\mu=0.5$ and $\mu=4$. For elastic collisions, $\Delta^*(1)=D_T^*(1)=0$, and
\beq
\label{5.8}
D_p^*(1)=\frac{x_1x_2}{\nu_D^*(1)}\frac{1-\mu}{1+(\mu-1)x_1}, \quad D^*(1)=\frac{1}{\nu_D^*(1)},
\eeq
where
\beq
\label{5.9}
\nu_D^*(1)=\frac{2\sqrt{2}\pi^{(d-1)/2}}{d\Gamma\left(\frac{d}{2}\right)}\frac{x_1\mu_{12}+x_2\mu_{21}}{\sqrt{\mu_{12}\mu_{21}}}.
\eeq
The thermal diffusion coefficient $D_T^*(\al)$ has not been normalized with its value in the elastic limit because this coefficient vanishes for elastic collisions \vicente{in the first Sonine approximation. \cite{FK72,KCLH87} Beyond the first Sonine solution, $D_T^*(1)\neq 0$ but its magnitude is very small}. We observe that in general the effect of inelasticity on mass transport is significant since the (reduced) coefficients $D^*$, $D_p^*$, and $D_T^*$ clearly deviate from their forms for elastic collisions. However, these deviations are in general smaller than those obtained in the conventional IHS model (see for instance, Figs.\ 1-3 of Ref.\ \onlinecite{GMD06}). This feature was already previously noted in the monodisperse case. \cite{GBS18} With respect to the dependence on the mass ratio, we see that there is a monotonic decrease of the diffusion coefficients with decreasing $\alpha$, except for the scaled diffusion $D^*(\al)/D^*(1)$ when $\mu=4$ since this coefficient exhibits in this case a non-monotonic dependence on inelasticity. Moreover, for sufficiently strong inelasticity (let's say, $\al \lesssim 0.75$), the impact of the coefficient of restitution on mass transport increases as the mass of the small particle increases. Figure \ref{fig3} also highlights that the thermal diffusion coefficient $D_T^*$ seems to be always positive, at least in the case $\Delta_{11}^*=\Delta_{22}^*=\Delta_{12}^*$ illustrated here. This feature contrasts to what happens in the conventional IHS model where this coefficient can be negative (see for instance, Fig.\ 3 of Ref.\ \onlinecite{GMD06}). The signature of the coefficient $D_T^*$ is relevant in problems such as granular segregation by thermal diffusion. \cite{JY02,BRM05,BRM06,SGNT06,G06,G08a,BEGS08,G09,G11}

The ratio $\eta^*(\al)/\eta^*(1)$ is plotted in Fig.\ \ref{fig4} as a function of the coefficient of restitution $\al$. As before, $\eta^*(1)$ refers to the shear viscosity coefficient for elastic collisions. As occurs in the case of the diffusion coefficients, we observe that the effect of inelasticity on the shear viscosity is less important than in the conventional IHS model (see for instance, Fig.\ 5 of Ref. \onlinecite{GM07}). In addition, depending on the mass ratio, the normalized shear viscosity decreases (increases) when decreasing $\al$ when the mass ratio is larger (smaller) than 1. The (reduced) coefficients $\varpi_1^*$ and $\zeta_U$ are plotted in Figs.\ \ref{fig5} and \ref{fig6}, respectively. We recall here that $\varpi_1^*=\zeta_U=0$ in the conventional IHS model. First, as expected, both coefficients vanish for elastic collisions. However, as inelasticity increases, the magnitude of both coefficients is not negligible. This means that $\varpi_1^*$ and $\zeta_U$ should be considered when one would solve the corresponding Navier--Stokes hydrodynamic equations. At a given value of $\mu$, while $\varpi_1^*$ increases with decreasing $\al$, $\zeta_U$ increases (decreases) with inelasticity when $\mu<1$ ($\mu>1$). This means that for the considered case of equal $\Delta$ and $\alpha$, $\zeta_U$ is always  positive (negative)
when the mass ratio is smaller (larger) than 1.

\section{Onsager's reciprocal relations}
\label{sec6}

Explicit knowledge of the Navier--Stokes transport coefficients for the $\Delta$-model of granular binary mixtures opens the possibility of several interesting applications. Among them, the quantification of the (possible) violation of the Onsager reciprocal relations is likely one of the most simple applications. This problem was already studied in the case of the conventional smooth IHS model. \cite{GMD06} Since time reversal symmetry is broken in granular gases (because collisions are inelastic), it is expected that Onsager's relations fail for finite degree of inelasticity. On the other hand, we think that the assessment of the expected violation and the influence of inelasticity on it is still an interesting problem.

\begin{figure}[!h]
\centering
\includegraphics[width=0.4\textwidth]{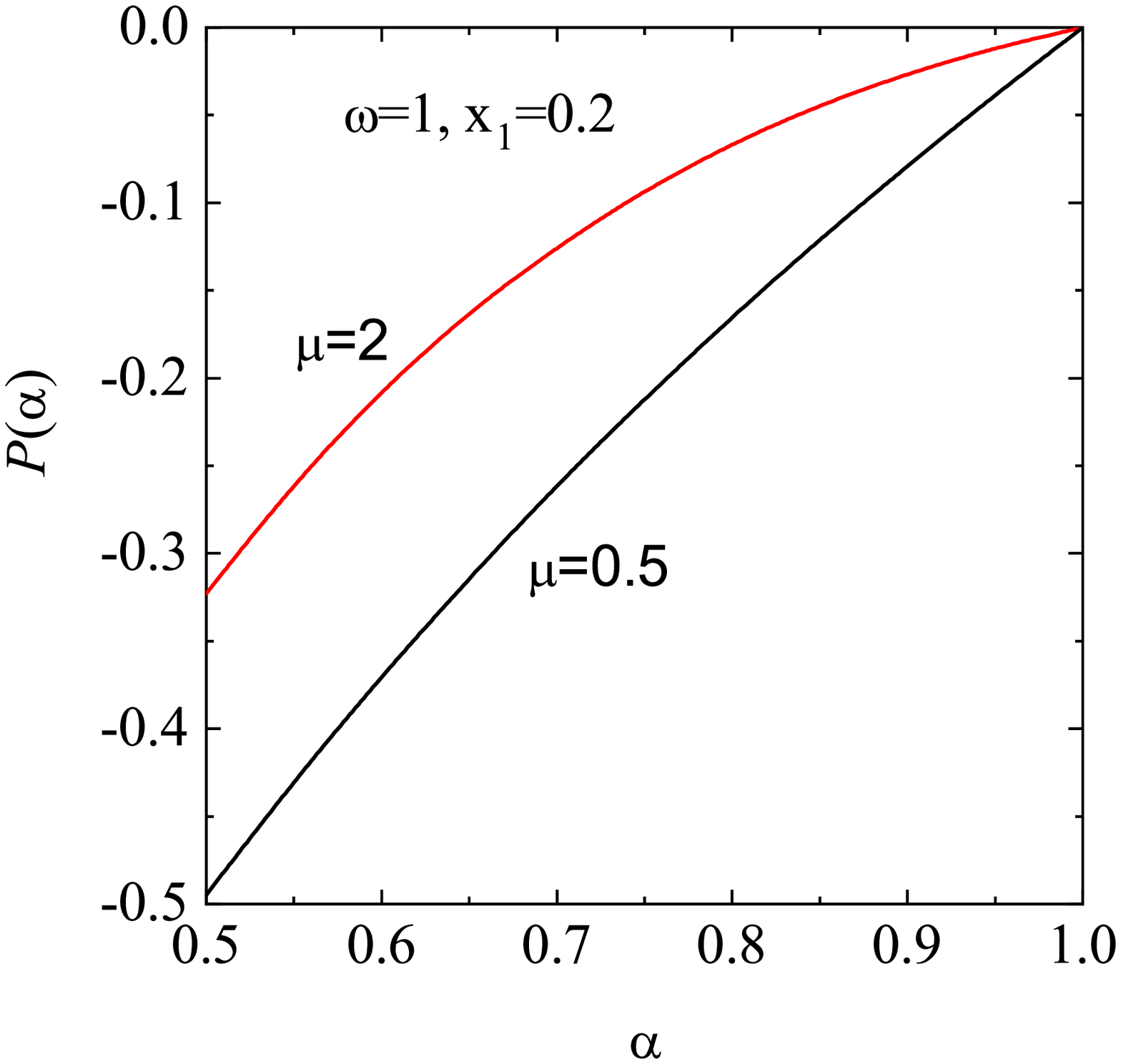}
\includegraphics[width=0.4\textwidth]{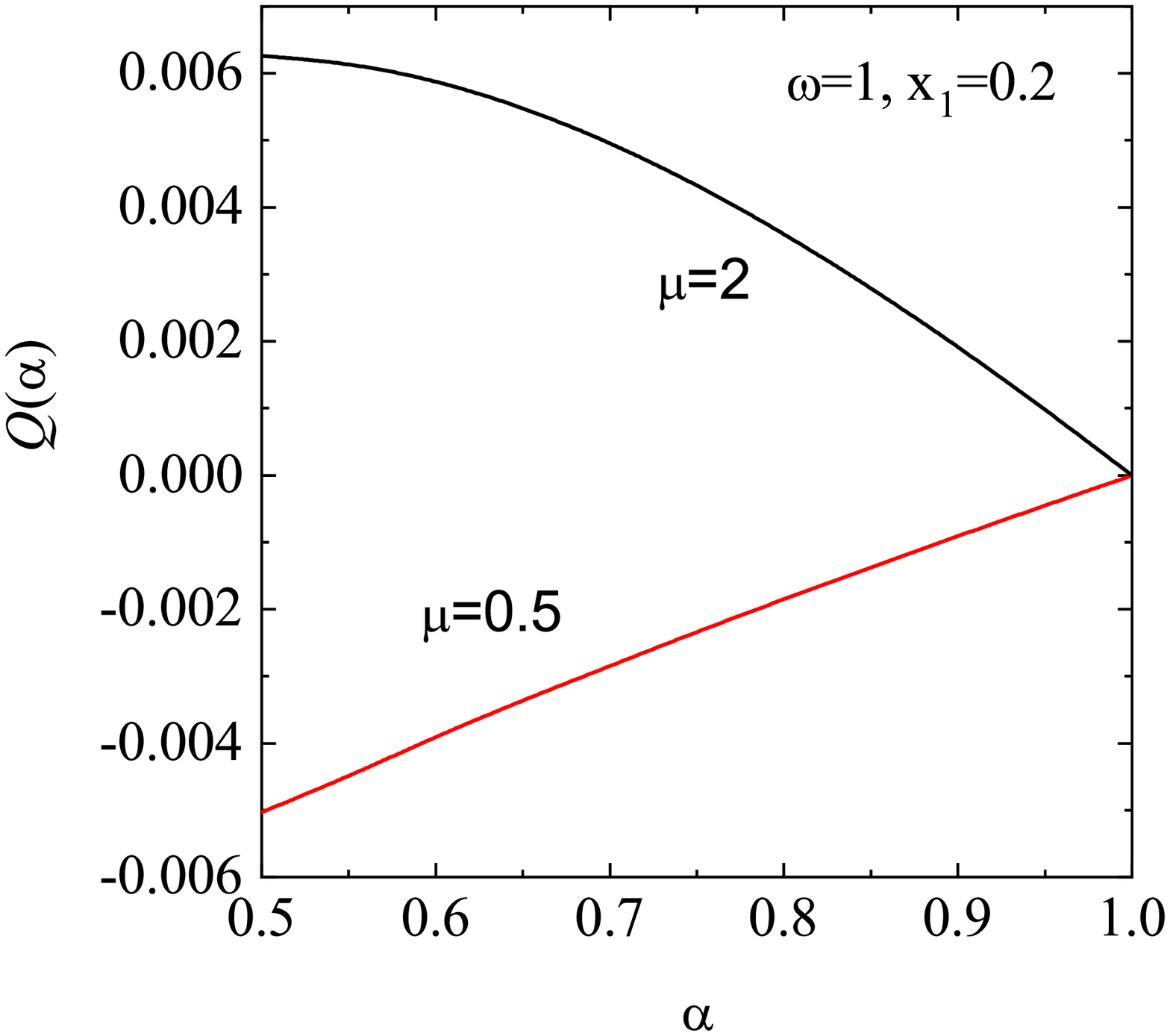}
\includegraphics[width=0.4\textwidth]{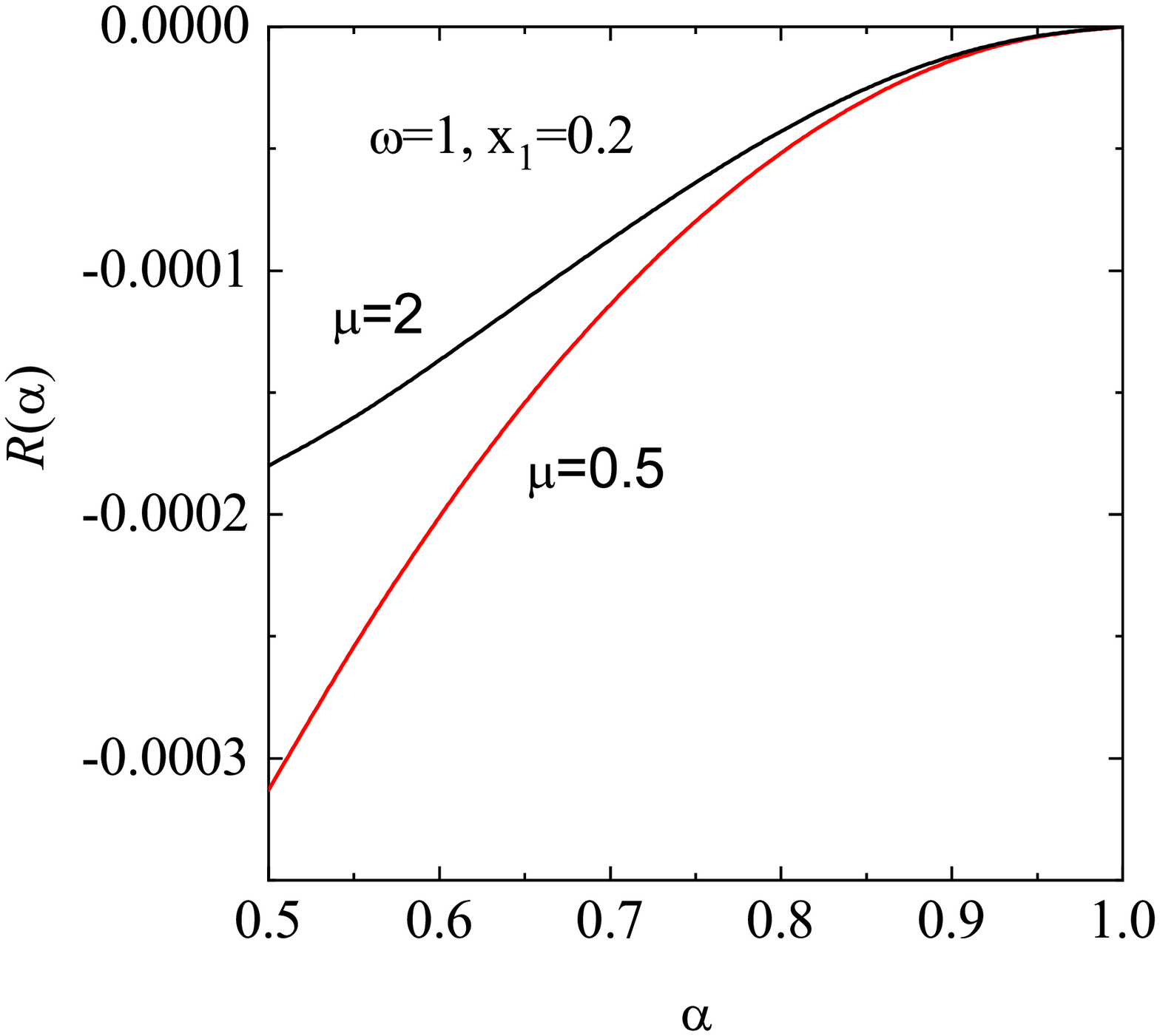}
\caption{Plot of the dimensionless coefficients $P(\al)$, $Q(\al)$, and $R(\al)$ versus the (common) coefficient of restitution $\al_{ij}\equiv \alpha$ for $d=2$, $\omega=\sigma_1/\sigma_2=1$, $x_1=0.2$, and two different values of the mass ratio $\mu=m_1/m_2$: $\mu=0.5$ and $\mu=2$.}
\label{fig7}
\end{figure}

In the usual language of the linear irreversible thermodynamics
for ordinary fluids, the constitutive equations for
the mass flux \eqref{3.2} and heat flow \eqref{4.2} of a binary mixture are written as \cite{GM84}
\begin{equation}
\mathbf{j}_{i}=-\sum_{i}L_{ij}\left( \frac{\nabla \mu _{j}}{T}\right)
_{T}-L_{iq}\frac{\nabla T}{T^{2}}-C_{p}\nabla p,  \label{7.1}
\end{equation}
\beqa
\mathbf{J}_{q}&=&\mathbf{q}^{(1)}-\frac{d+2}{2}T\frac{m_2-m_1}{m_1m_2}\mathbf{j}_1^{(1)}\nonumber\\
 &=&-L_{qq}\nabla T-\sum_{i}L_{qi}\left( \frac{\nabla \mu _{i}}{T}
\right)_{T}-C_{p}^{\prime}\nabla p,  \label{7.2}
\eeqa
where $\mu_{i}$ is the chemical potential of the component $i$. In the low-density regime,
\begin{equation}
\left( \frac{\nabla \mu_{i}/m_i}{T}\right)_{T}=\frac{1}{m_{i}}\nabla \ln
(x_{i}p).  \label{7.3}
\end{equation}
The coefficients $L_{ij}$ are the so-called Onsager phenomenological
coefficients. For ordinary or molecular fluids ($\al_{ij}=1$), Onsager showed that time
reversal invariance of the underlying microscopic equations of motion leads to the restrictions:
\begin{equation}
L_{ij}=L_{ji},\quad L_{iq}=L_{qi},\quad C_{p}=C_{p}^{\prime}=0.  \label{7.4}
\end{equation}
The first two symmetries are called reciprocal relations as they relate
transport coefficients for different processes. Thus, the coefficients $L_{qi}$ link the mass flux to the thermal gradient while the coefficients $L_{iq}$ link the heat flux to the gradient of the chemical potentials. The last two identities ($C_p=0$ and $C_p'=0$) are statements
that the pressure gradient does not appear in any of the fluxes even though it is admitted by symmetry. In particular, the condition $C_p'=0$ is important for monocomponent elastic gases since it yields Fourier's law for heat flux ($\mathbf{q}^{(1)}\propto \nabla T$) and hence, there is no any contribution proportional to the heat flux coming from the density gradient $\nabla n$. On the contrary, for the IHS model, $C_p'\neq0$ and there is an additional contribution to the heat flux proportional to $\nabla n$. \cite{SMR99,BDKS98,G19}

In order to analyze the violation of Onsager's relations, one has first to express the Onsager coefficients ($L_{ij}$, $L_{1q}$, $C_p$, $L_{qq}$, $L_{q1}$, and $C_p'$) in terms of both the diffusion ($D$, $D_p$, $D_T$) and heat flux ($D''$, $L$, $\lambda$) transport coefficients. To make this connection, since $\nabla x_{1}=-\nabla x_{2}$, then Eq.~\eqref{7.3} yields
\begin{equation}
\frac{(\nabla \mu_{1})_{T}-(\nabla \mu_{2})_{T}}{T}=\frac{n\rho }{\rho
_{1}\rho _{2}}\left[ \nabla x_{1}+\frac{n_{1}n_{2}}{n\rho }
(m_{2}-m_{1})\nabla \ln p\right] .  \label{7.5}
\end{equation}
The relationships between the Onsager coefficients $L_{ij}$ and those appearing in Eqs.\ \eqref{3.2} and \eqref{4.2} are
\begin{equation}
L_{11}=-L_{12}=-L_{21}=\frac{m_{1}m_{2}\rho _{1}\rho _{2}}{\rho ^{2}}D,\quad
L_{1q}=\rho T D_T,  \label{7.6}
\end{equation}
\begin{equation}
\label{7.7}
L_{q1}=-L_{q2}=\frac{T^{2}\rho _{1}\rho_{2}}{n\rho}D^{\prime \prime}-
\frac{d+2}{2}\frac{T\rho_{1}\rho_{2}}{\rho^{2}}(m_{2}-m_{1})D,
\eeq
\beq
\label{7.7.1}
L_{qq}=\lambda -\frac{d+2}{2}\rho \frac{m_{2}-m_{1}}{m_{1}m_{2}}D_T,
\end{equation}
\begin{equation}
C_{p}\equiv \frac{\rho}{p} D_{p}-\frac{\rho_{1}\rho_{2}}{p\rho^{2}}
(m_{2}-m_{1})D,  \label{7.8}
\end{equation}
\begin{equation}
C_{p}^{\prime}\equiv L-\frac{d+2}{2}T\frac{m_{2}-m_{1}}{m_{1}m_{2}}
C_{p}- \frac{n_{1}n_{2}}{np\rho }T^{2}(m_2-m_1)D^{\prime \prime}.
\label{7.9}
\end{equation}
As said before, since $D$ is symmetric under the change $1\leftrightarrow 2$, then Onsager's relation $L_{12}=L_{21}$ trivially holds. To analyze the other relations, we define the dimensionless function
\beqa
\label{7.13}
P(\al_{ij})&\equiv& \left(\frac{\gamma_1}{\mu_{12}}-\frac{\gamma_2}{\mu_{21}}-\frac{m_2^2-m_1^2}{m_1m_2}\right)D^*\nonumber\\
& & -
\frac{2}{d+2}\frac{(m_1+m_2)n\rho}{\rho_1\rho_2}D_T^*,
\eeqa
which vanishes when $L_{1q}=L_{q1}$. Similarly,
\beq
\label{7.14}
Q(\al_{ij})\equiv D_p^*-\frac{\rho_1\rho_2}{n \rho}\frac{m_2-m_1}{m_1m_2}D^*
\eeq
vanishes when $C_p=0$. Finally, when $C_p=0$ and $C_p'=0$, the function
\beq
\label{7.15}
R(\al_{ij})\equiv \Big[\mu_{21}(1-\gamma_1)-\mu_{12}(1-\gamma_2)\Big]Q(\al_{ij})
\eeq
equals zero.

For elastic collisions, $D_T^*=0$ and $D_p^*$ and $D^*$ are given by Eqs.\ \eqref{5.8} and \eqref{5.9}; this leads to $P(1)=Q(1)=R(1)=0$. Also, for mechanically equivalent particles with arbitrary $\alpha$, $D_p^*=D_T^*=0$ so that $P$, $Q$, and $R$ vanish. However, beyond these limit cases, Onsager's relations do not apply as
expected. At this macroscopic level the origin of this failure is due to the homogeneous reference state [which gives contributions to diffusion coefficients coming from the derivatives $(\partial_{x_1}\zeta_i^*)_s$ and $(\partial_{\Delta^*}\zeta_i^*)_s$] as well as the occurrence of different kinetic temperatures for both components ($\gamma_1\neq \gamma_2$). Figure \ref{fig7} shows the dependence of the quantities $P$, $Q$, and $R$ on the (common) coefficient of restitution $\alpha_{ij}\equiv \alpha$ for mass ratios $\mu =0.5$ and $2$. Violation of Onsager's relations is especially relevant in the case of the function $P$. We see that the departure from zero in the cases of $Q$ and $R$ is very small even for strong dissipation, implying that $C_p$ and $C_p'$ are small. In fact, the deviations from Onsager's relations are significantly much smaller than those found in the IHS model for the same systems (see Figs.\ 7, 8, and 9 of Ref.~\onlinecite{GMD06}).

\section{Discussion}
\label{sec7}

The present paper has been focused on the derivation of the Navier--Stokes hydrodynamic equations for a granular binary mixture of inelastic hard spheres in the context of the so-called $\Delta$-model. This model was originally proposed by Brito \emph{et al.} \cite{BRS13} to mimic the transfer of energy from the vertical to horizontal degrees of freedom in a quasi-two dimensional-dimensional geometry. \cite{OU98,PMEU04,MVPRKEU05,CCDHMRV08,PCR09,RPGRSCM11,GSVP11,CMS12,PGGSV12} Beyond its possible connection with this sort of experiments, the $\Delta$-model can be also seen as a nice and reliable alternative to the use of external driving forces to achieve a nonequilibrium steady state in a granular gas when collisional cooling is compensated for by the injected energy.

Although this collisional model has been widely employed by several groups \cite{BGMB13,BMGB14,BBMG15,BBGM16,SRB14,GBS18} for studying dynamic properties (kurtosis in homogeneous states, transport coefficients, $\dots$) for \emph{monocomponent} granular gases, we are not aware of any previous attempt for extending the previous efforts to the interesting case of \emph{multicomponent} granular gases, except for our previous analysis on the lack of equipartition in homogeneous binary mixtures.~\cite{BSG20} Needless to say, the determination of the complete set of Navier--Stokes transport coefficients of granular mixtures is challenging not only from a fundamental point of view, but also from a more practical view since granular matter is usually presented in nature as a collection of particles of different sizes, shapes, masses, and/or coefficients of restitution. Thus, given the high number of parameters involved in the description of these systems, one usually considers simple systems to gain some insight. For this reason, the low-density regime has been considered here  where the set of Boltzmann kinetic equations for the mixture provides an accurate framework for analyzing transport properties.

As in previous works on granular mixtures, \cite{GD02,GM07} the constitutive equations for the mass, momentum, and heat fluxes and the cooling rate have been obtained by solving the Boltzmann equation by means of the Chapman--Enskog expansion up to first order in the spatial gradients. The constitutive equation of the mass flux is given by Eq.\ \eqref{3.2} where the diffusion transport coefficients $D$, $D_p$, and $D_T$ are defined by Eqs.\  \eqref{3.3}--\eqref{3.5}, respectively. The pressure tensor is given by Eq.\ \eqref{4.1} where the shear viscosity coefficient $\eta$ is defined by Eq.\ \eqref{4.3}. The heat flux is given by Eq.\ \eqref{4.2} where the heat flux transport coefficients $D''$, $L$, and $\lambda$ are defined by Eqs.\ \eqref{4.4}--\eqref{4.6}, respectively. Apart from the above transport coefficients, there are non-vanishing first-order contributions $T_i^{(1)}$ to the partial temperatures and the cooling rate $\zeta_U$; they are given by Eqs.\ \eqref{4.7}--\eqref{4.8} and \eqref{4.26a}, respectively. This latter result contrasts with the one previously obtained in the conventional IHS model where $T_i^{(1)}=\zeta_U=0$ at low-density. \cite{GD02,GM07}

Explicit forms of the above transport coefficients have been obtained by considering the leading terms in a Sonine polynomial expansion of the first-order distribution function. This is the usual way for determining these quantities for elastic \cite{CC70} and inelastic \cite{G19} gases. On the other hand, given that the evaluation of the heat flux transport coefficients requires to consider the second Sonine approximation, here we have addressed the computation of the diffusion transport coefficients, the shear viscosity, and the quantities $T_i^{(1)}$, and $\zeta_U$. In the general time-dependent problem, the differential equations obeying the diffusion coefficients are given by Eqs.\ \eqref{4.14}--\eqref{4.16}, the viscosity obeys Eq.\ \eqref{4.22}, and the first-order contributions to the partial temperatures are given in terms of the solution of Eq.\ \eqref{4.26b}. The numerical solutions of the above differential equations provide the dependence of the transport coefficients on the parameter space of the system.

Considering that the zeroth-order distribution functions $f_i^{(0)}$ are involved in the evaluation of transport coefficients, one has to characterize first these distributions before computing transport. This study has been previously made in Ref.\ \onlinecite{BSG20} where it has been shown that $f_i^{(0)}$ has the scaling form \eqref{2.7} and the temperature ratios $\gamma_i$ obey the set of coupled equations \eqref{2.12.1}. In the steady state ($\zeta_1^{(0)}=\zeta_2^{(0)}=\zeta^{(0)}=0$), the dependence of $\gamma_i$ on the parameters of the mixture has been explicitly obtained by approximating the scaled distributions $\varphi_i$ by Maxwellian distributions at $T_i^{(0)}$ [see Eq.\ \eqref{b2}]. In spite of this  approximation, the theoretical results for the temperature ratio compare quite well with computer simulations, specially for low-density mixtures. \cite{BSG20}

Once the reference state is well characterized, the forms of the above set of transport coefficients under steady state conditions have been explicitly determined; their expressions have been displayed in Eqs.\ \eqref{5.1}--\eqref{5.2.1}, \eqref{5.3}, and \eqref{5.4}. It is apparent that in general they exhibit a quite complex dependence on the coefficients of restitution and the remaining parameters of the mixture. An interesting point is that their expressions not only depend on the hydrodynamic fields in the steady state, but, in addition, there are contributions to them coming from the derivatives of both the temperature ratio and the cooling rate in the vicinity of the steady state. These contributions measure the distance of the perturbed state from the steady reference state. This sort of contributions are also present in the case of \emph{driven} granular mixtures \cite{KG13,KG18,GGKG20} but they absent in the conventional IHS model for \emph{undriven} granular mixtures. \cite{GD02,GM07,GMV13a}

To illustrate the dependence of transport on the (common) coefficient of restitution $\al$, the simplest case $\Delta_{11}=\Delta_{22}=\Delta_{12}$ has been considered. Figures \ref{fig1}--\ref{fig6} highlights the significant effect of inelasticity on mass and momentum transport as well as on the partial temperatures. However, at a more quantitative level, the influence of $\al$ on the transport coefficients is smaller than that of the conventional collisional model. \cite{GD02,GM07,GMV13a}

It is well known that the hydrodynamic equations for granular mixtures are the same as for a molecular mixture, except for (i) a sink in the energy equation due to granular cooling, and (ii) additional transport coefficients in the
mass and heat flux constitutive equations. These additional contributions arise because Onsager reciprocal relations \cite{GM84} among various transport coefficients are expected to fail. Here, as an application of the previous results, we have assessed in Sec.\ \ref{sec6} the violation of Onsager's relations as inelasticity increases. Notably,  as Fig.~\ref{fig7} shows, the failure of these relations are much smaller than those reported in Ref.\ \onlinecite{GMD06} for the same systems.

As mentioned in Sec.\ \ref{sec1}, the $\Delta$-model was originally proposed to mimic the quasi-two-dimensional geometry of a confined granular gas .\cite{OU98,PMEU04,MVPRKEU05,CCDHMRV08,PCR09,RPGRSCM11,CMS12} However, although this collisional model is able to describe quite well \cite{BBGM17} the homogeneous evolution observed in the experiments, it fails to predict the existence of non-equilibrium phase transitions. For this reason,  modified Boltzmann kinetic equations for this special quasi-two-dimensional confinement have been proposed \cite{BMG16,RSG18} and the inhomogeneous cooling state has been widely analyzed. \cite{BGM19,MGB19a,MGB19}

On the other hand, there are still some interesting open problems in the $\Delta$-model. Among them, the evaluation of the heat flux transport coefficients by considering the second Sonine approximation is a challenging issue. The knowledge of these coefficients will allow us to perform a \emph{linear} stability analysis of the homogeneous steady state. A previous study for monocomponent gases \cite{BBGM16} has shown the stability of the homogeneous state for small spatial perturbations and it is relevant to determine if the stability of the homogeneous steady state is extended for granular mixtures. Moreover, given that most of the theoretical results found here have been obtained under certain approximations (Maxwellian distribution functions for the reference states $f_i^{(0)}$, leading Sonine approximations for the diffusion transport coefficients and the shear viscosity), a natural project is to undertake simulations to gauge the reliability of the present results. In particular, we plan to carry out computer simulations to measure the tracer diffusion coefficient (namely, a binary mixture where the concentration of one of the components is negligible) in a similar way as those simulations performed in the conventional IHS model. \cite{BRCG00,GV09,GV12} An additional challenging problem is to measure the Navier--Stokes shear viscosity by studying the decay of a small perturbation to the transversal component of the velocity field. \cite{BRC99} Finally, another possible project for the next future is to analyze thermal diffusion segregation. \cite{G08a,BEGS08,G09,G11} Works on the above issues will be developed in the near future.

\acknowledgments

The research of V.G.~has been supported by the Spanish Ministerio de Econom{\'i}a y Competitividad through Grant No.~FIS2016-76359-P and by the Junta de Extremadura (Spain) Grant Nos.~IB16013 and GR18079, partially financed by ``Fondo Europeo de Desarrollo Regional'' funds. The work of R.B.~has been supported by the Spanish Ministerio de Econom{\'i}a y Competitividad through Grant No.~FIS2017-83709-R. The research of R.S.~has
been supported by the Fondecyt Grant No.~1180791 of ANID (Chile).


\begin{widetext}

\appendix
\section{Derivatives of the temperature ratio in the vicinity of the steady state}
\label{appA}

In this Appendix, the derivatives of the temperature ratio $\gamma_1=T_1^{(0)}/T$ with respect to $\Delta_{11}^*$, $\Delta_{22}^*$, $\Delta_{12}^*$,  and $x_1$ in the vicinity of the steady state are evaluated. The derivatives of $\gamma_2=T_2^{(0)}/T$ can be easily obtained by taking into account the relation $\gamma_2=x_2^{-1}(1-x_1\gamma_1)$.

Let us consider first the derivative of $\gamma_1$ with respect to $\Delta_{11}^*$. To determine it, we consider Eq.\ \eqref{2.12.2} for $i=1$:
\beq
\label{b1}
\frac{1}{2}\zeta_0^*\Delta^*\frac{\partial \gamma_1}{\partial \Delta^*}=\gamma_1\left(\zeta_0^*-\zeta_1^*\right).
\eeq
The (reduced) partial cooling rates $\zeta_1^*$ are functionals of the (scaled) distributions $\varphi_1$ and $\varphi_2$ whose exact forms are not known. \vicente{However, recent results \cite{BSG20} have clearly shown that the quantities $\zeta_i^*$ can be well estimated by using Maxwellian distributions at different temperatures. This is justified by the good agreement found between theory (based on the above assumption) and simulations for the global temperature and the temperature ratio even for strong dissipation and/or disparate values of the mass and size ratios. Thus, to estimate $\zeta_1^*$ we consider the approximation}
\beq
\label{b2}
\varphi_i(\mathbf{c})\to \pi^{-d/2}\theta_i^{d/2}e^{-\theta_i c^2}.
\eeq
In the Maxwellian approximation \eqref{b2}, the cooling rate $\zeta_1^*$ is \cite{BSG20}
\beqa
\label{b3}
\zeta_1^*&=&\frac{\sqrt{2}\pi^{(d-1)/2}}{d\Gamma\left(\frac{d}{2}\right)}x_1 \left(\frac{\sigma_{1}}{\sigma_{12}}\right)^{d-1}\theta_1^{-1/2}\left(1-\al_{11}^2-2\Delta_{11}^{*2}\theta_1-\sqrt{2\pi \theta_1}
\Delta_{11}^*\al_{11}\right)\nonumber\\
& & +\frac{4\pi^{(d-1)/2}}{d\Gamma\left(\frac{d}{2}\right)}
x_2\mu_{21}(1+\al_{12})\theta_1^{-1/2}
\left(1+\theta_{12}\right)^{1/2}
\left[1-\frac{1}{2}\mu_{21}(1+\alpha_{12})(1+\theta_{12}) \right]\nonumber\\
& &-\frac{4\pi^{d/2}}{d\Gamma\left(\frac{d}{2}\right)}x_2\mu_{21}\Delta_{12}^*\left[
\frac{2\mu_{21}\Delta_{12}^*}{\sqrt{\pi}}\theta_1^{1/2}\left(1+\theta_{12}\right)^{1/2}
-1+\mu_{21}(1+\al_{12})\left(1+\theta_{12}\right)\right],
\eeqa
where $\theta_{ij}=\theta_i/\theta_j$. The expression of $\zeta_2^*$ can be obtained from Eq.\ \eqref{b3} by making the change $1\leftrightarrow 2$. The total cooling rate is given by $\zeta_0^*=x_1\gamma_1 \zeta_1^*+x_2\gamma_2 \zeta_2^*=\zeta_2^*+x_1\gamma_1(\zeta_1^*-\zeta_2^*)$, where use has been made of the relation $x_1\gamma_1+x_2\gamma_2=1$. In the steady state, $\zeta_1^*=\zeta_2^*=\zeta_0^*=0$ and hence, Eq.\ \eqref{b1} is trivially verified. To determine the derivative $\partial \gamma_1/\partial \Delta_{11}^*$ at the steady state, we take first the derivative with respect to $\Delta_{11}^*$ in both sides of Eq.\ \eqref{b1} and then take the steady-state limit. After some algebra, one gets the result
\beq
\label{b4}
\frac{1}{2}\Bigg[x_1 \gamma_{1s}\left(\frac{\partial \zeta_1^*}{\partial \Delta_{11}^*}\right)_s+x_2 \gamma_{2s}\left(\frac{\partial \zeta_2^*}{\partial \Delta_{11}^*}\right)_s\Bigg]\Bigg[\Delta_{11}^* \left(\frac{\partial \gamma_1}{\partial \Delta_{11}^*}\right)_s
+\Delta_{22}^* \left(\frac{\partial \gamma_1}{\partial \Delta_{22}^*}\right)_s+\Delta_{12}^* \left(\frac{\partial \gamma_1}{\partial \Delta_{12}^*}\right)_s-2\gamma_{1s}\Bigg]=-\gamma_{1s}\left(\frac{\partial \zeta_1^*}{\partial \Delta_{11}^*}\right)_s,
\eeq
where the subscript $s$ means that all the quantities are evaluated in the steady state. According to Eq.\ \eqref{b3}, $\zeta_i^*$ depends on $\Delta_{11}^*$ explicitly and also through its dependence on $\gamma_1$. Thus,
\beq
\label{b5}
\left(\frac{\partial \zeta_i^*}{\partial \Delta_{11}^*}\right)_s=\left(\frac{\partial \zeta_i^*}{\partial \Delta_{11}^*}\right)_{\gamma_1}+\left(\frac{\partial \zeta_i^*}{\partial \gamma_1}\right)\left(\frac{\partial \gamma_1}{\partial \Delta_{11}^*}\right)_s.
\eeq
Equations defining the remaining derivatives $\partial \gamma_1/\partial \Delta_{22}^*$ and $\partial \gamma_1/\partial \Delta_{12}^*$ can be easily obtained by following identical steps. They are given
\beq
\label{b6}
\frac{1}{2}\Bigg[x_1 \gamma_{1s}\left(\frac{\partial \zeta_1^*}{\partial \Delta_{22}^*}\right)_s+x_2 \gamma_{2s}\left(\frac{\partial \zeta_2^*}{\partial \Delta_{22}^*}\right)_s\Bigg]\Bigg[\Delta_{11}^* \left(\frac{\partial \gamma_1}{\partial \Delta_{11}^*}\right)_s
+\Delta_{22}^* \left(\frac{\partial \gamma_1}{\partial \Delta_{22}^*}\right)_s+\Delta_{12}^* \left(\frac{\partial \gamma_1}{\partial \Delta_{12}^*}\right)_s-2\gamma_{1s}\Bigg]=-\gamma_{1s}\left(\frac{\partial \zeta_1^*}{\partial \Delta_{22}^*}\right)_s,
\eeq
\beq
\label{b7}
\frac{1}{2}\Bigg[x_1 \gamma_{1s}\left(\frac{\partial \zeta_1^*}{\partial \Delta_{12}^*}\right)_s+x_2 \gamma_{2s}\left(\frac{\partial \zeta_2^*}{\partial \Delta_{12}^*}\right)_s\Bigg]\Bigg[\Delta_{11}^* \left(\frac{\partial \gamma_1}{\partial \Delta_{11}^*}\right)_s
+\Delta_{22}^* \left(\frac{\partial \gamma_1}{\partial \Delta_{22}^*}\right)_s+\Delta_{12}^* \left(\frac{\partial \gamma_1}{\partial \Delta_{12}^*}\right)_s-2\gamma_{1s}\Bigg]=-\gamma_{1s}\left(\frac{\partial \zeta_1^*}{\partial \Delta_{12}^*}\right)_s.
\eeq
The solution to the set of algebraic nonlinear equations \eqref{b4}, \eqref{b6}, and \eqref{b7} provides the derivatives $\partial \gamma_1/\partial \Delta_{11}^*$, $\partial \gamma_1/\partial \Delta_{22}^*$, and $\partial \gamma_1/\partial \Delta_{12}^*$ at the steady state. A more simple expression can be obtained for the particular case $\Delta_{11}^*=\Delta_{22}^*=\Delta_{12}^*=\Delta^*$. In this situation, it is easy to see that the derivative $\Lambda_{\gamma_1,\Delta}\equiv (\partial \gamma_1/\partial \Delta^*)_s$ obeys the following quadratic equation:
\beq
\label{b8}
B \Delta^* \Lambda_{\gamma_1,\Delta}^2+\left(A \Delta^*-2B\gamma_{1s}+\gamma_{1s}\frac{\partial \zeta_1^*}{\partial \gamma_1}\right)\Lambda_{\gamma_1,\Delta}-2A \gamma_{1s}+\gamma_{1s}\frac{\partial \zeta_1^*}{\partial \gamma_1}=0,
\eeq
where
\beq
\label{b9}
A=\frac{1}{2}\Bigg[x_1 \gamma_{1s}\left(\frac{\partial \zeta_1^*}{\partial \Delta^*}\right)_{\gamma_1}+x_2 \gamma_{2s}\left(\frac{\partial \zeta_2^*}{\partial \Delta^*}\right)_{\gamma_1}\Bigg], \quad B=\frac{1}{2}\Bigg(x_1 \gamma_{1s}\frac{\partial \zeta_1^*}{\partial \gamma_1}+x_2 \gamma_{2s}\frac{\partial \zeta_2^*}{\partial \gamma_1}\Bigg).
\eeq
An analysis of the solutions to Eq. \ \eqref{b8} shows that in general
one of the roots leads to unphysical behavior of the diffusion coefficients for nearly elastic spheres. We take the other root as
the physical root of the quadratic equation.

The derivative $\Lambda_{\gamma_1,x_1}\equiv (\partial \gamma_1/\partial x_1)_s$ at the steady state can be determined in a similar way by taking first the derivative with respect to $x_1$ in both sides of Eq.\ \eqref{b1} and then taking the steady-state limit. The result is
\beq
\label{b10}
\Lambda_{\gamma_1,x_1}=-\frac{\gamma_{1s}\frac{\partial \zeta_1^*}{\partial x_1}+\frac{1}{2}\left(x_1\gamma_{1s}\frac{\partial \zeta_1^*}{\partial x_1}+x_2\gamma_{2s}\frac{\partial \zeta_2^*}{\partial x_1}\right)\Bigg[\Delta_{11}^* \left(\frac{\partial \gamma_1}{\partial \Delta_{11}^*}\right)_s
+\Delta_{22}^* \left(\frac{\partial \gamma_1}{\partial \Delta_{22}^*}\right)_s+\Delta_{12}^* \left(\frac{\partial \gamma_1}{\partial \Delta_{12}^*}\right)_s-2\gamma_{1s}\Bigg]}
{\gamma_{1s}\frac{\partial \zeta_1^*}{\partial \gamma_1}+\frac{1}{2}\left(x_1\gamma_{1s}\frac{\partial \zeta_1^*}{\partial \gamma_1}+x_2\gamma_{2s}\frac{\partial \zeta_2^*}{\partial \gamma_1}\right)\Bigg[\Delta_{11}^* \left(\frac{\partial \gamma_1}{\partial \Delta_{11}^*}\right)_s
+\Delta_{22}^* \left(\frac{\partial \gamma_1}{\partial \Delta_{22}^*}\right)_s+\Delta_{12}^* \left(\frac{\partial \gamma_1}{\partial \Delta_{12}^*}\right)_s-2\gamma_{1s}\Bigg]}.
\eeq
In Eq.\ \eqref{b10}, it is understood that the derivative $\partial_{x_1}\zeta_i^*$ is taken at $\gamma_1\equiv \text{const.}$

\section{Some technical details on the first-order Chapman--Enskog solution}
\label{appB}

To first order in the gradients, the equation for $f_{i}^{(1)}$ is
\begin{equation}
\left( \partial_{t}^{(0)}+{\cal L}_{i}\right) f_{i}^{(1)}+{\cal M}
_{i}f_{j}^{(1)}=-\left(D_{t}^{(1)}+{\bf V}\cdot \nabla \right)f_{i}^{(0)},  \label{2.13}
\end{equation}
where $D_{t}^{(1)}=\partial_{t}^{(1)}+\mathbf{U}\cdot \nabla$. In Eq.\ \eqref{2.13} it is understood that $i\neq j$ and the linear operators ${\cal L}_{i}$ and ${\cal M}_{i}$ are
\begin{equation}
{\cal L}_{i}f_{i}^{(1)}=-\left(
J_{ii}[f_{i}^{(0)},f_{i}^{(1)}]+J_{ii}[f_{i}^{(1)},f_{i}^{(0)}]+
J_{ij}[f_{i}^{(1)},f_{j}^{(0)}]\right),
\label{2.14}
\end{equation}
\begin{equation}
{\cal M}_{i}f_{j}^{(1)}=-J_{ij}[f_{i}^{(0)},f_{j}^{(1)}].  \label{2.15}
\end{equation}
The action of the time derivatives $D_{t}^{(1)}$ on the hydrodynamic
fields is
\begin{equation}
D_{t}^{(1)}x_{1}=0,  \quad D_{t}^{(1)}p=-\frac{d+2}{d}p\nabla \cdot {\bf U}-p\zeta^{(1)}, \quad D_{t}^{(1)}T=-\frac{2}{d}T\nabla \cdot {\bf U}-T\zeta^{(1)}, \quad D_{t}^{(1)}{\bf U}=-\rho^{-1}\nabla p, \label{2.16}
\end{equation}
where use has been made of the results ${\bf j}_{i}^{(0)}={\bf q}^{(0)}=\textbf{0}$. Here, $\zeta^{(1)}$ is the first-order contribution to the cooling rate. The right-hand side of Eq.\ (\ref{2.13}) can be explicitly evaluated from the relations \eqref{2.16} with the result
\begin{eqnarray}
&&-\left(D_{t}^{(1)}+{\bf V}\cdot \nabla \right) f_{i}^{(0)}
=-\left( \frac{\partial }{\partial x_{1}}f_{i}^{(0)}\right)_{p,T}{\bf V}
\cdot \nabla x_{1}-\left( \frac{f_{i}^{(0)}}{p}{\bf V}+\rho^{-1}
\frac{\partial f_{i}^{(0)}}{\partial {\bf V}}\right) \cdot \nabla p
\nonumber \\
&&-\frac{\partial f_{i}^{(0)}}{\partial T}{\bf V}\cdot \nabla T+
V_\beta \frac{\partial f_i^{(0)}}{\partial V_\lambda}\frac{1}{2}\left(\frac{\partial U_\beta}{\partial r_\lambda}+
\frac{\partial U_\lambda}{\partial r_\beta}-\frac{2}{d}\delta_{\beta\lambda}\nabla\cdot \mathbf{U}\right)
\nonumber\\
& & +\left(\frac{d+2}{d}p\frac{\partial f_{i}^{(0)}}{\partial p}+\frac{2}{d}T\frac{\partial f_{i}^{(0)}}{\partial T}+\frac{1}{d}\mathbf{V}\cdot \frac{\partial f_{i}^{(0)}}{\partial {\bf V}}\right)\nabla \cdot \mathbf{U}+\left(p\frac{\partial f_{i}^{(0)}}{\partial p}+T\frac{\partial f_{i}^{(0)}}{\partial T}\right)\zeta^{(1)}.
\label{2.20}
\end{eqnarray}

The kinetic equation for $f_{i}^{(1)}$ can be easily written when one takes into account Eq.\ \eqref{2.20}:
\beqa
\left( \partial_{t}^{(0)}+{\cal L}_{i}\right) f_{i}^{(1)}+{\cal M}
_{i}f_{j}^{(1)}&-&\left(p\frac{\partial f_{i}^{(0)}}{\partial p}+T\frac{\partial f_{i}^{(0)}}{\partial T}\right)\zeta^{(1)}=
{\bf A}_{i}\cdot \nabla x_{1}+{\bf B}_{i}\cdot \nabla p+
{\bf C}_{i}\cdot \nabla T\nonumber\\
& & +D_{i,\beta \lambda}\frac{1}{2}\left(\frac{\partial U_\beta}{\partial r_\lambda}+
\frac{\partial U_\lambda}{\partial r_\beta}-\frac{2}{d}\delta_{\beta\lambda}\nabla\cdot \mathbf{U}\right)+
E_i' \nabla \cdot \mathbf{U}.
\label{2.21}
\eeqa
The coefficients of the field gradients on the right side are functions of ${\bf V}$ and the hydrodynamic fields. They are given by
\begin{equation}
{\bf A}_{i}({\bf V})=-\left(\frac{\partial}{\partial x_{1}}
f_{i}^{(0)}\right)_{p,T}{\bf V}, \quad  {\bf B}_{i}({\bf V})=-\frac{\partial f_{i}^{(0)}}{\partial p}{\bf V}-\rho^{-1}
\frac{\partial f_{i}^{(0)}}{\partial {\bf V}}, \label{2.22}
\end{equation}
\begin{equation}
{\bf C}_{i}({\bf V})=-\frac{\partial f_{i}^{(0)}}{\partial T}{\bf V}, \quad D_{i,\beta\lambda}({\bf V})=V_\beta \frac{\partial f_i^{(0)}}{\partial V_\lambda},
\label{2.23}
\end{equation}
\beq
\label{2.24}
E_i'(\mathbf{V})=\frac{d+2}{d}p\frac{\partial f_{i}^{(0)}}{\partial p}+\frac{2}{d}T\frac{\partial f_{i}^{(0)}}{\partial T}+\frac{1}{d}\mathbf{V}\cdot \frac{\partial f_{i}^{(0)}}{\partial {\bf V}}.
\eeq
Note that that $\zeta^{(1)}$ is given in terms of the unknown distribution function $f^{(1)}$. In addition, since $\zeta^{(1)}$ is a scalar then it must be proportional to $\nabla \cdot \mathbf{U}$ since $\nabla x_1$, $\nabla p$, and $\nabla T$ are vectors and the tensor $\partial_\lambda U_\beta+\partial_\beta U_\lambda-(2/d)\delta_{\lambda\beta}\nabla \cdot \mathbf{U}$ is a traceless tensor. Therefore, the term $\zeta^{(1)}$ can be written as
\beq
\label{2.24.1}
\zeta^{(1)}=\zeta_U \nabla \cdot \mathbf{U},
\eeq
and Eq.\ \eqref{2.21} reads
\beqa
\left(\partial_{t}^{(0)}+{\cal L}_{i}\right) f_{i}^{(1)}+{\cal M}
_{i}f_{j}^{(1)}&=&
{\bf A}_{i}\cdot \nabla x_{1}+{\bf B}_{i}\cdot \nabla p+
{\bf C}_{i}\cdot \nabla T\nonumber\\
& & +D_{i,\beta \lambda}\frac{1}{2}\left(\frac{\partial U_\beta}{\partial r_\lambda}+
\frac{\partial U_\lambda}{\partial r_\beta}-\frac{2}{d}\delta_{\beta\lambda}\nabla\cdot \mathbf{U}\right)+
E_i \nabla \cdot \mathbf{U},
\label{2.24.2}
\eeqa
where
\beq
\label{2.24.3}
E_i(\mathbf{V})=E_i'(\mathbf{V})+\left(p\frac{\partial f_{i}^{(0)}}{\partial p}+T\frac{\partial f_{i}^{(0)}}{\partial T}\right)\zeta_U
=-\frac{1}{d}\Delta^*\frac{\partial f_i^{(0)}}{\partial \Delta^*}-\frac{1}{2}\zeta_U \left[\frac{\partial}
{\partial \mathbf{V}}\cdot \left(\mathbf{V}f_i^{(0)}\right)+\Delta^*\frac{\partial f_i^{(0)}}{\partial \Delta^*}\right].
\eeq
Upon obtaining Eq.\ \eqref{2.24.3} use has been made of the relations \eqref{2.9} and \eqref{2.10}.

The solution to Eq.\ (\ref{2.21}) is of the form
\begin{equation}
f_{i}^{(1)}={\boldsymbol {\cal A}}_{i}\cdot \nabla x_{1}+{\boldsymbol {\cal B}}_{i}\cdot
\nabla p+{\boldsymbol {\cal C}}_{i}\cdot \nabla T+\mathcal{D}_{i,\beta\lambda
}\frac{1}{2}\left(\frac{\partial U_\beta}{\partial r_\lambda}+
\frac{\partial U_\lambda}{\partial r_\beta}-\frac{2}{d}\delta_{\beta\lambda}\nabla\cdot \mathbf{U}\right)
+\mathcal{E}_i \nabla\cdot \mathbf{U}\;.
\label{2.27}
\end{equation}
The coefficients ${\boldsymbol {\cal A}}_{i}$, ${\boldsymbol {\cal B}}_{i}$, ${\boldsymbol {\cal C}}_{i}$, $\mathcal{D}_{i,\beta\lambda}$, and $\mathcal{E}_i$ are functions of the peculiar velocity ${\bf V}$ and the hydrodynamic fields. The cooling rate
depends on space through its dependence on $x_{1}$, $p$, and $T$. The time
derivative $\partial_{t}^{(0)}$ acting on these quantities can be evaluated
by the replacement $\partial _{t}^{(0)}\rightarrow -\zeta ^{(0)}\left(
T\partial _{T}+p\partial _{p}\right) $. In addition, there are contributions coming from the action of the operator
$\partial_{t}^{(0)}$ on the temperature and pressure gradients given by
\begin{eqnarray}
\partial_{t}^{(0)}\nabla T &=&-\nabla \left( T\zeta ^{(0)}\right) =-\zeta
^{(0)}\nabla T-T\nabla \zeta ^{(0)}  \nonumber \\
&=&-\frac{\zeta ^{(0)}}{2}\left(1-\Delta^*\frac{\partial \ln \zeta_0^*}{\partial \Delta^*}\right)
\nabla T-T\left[ \left( \frac{\partial \zeta
^{(0)}}{\partial x_{1}}\right) _{p,T}\nabla x_{1}+\frac{\zeta ^{(0)}}{p}
\nabla p\right] ,  \label{2.28}
\end{eqnarray}
\begin{eqnarray}
\partial_{t}^{(0)}\nabla p &=&-\nabla \left( p\zeta ^{(0)}\right) =-\zeta
^{(0)}\nabla p-p\nabla \zeta ^{(0)}  \nonumber \\
&=&-2\zeta ^{(0)}\nabla p-p\left(\frac{\partial \zeta ^{(0)}}{
\partial x_{1}}\right)_{p,T}\nabla x_{1}+\frac{p\zeta ^{(0)}}{2T}\left(1+\Delta^*\frac{\partial \ln \zeta_0^*}
{\partial \Delta^*}\right)\nabla T,
\label{2.29}
\end{eqnarray}
where we recall that $\zeta_0^*=\zeta^{(0)}/\nu$ and $\nu=n\sigma_{12}^{d-1}v_\text{th}$.

The integral equations for ${\boldsymbol {\cal A}}_{i}$, ${\boldsymbol {\cal B}}_{i}$, ${\boldsymbol {\cal C}}_{i}$, $\mathcal {D}_{i,\beta\lambda}$, and $\mathcal{E}_i$ are identified as coefficients of the independent gradients in Eq.\ (\ref{2.21}):
\begin{equation}
\left[-\zeta ^{(0)}\left( T\partial _{T}+p\partial _{p}\right) +{\cal L}_{i}
\right] {\boldsymbol {\cal A}}_{i}+{\cal M}_{i}{\boldsymbol {\cal A}}_{j}={\bf A}_{i}+\left(
\frac{\partial \zeta ^{(0)}}{\partial x_{1}}\right) _{p,T}\left( p{\boldsymbol {\cal B}}_{i}+T{\boldsymbol {\cal C}}_{i}\right),
\label{2.30}
\end{equation}
\begin{equation}
\left[-\zeta ^{(0)}\left( T\partial _{T}+p\partial _{p}\right) +{\cal L}
_{i}-2\zeta ^{(0)}\right] {\boldsymbol {\cal B}}_{i}+{\cal M}_{i}{\boldsymbol {\cal B}}_{j}={\bf B}_{i}+
\frac{T\zeta ^{(0)}}{p}{\boldsymbol {\cal C}}_{i},  \label{2.31}
\end{equation}
\begin{equation}
\left[-\zeta ^{(0)}\left( T\partial _{T}+p\partial _{p}\right) +{\cal L}
_{i}-\frac{1}{2}\zeta ^{(0)}\left(1-\Delta^*\frac{\partial \ln \zeta_0^*}{\partial \Delta^*}\right)\right] {\boldsymbol {\cal C}}_{i}+{\cal M}_{i}{\boldsymbol {\cal C}}_{j}={\bf C}_{i}-\frac{p\zeta ^{(0)}}{2T}\left(1+\Delta^*\frac{\partial \ln \zeta_0^*}{\partial \Delta^*}\right)
{\boldsymbol {\cal B}}_{i},  \label{2.32}
\end{equation}
\begin{equation}
\left[-\zeta ^{(0)}\left( T\partial_{T}+p\partial_{p}\right) +{\cal
L}_{i}\right] \mathcal{D}_{i,\beta \lambda}+{\cal M}_{i}\mathcal{D}_{j,\beta\lambda}=D_{i,\beta\lambda},  \label{2.33}
\end{equation}
\begin{equation}
\left[-\zeta ^{(0)}\left( T\partial_{T}+p\partial_{p}\right) +{\cal
L}_{i}\right] \mathcal {E}_{i}+{\cal M}_{i}\mathcal {E}_{j}=E_{i}.  \label{2.34}
\end{equation}
In Eqs.\ \eqref{2.30}--\eqref{2.34}, as said before, it is understood that $i\neq j$.

Since $\zeta_U$ is coupled with the unknowns $\mathcal{E}_i$, its explicit form can be identified after expanding the expression \eqref{1.11} of the cooling rate to first order in spatial gradients. After some algebra, $\zeta_U$ can be written as $\zeta_U=\zeta_U^{(0)}+\zeta_U^{(1)}$, where
\beq
\label{2.35}
\zeta_U^{(0)}=\frac{1}{nT}\frac{\pi^{(d-1)/2}}{d\Gamma\left(\frac{d+3}{2}\right)}\sum_{i=1}^2\sum_{j=1}^2 \sigma_{ij}^{d-1}m_{ij} (1-\al_{ij}^2)\int d \mathbf{v}_1\int d \mathbf{v}_2 \; g^3\; f_i^{(0)}(\mathbf{V}_1)\mathcal{E}_j (\mathbf{V}_2),
\eeq
\beqa
\label{2.35.1}
\zeta_U^{(1)}&=&-\frac{1}{d n T}\sum_{i=1}^2\sum_{j=1}^2 \sigma_{ij}^{d-1}m_{ij}\int d \mathbf{v}_1\int d \mathbf{v}_2 \left[f_i^{(0)}(\mathbf{V}_1)\mathcal{E}_j (\mathbf{V}_2)+f_j^{(0)}(\mathbf{V}_2)\mathcal{E}_i (\mathbf{V}_1)\right]\Bigg[-4B_2 \Delta_{ij}(\mathbf{g}\cdot \mathbf{V}_1)\nonumber\\
& &+4B_2 \mu_{ji}\Delta_{ij}(1+\al_{ij})g^2+4B_1\mu_{ji}\Delta_{ij}^2 g\Bigg].
\eeqa
Here, $m_{ij}=m_im_j/(m_i+m_j)$ and
\beq
\label{2.35.2}
B_k=\pi^{(d-1)/2}\frac{\Gamma\left(\frac{k+1}{2}\right)}{\Gamma\left(\frac{k+d}{2}\right)}.
\eeq

\section{Collision frequencies $\nu_D$, $\tau_{ij}$, and $\omega_{ij}$}
\label{appC}

In this Appendix, we give some technical details for the evaluation of the collision frequencies $\nu_D$, $\tau_{ij}$, and $\omega_{ij}$. To obtain them, the property \eqref{1.6.1} is used. Let us consider first $\nu_D$. According to Eqs. \ \eqref{1.1}--\eqref{1.2} and \eqref{1.6.1},  the quantity $\nu_D$ can be split in two parts: one of them already computed in the conventional IHS model (i.e., when $\Delta_{ij}=0$) and the other part involving terms proportional to $\Delta_{ij}$. Thus, the collision frequency $\nu_D$ reads
\beq
\label{a2}
\nu_D=\nu_D^{(0)}+\nu_D^{(1)},
\eeq
where $\nu_D^{(0)}$ was determined in Ref.\ \onlinecite{GM07} and its expression is
\beq
\label{a3}
\nu_D^{(0)}=\frac{2\pi^{(d-1)/2}}{d\Gamma\left(\frac{d}{2}\right)}(1+\al_{12})\left(\frac{\theta_1+\theta_2}{\theta_1\theta_2}\right)^{1/2}
\left(x_1\mu_{12}+x_2\mu_{21}\right)n\sigma_{12}^{d-1}v_\text{th}.
\eeq
The quantity $\nu_D^{(1)}$ is
\beq
\label{a4}
\nu_D^{(1)}=\frac{2\pi^{d/2}}{d\Gamma\left(\frac{d}{2}\right)}\frac{\mu_{21}\Delta_{12}}{dn_1T_1^{(0)}}m_1\sigma_{12}^{d-1}\int d\mathbf{V}_1\int d\mathbf{V}_2
\left[f_{1,M}(\mathbf{V}_1)f_2^{(0)}(\mathbf{V}_2)(\mathbf{g}\cdot \mathbf{V}_1)-\frac{x_1T_1^{(0)}}{x_2T_2^{(0)}}f_{2,M}(\mathbf{V}_2)f_1^{(0)}(\mathbf{V}_1)(\mathbf{g}\cdot \mathbf{V}_2)\right],
\eeq
where use has been made of the result
\beq
\label{a5}
\int d\widehat{\boldsymbol{\sigma}}\Theta (\widehat{{\boldsymbol {\sigma }}}\cdot {\bf g})
(\widehat{\boldsymbol {\sigma }}\cdot {\bf g})\widehat{\boldsymbol {\sigma }}=B_2 \mathbf{g}=\frac{\pi^{d/2}}{d\Gamma\left(\frac{d}{2}\right)}\mathbf{g}.
\eeq
To integrate over $\mathbf{V}_1$ and $\mathbf{V}_2$ in Eq.\ \eqref{a4} we substitute the zeroth-order distributions $f_i^{(0)}$ ($i=1,2$) by their Maxwellian distributions $f_{i,\text{M}}$ defined by Eq.\ \eqref{4.13}. With these replacements, $\nu_D^{(1)}$ is finally given by
\beq
\label{a6}
\nu_D^{(1)}=\frac{2\pi^{d/2}}{d\Gamma\left(\frac{d}{2}\right)}\Delta_{12}^*\left(x_1\mu_{12}+x_2\mu_{21}\right)n\sigma_{12}^{d-1}v_\text{th}.
\eeq
The expression \eqref{4.18.1} for $\nu_D$ can be easily obtained from Eqs.\ \eqref{a3} and \eqref{a6}.

The collision frequencies $\tau_{11}$ and $\tau_{12}$ are defined by Eqs.\ \eqref{4.23} and \eqref{4.24}, respectively. To obtain them, as before, we replace $f_i^{(0)}$ by its Maxwellian distribution $f_{i,\text{M}}$. As in the case of $\nu_D$, the forms of $\tau_{11}$ and $\tau_{12}$ can be written as
\beq
\label{a7}
\tau_{11}=\tau_{11}^{(0)}+\tau_{11}^{(1)}, \quad \tau_{12}=\tau_{12}^{(0)}+\tau_{12}^{(1)},
\eeq
where the contributions $\tau_{11}^{(0)}$ and $\tau_{12}^{(0)}$ (i.e., when $\Delta_{ij}=0$) are \cite{GM07}
\begin{eqnarray}
 \label{a8}
\tau_{11}^{(0)} &=&\frac{2\pi ^{(d-1)/2}}{d(d+2)\Gamma \left( \frac{d}{2}\right) } \upsilon_{\text{th}}\Bigg\{ n_{1}\sigma
_{1}^{d-1}(2\theta_{1})^{-1/2}(3+2d-3\alpha_{11})(1+\alpha_{11})+2n_{2}\sigma_{12}^{d-1}\mu_{21}(1+\alpha_{12})\theta_{1}^{3/2}
\theta_{2}^{-1/2}\nonumber\\
& & \times \left[(d+3)(\mu_{12}\theta_2-\mu_{21}\theta_1)\theta_{1}^{-2}(\theta_{1}+\theta_{2})^{-1/2}+\frac{3+2d-3\alpha_{12}}{2}\mu_{21}
\theta_{1}^{-2}(\theta_{1}+\theta_{2})^{1/2}\right.\nonumber\\
& &\left.+\frac{2d(d+1)-4}{2(d-1)}\theta
_{1}^{-1}(\theta_{1}+\theta_{2})^{-1/2}\right] \Bigg\},
\end{eqnarray}
\begin{eqnarray}
 \label{a9}
\tau_{12}^{(0)}&=&\frac{4\pi ^{(d-1)/2}}{d(d+2)\Gamma \left( \frac{d}{2}\right) } \upsilon_{\text{th}} n_{1}
\sigma_{12}^{d-1}\mu_{12}\theta_{2}^{3/2}\theta_{1}^{-1/2}(1+\alpha_{12})\Big[ (d+3)(\mu_{12}\theta_2-\mu_{21}\theta_1)\theta_{2}^{-2}(\theta_{1}
+\theta_{2})^{-1/2}\nonumber\\
& & +\frac{3+2d-3\alpha_{12}}{2}\mu_{21}\theta_{2}^{-2}(\theta_{1}+\theta_{2})^{1/2}-\frac{2d(d+1)-4}{2(d-1)}\theta_{2}^{-1}(\theta_{1}+\theta
_{2})^{-1/2}\Big].
\end{eqnarray}
The evaluation of $\tau_{11}^{(1)}$ and $\tau_{12}^{(2)}$ follows similar mathematical steps as those made in the evaluation of $\nu_D^{(1)}$. Only the final expressions are provided here. They are given by
\beqa
\label{a10}
\tau_{11}^{(1)}&=&\frac{\sqrt{2}\pi^{(d-1)/2}}{d(d+2)\Gamma\left(\frac{d}{2}\right)}\upsilon_{\text{th}}n_1\sigma_1^{d-1}\Delta_{11}^*
\left[\sqrt{2\pi}(d-2\al_{11})-2\theta_1^{-1/2}\Delta_{11}^*\right]
\nonumber\\
& &
-\frac{8\pi^{(d-1)/2}}{d(d+2)\Gamma\left(\frac{d}{2}\right)}\upsilon_{\text{th}}n_2\sigma_{12}^{d-1}
\mu_{21}^2\theta_1^2 \Delta_{12}^* \Big[\sqrt{\pi}\theta_2^{-2}(1+\al_{12})+\left(\theta_1+\theta_2\right)^{-1/2}\theta_1^{1/2}\theta_2^{-3/2}\Delta_{12}^*\Big],
\eeqa
\beq
\label{a11}
\tau_{12}^{(1)}=-\frac{4\pi^{(d-1)/2}}{d(d+2)\Gamma\left(\frac{d}{2}\right)}\upsilon_{\text{th}}n_1\sigma_{12}^{d-1}
\mu_{12}\theta_2^2 \Delta_{12}^* \Big[2\sqrt{\pi}\mu_{21}\theta_1^{-2}(1+\al_{12})-(d+2)\sqrt{\pi}\theta_1^{-2}
+2\left(\theta_1+\theta_2\right)^{-1/2}\theta_2^{1/2}\theta_1^{-3/2}\Delta_{12}^*\Big].
\eeq
The expressions of $\tau_{22}$ and $\tau_{21}$ can be easily inferred from Eqs.\ \eqref{a8}--\eqref{a11} by making the change $1\leftrightarrow 2$. In the case of mechanically equivalent particles ($m_1=m_2$, $\sigma_1=\sigma_2$, $\al_{ij}=\al$, and $\Delta_{ij}^*=\Delta^*$), Eqs.\ \eqref{a10} and \eqref{a11} yield
\beq
\label{a12}
\tau_{11}^{(1)}+\tau_{12}^{(1)}=\frac{\sqrt{2}\pi^{(d-1)/2}}{d(d+2)\Gamma\left(\frac{d}{2}\right)}\upsilon_{\text{th}}n\sigma^{d-1}\Delta^*
\left[\sqrt{2\pi}(d-2\al)-2\Delta^*\right].
\eeq
This result is consistent with the one previously found for monocomponent granular gases in the $\Delta$-model. \cite{BBMG15}

Finally, the expressions of the collision frequencies $\omega_{11}$ and $\omega_{12}$ are
\beqa
\label{a13}
\omega_{11}&=&-\frac{\pi^{(d-1)/2}}{d\Gamma\left(\frac{d}{2}\right)}\upsilon_{\text{th}}\theta_1^{-1/2}\Bigg\{\frac{3}{\sqrt{2}}
n_1\sigma_1^{d-1}\left(1-\alpha_{11}^2\right)-n_2 \sigma_{12}^{d-1} \mu_{21} \left(1+\alpha_{12}\right)\left(\theta_1+\theta_2\right)^{-1/2}\theta_2^{-1/2}\nonumber\\
& & \times
\Big[3\mu_{21}\left(1+\alpha_{12}\right)\left(\theta_1+\theta_2\right)-2\left(2\theta_1+3\theta_2\right)\Big]\Bigg\}
+\frac{2\pi^{(d-1)/2}}{d\Gamma\left(\frac{d}{2}\right)}\upsilon_{\text{th}}n_1\sigma_1^{d-1}\Delta_{11}^*\left(\sqrt{\pi}\alpha_{11}+
\sqrt{\frac{\theta_1}{2}}\Delta_{11}^*\right)\nonumber\\
& & +
\frac{4\pi^{(d-1)/2}}{d\Gamma\left(\frac{d}{2}\right)}\upsilon_{\text{th}}n_2 \sigma_{12}^{d-1} \mu_{21}^2 \Delta_{12}^*\Bigg\{\sqrt{\pi}\left(1+\al_{12}-\mu_{21}^{-1}\right)+\left(\theta_1+\theta_2\right)^{-1/2}\theta_1^{3/2}\theta_2^{-1/2}
\Delta_{12}^*\nonumber\\
& & \times\Big[d-d\left(\theta_1+\theta_2\right)\theta_1^{-1}+(d+1)
\theta_1\theta_2^{-1}\Big]\Bigg\},
\eeqa
\beqa
\label{a14}
\omega_{12}&=&\frac{\pi^{(d-1)/2}}{d\Gamma\left(\frac{d}{2}\right)}\upsilon_{\text{th}} n_2 \sigma_{12}^{d-1}\mu_{12}
\left(1+\alpha_{12}\right)\left(\theta_1+\theta_2\right)^{-1/2}\theta_1^{-1/2}\theta_2^{-1/2}\Big[3\mu_{21}\left(1+\alpha_{12}\right)
\left(\theta_1+\theta_2\right)-2\theta_2\Big]\nonumber\\
& & +\frac{4\pi^{(d-1)/2}}{d\Gamma\left(\frac{d}{2}\right)}\upsilon_{\text{th}}n_2 \sigma_{12}^{d-1} \mu_{12} \mu_{21} \Delta_{12}^*\Bigg\{\sqrt{\pi}\left(1+\al_{12}\right)+
\left(\theta_1+\theta_2\right)^{-1/2}\theta_1^{-1/2}\theta_2^{3/2}
\Delta_{12}^*\nonumber\\
& & \times\Big[d-d\left(\theta_1+\theta_2\right)\theta_2^{-1}+(d+1)
\theta_1^{-1}\theta_2\Big]\Bigg\}.
\eeqa
As before, the expressions of $\omega_{22}$ and $\omega_{21}$ can be easily obtained from Eqs.\ \eqref{a13} and \eqref{a14}, respectively, by making the change $1\leftrightarrow 2$.
\end{widetext}

\textbf{DATA AVAILABILITY}

The data that support the findings of this study are available from the corresponding author upon reasonable request.


\begin{thebibliography}{82}%
\makeatletter
\providecommand \@ifxundefined [1]{%
 \@ifx{#1\undefined}
}%
\providecommand \@ifnum [1]{%
 \ifnum #1\expandafter \@firstoftwo
 \else \expandafter \@secondoftwo
 \fi
}%
\providecommand \@ifx [1]{%
 \ifx #1\expandafter \@firstoftwo
 \else \expandafter \@secondoftwo
 \fi
}%
\providecommand \natexlab [1]{#1}%
\providecommand \enquote  [1]{``#1''}%
\providecommand \bibnamefont  [1]{#1}%
\providecommand \bibfnamefont [1]{#1}%
\providecommand \citenamefont [1]{#1}%
\providecommand \href@noop [0]{\@secondoftwo}%
\providecommand \href [0]{\begingroup \@sanitize@url \@href}%
\providecommand \@href[1]{\@@startlink{#1}\@@href}%
\providecommand \@@href[1]{\endgroup#1\@@endlink}%
\providecommand \@sanitize@url [0]{\catcode `\\12\catcode `\$12\catcode
  `\&12\catcode `\#12\catcode `\^12\catcode `\_12\catcode `\%12\relax}%
\providecommand \@@startlink[1]{}%
\providecommand \@@endlink[0]{}%
\providecommand \url  [0]{\begingroup\@sanitize@url \@url }%
\providecommand \@url [1]{\endgroup\@href {#1}{\urlprefix }}%
\providecommand \urlprefix  [0]{URL }%
\providecommand \Eprint [0]{\href }%
\providecommand \doibase [0]{http://dx.doi.org/}%
\providecommand \selectlanguage [0]{\@gobble}%
\providecommand \bibinfo  [0]{\@secondoftwo}%
\providecommand \bibfield  [0]{\@secondoftwo}%
\providecommand \translation [1]{[#1]}%
\providecommand \BibitemOpen [0]{}%
\providecommand \bibitemStop [0]{}%
\providecommand \bibitemNoStop [0]{.\EOS\space}%
\providecommand \EOS [0]{\spacefactor3000\relax}%
\providecommand \BibitemShut  [1]{\csname bibitem#1\endcsname}%
\let\auto@bib@innerbib\@empty
\bibitem [{\citenamefont {Brilliantov}\ and\ \citenamefont
  {P\"oschel}(2004)}]{BP04}%
  \BibitemOpen
  \bibfield  {author} {\bibinfo {author} {\bibfnamefont {N.}~\bibnamefont
  {Brilliantov}}\ and\ \bibinfo {author} {\bibfnamefont {T.}~\bibnamefont
  {P\"oschel}},\ }\href@noop {} {\emph {\bibinfo {title} {Kinetic Theory of
  Granular Gases}}}\ (\bibinfo  {publisher} {Oxford University Press, Oxford},\
  \bibinfo {year} {2004})\BibitemShut {NoStop}%
\bibitem [{\citenamefont {Garz\'o}(2019)}]{G19}%
  \BibitemOpen
  \bibfield  {author} {\bibinfo {author} {\bibfnamefont {V.}~\bibnamefont
  {Garz\'o}},\ }\href@noop {} {\emph {\bibinfo {title} {Granular Gaseous
  Flows}}}\ (\bibinfo  {publisher} {Springer Nature, Cham},\
  \bibinfo {year} {2019})\BibitemShut {NoStop}%
\bibitem [{\citenamefont {Yang}\ \emph {et~al.}(2002)\citenamefont {Yang},
  \citenamefont {Huan}, \citenamefont {Candela}, \citenamefont {Mair},\ and\
  \citenamefont {Walsworth}}]{YHCMW02}%
  \BibitemOpen
  \bibfield  {author} {\bibinfo {author} {\bibfnamefont {X.}~\bibnamefont
  {Yang}}, \bibinfo {author} {\bibfnamefont {C.}~\bibnamefont {Huan}}, \bibinfo
  {author} {\bibfnamefont {D.}~\bibnamefont {Candela}}, \bibinfo {author}
  {\bibfnamefont {R.~W.}\ \bibnamefont {Mair}}, \ and\ \bibinfo {author}
  {\bibfnamefont {R.~L.}\ \bibnamefont {Walsworth}},\ }\bibfield  {title}
  {\enquote {\bibinfo {title} {Measurements of grain motion in a dense,
  three-dimensional granular fluid},}\ }\href@noop {} {\bibfield  {journal}
  {\bibinfo  {journal} {Phys. Rev. Lett.}\ }\textbf {\bibinfo {volume} {88}},\
  \bibinfo {pages} {{044}{301}} (\bibinfo {year} {2002})}\BibitemShut {NoStop}%
\bibitem [{\citenamefont {Huan}\ \emph {et~al.}(2004)\citenamefont {Huan},
  \citenamefont {Yang}, \citenamefont {Candela}, \citenamefont {Mair},\ and\
  \citenamefont {Walsworth}}]{HYCMW04}%
  \BibitemOpen
  \bibfield  {author} {\bibinfo {author} {\bibfnamefont {C.}~\bibnamefont
  {Huan}}, \bibinfo {author} {\bibfnamefont {X.}~\bibnamefont {Yang}}, \bibinfo
  {author} {\bibfnamefont {D.}~\bibnamefont {Candela}}, \bibinfo {author}
  {\bibfnamefont {R.~W.}\ \bibnamefont {Mair}}, \ and\ \bibinfo {author}
  {\bibfnamefont {R.~L.}\ \bibnamefont {Walsworth}},\ }\bibfield  {title}
  {\enquote {\bibinfo {title} {{N}{M}{R} experiments on a three-dimensional
  vibrofluidized granular medium},}\ }\href@noop {} {\bibfield  {journal}
  {\bibinfo  {journal} {Phys. Rev. E}\ }\textbf {\bibinfo {volume} {69}},\
  \bibinfo {pages} {{041}{302}} (\bibinfo {year} {2004})}\BibitemShut {NoStop}%
\bibitem [{\citenamefont {Abate}\ and\ \citenamefont {Durian}(2006)}]{AD06}%
  \BibitemOpen
  \bibfield  {author} {\bibinfo {author} {\bibfnamefont {A.~R.}\ \bibnamefont
  {Abate}}\ and\ \bibinfo {author} {\bibfnamefont {D.~J.}\ \bibnamefont
  {Durian}},\ }\bibfield  {title} {\enquote {\bibinfo {title} {Approach to
  jamming in an air-fluidized granular bed},}\ }\href@noop {} {\bibfield
  {journal} {\bibinfo  {journal} {Phys. Rev. E}\ }\textbf {\bibinfo {volume}
  {74}},\ \bibinfo {pages} {{031}{308}} (\bibinfo {year} {2006})}\BibitemShut
  {NoStop}%
\bibitem [{\citenamefont {Schr\"oter}, \citenamefont {Goldman},\ and\
  \citenamefont {Swinney}(2005)}]{SGS05}%
  \BibitemOpen
  \bibfield  {author} {\bibinfo {author} {\bibfnamefont {M.}~\bibnamefont
  {Schr\"oter}}, \bibinfo {author} {\bibfnamefont {D.~I.}\ \bibnamefont
  {Goldman}}, \ and\ \bibinfo {author} {\bibfnamefont {H.~L.}\ \bibnamefont
  {Swinney}},\ }\bibfield  {title} {\enquote {\bibinfo {title} {Stationary
  state volume fluctuations in a granular medium},}\ }\href@noop {} {\bibfield
  {journal} {\bibinfo  {journal} {Phys. Rev. E}\ }\textbf {\bibinfo {volume}
  {71}},\ \bibinfo {pages} {{030}{301}(R)} (\bibinfo {year}
  {2005})}\BibitemShut {NoStop}%
\bibitem [{\citenamefont {Puglisi}\ \emph {et~al.}(1999)\citenamefont
  {Puglisi}, \citenamefont {Loreto}, \citenamefont {Marconi},\ and\
  \citenamefont {Vulpiani}}]{PLMV99}%
  \BibitemOpen
  \bibfield  {author} {\bibinfo {author} {\bibfnamefont {A.}~\bibnamefont
  {Puglisi}}, \bibinfo {author} {\bibfnamefont {V.}~\bibnamefont {Loreto}},
  \bibinfo {author} {\bibfnamefont {U.~M.~B.}\ \bibnamefont {Marconi}}, \ and\
  \bibinfo {author} {\bibfnamefont {A.}~\bibnamefont {Vulpiani}},\ }\bibfield
  {title} {\enquote {\bibinfo {title} {Kinetic approach to granular gases},}\
  }\href@noop {} {\bibfield  {journal} {\bibinfo  {journal} {Phys. Rev. E}\
  }\textbf {\bibinfo {volume} {59}},\ \bibinfo {pages} {5582--5595} (\bibinfo
  {year} {1999})}\BibitemShut {NoStop}%
\bibitem [{\citenamefont {Cafiero}, \citenamefont {Luding},\ and\ \citenamefont
  {Herrmann}(2000)}]{CLH00}%
  \BibitemOpen
  \bibfield  {author} {\bibinfo {author} {\bibfnamefont {R.}~\bibnamefont
  {Cafiero}}, \bibinfo {author} {\bibfnamefont {S.}~\bibnamefont {Luding}}, \
  and\ \bibinfo {author} {\bibfnamefont {H.~J.}\ \bibnamefont {Herrmann}},\
  }\bibfield  {title} {\enquote {\bibinfo {title} {Two-dimensional granular gas
  of inelastic spheres with multiplicative driving},}\ }\href@noop {}
  {\bibfield  {journal} {\bibinfo  {journal} {Phys. Rev. Lett.}\ }\textbf
  {\bibinfo {volume} {84}},\ \bibinfo {pages} {6014--6017} (\bibinfo {year}
  {2000})}\BibitemShut {NoStop}%
\bibitem [{\citenamefont {Prevost}, \citenamefont {Egolf},\ and\ \citenamefont
  {Urbach}(2002)}]{PEU02}%
  \BibitemOpen
  \bibfield  {author} {\bibinfo {author} {\bibfnamefont {A.}~\bibnamefont
  {Prevost}}, \bibinfo {author} {\bibfnamefont {D.~A.}\ \bibnamefont {Egolf}},
  \ and\ \bibinfo {author} {\bibfnamefont {J.~S.}\ \bibnamefont {Urbach}},\
  }\bibfield  {title} {\enquote {\bibinfo {title} {Forcing and velocity
  correlations in a vibrated granular monolayer},}\ }\href@noop {} {\bibfield
  {journal} {\bibinfo  {journal} {Phys. Rev. Lett.}\ }\textbf {\bibinfo
  {volume} {89}},\ \bibinfo {pages} {{084}{301}} (\bibinfo {year}
  {2002})}\BibitemShut {NoStop}%
\bibitem [{\citenamefont {Paganobarraga}\ \emph {et~al.}(2002)\citenamefont
  {Paganobarraga}, \citenamefont {Trizac}, \citenamefont {van Noije},\ and\
  \citenamefont {Ernst}}]{PTNE02}%
  \BibitemOpen
  \bibfield  {author} {\bibinfo {author} {\bibfnamefont {I.}~\bibnamefont
  {Paganobarraga}}, \bibinfo {author} {\bibfnamefont {E.}~\bibnamefont
  {Trizac}}, \bibinfo {author} {\bibfnamefont {T.~P.~C.}\ \bibnamefont {van
  Noije}}, \ and\ \bibinfo {author} {\bibfnamefont {M.~H.}\ \bibnamefont
  {Ernst}},\ }\bibfield  {title} {\enquote {\bibinfo {title} {Randomly driven
  granular fluids: {C}ollisional statistics and short scale structure},}\
  }\href@noop {} {\bibfield  {journal} {\bibinfo  {journal} {Phys. Rev. E}\
  }\textbf {\bibinfo {volume} {65}},\ \bibinfo {pages} {{011}{303}} (\bibinfo
  {year} {2002})}\BibitemShut {NoStop}%
\bibitem [{\citenamefont {Puglisi}, \citenamefont {Baldassarri},\ and\
  \citenamefont {Loreto}(2002)}]{PBL02}%
  \BibitemOpen
  \bibfield  {author} {\bibinfo {author} {\bibfnamefont {A.}~\bibnamefont
  {Puglisi}}, \bibinfo {author} {\bibfnamefont {A.}~\bibnamefont
  {Baldassarri}}, \ and\ \bibinfo {author} {\bibfnamefont {V.}~\bibnamefont
  {Loreto}},\ }\bibfield  {title} {\enquote {\bibinfo {title}
  {Fluctuation-dissipation relations in driven granular gases},}\ }\href@noop
  {} {\bibfield  {journal} {\bibinfo  {journal} {Phys. Rev. E}\ }\textbf
  {\bibinfo {volume} {66}},\ \bibinfo {pages} {{061}{305}} (\bibinfo {year}
  {2002})}\BibitemShut {NoStop}%
\bibitem [{\citenamefont {Fiege}, \citenamefont {Aspelmeier},\ and\
  \citenamefont {Zippelius}(2009)}]{FAZ09}%
  \BibitemOpen
  \bibfield  {author} {\bibinfo {author} {\bibfnamefont {A.}~\bibnamefont
  {Fiege}}, \bibinfo {author} {\bibfnamefont {T.}~\bibnamefont {Aspelmeier}}, \
  and\ \bibinfo {author} {\bibfnamefont {A.}~\bibnamefont {Zippelius}},\
  }\bibfield  {title} {\enquote {\bibinfo {title} {Long-time tails and cage
  effect in driven granular fluids},}\ }\href@noop {} {\bibfield  {journal}
  {\bibinfo  {journal} {Phys. Rev. Lett.}\ }\textbf {\bibinfo {volume} {102}},\
  \bibinfo {pages} {{098}{001}} (\bibinfo {year} {2009})}\BibitemShut {NoStop}%
\bibitem [{\citenamefont {Kranz}, \citenamefont {Sperl},\ and\ \citenamefont
  {Zippelius}(2010)}]{KSZ10}%
  \BibitemOpen
  \bibfield  {author} {\bibinfo {author} {\bibfnamefont {W.~T.}\ \bibnamefont
  {Kranz}}, \bibinfo {author} {\bibfnamefont {M.}~\bibnamefont {Sperl}}, \ and\
  \bibinfo {author} {\bibfnamefont {A.}~\bibnamefont {Zippelius}},\ }\bibfield
  {title} {\enquote {\bibinfo {title} {Glass transition for driven granular
  fluids},}\ }\href@noop {} {\bibfield  {journal} {\bibinfo  {journal} {Phys.
  Rev. Lett.}\ }\textbf {\bibinfo {volume} {104}},\ \bibinfo {pages}
  {{225}{701}} (\bibinfo {year} {2010})}\BibitemShut {NoStop}%
\bibitem [{\citenamefont {Vollmayr-Lee}, \citenamefont {Aspelmeier},\ and\
  \citenamefont {Zippelius}(2011)}]{VAZ11}%
  \BibitemOpen
  \bibfield  {author} {\bibinfo {author} {\bibfnamefont {K.}~\bibnamefont
  {Vollmayr-Lee}}, \bibinfo {author} {\bibfnamefont {T.}~\bibnamefont
  {Aspelmeier}}, \ and\ \bibinfo {author} {\bibfnamefont {A.}~\bibnamefont
  {Zippelius}},\ }\bibfield  {title} {\enquote {\bibinfo {title} {Hydrodynamic
  correlation functions of a driven granular fluid in steady state},}\
  }\href@noop {} {\bibfield  {journal} {\bibinfo  {journal} {Phys. Rev. E}\
  }\textbf {\bibinfo {volume} {83}},\ \bibinfo {pages} {011301} (\bibinfo
  {year} {2011})}\BibitemShut {NoStop}%
\bibitem [{\citenamefont {Gradenigo}\ \emph {et~al.}(2011)\citenamefont
  {Gradenigo}, \citenamefont {Sarracino}, \citenamefont {Villamaina},\ and\
  \citenamefont {Puglisi}}]{GSVP11}%
  \BibitemOpen
  \bibfield  {author} {\bibinfo {author} {\bibfnamefont {G.}~\bibnamefont
  {Gradenigo}}, \bibinfo {author} {\bibfnamefont {A.}~\bibnamefont
  {Sarracino}}, \bibinfo {author} {\bibfnamefont {D.}~\bibnamefont
  {Villamaina}}, \ and\ \bibinfo {author} {\bibfnamefont {A.}~\bibnamefont
  {Puglisi}},\ }\bibfield  {title} {\enquote {\bibinfo {title} {Non-equilibrium
  length in granular fluids: {F}rom experiment to fluctuating hydrodynamics},}\
  }\href@noop {} {\bibfield  {journal} {\bibinfo  {journal} {Europhys. Lett.}\
  }\textbf {\bibinfo {volume} {96}},\ \bibinfo {pages} {{14}{004}} (\bibinfo
  {year} {2011})}\BibitemShut {NoStop}%
\bibitem [{\citenamefont {Puglisi}\ \emph {et~al.}(2012)\citenamefont
  {Puglisi}, \citenamefont {Gnoli}, \citenamefont {Gradenigo}, \citenamefont
  {Sarracino},\ and\ \citenamefont {Villamaina}}]{PGGSV12}%
  \BibitemOpen
  \bibfield  {author} {\bibinfo {author} {\bibfnamefont {A.}~\bibnamefont
  {Puglisi}}, \bibinfo {author} {\bibfnamefont {A.}~\bibnamefont {Gnoli}},
  \bibinfo {author} {\bibfnamefont {G.}~\bibnamefont {Gradenigo}}, \bibinfo
  {author} {\bibfnamefont {A.}~\bibnamefont {Sarracino}}, \ and\ \bibinfo
  {author} {\bibfnamefont {D.}~\bibnamefont {Villamaina}},\ }\bibfield  {title}
  {\enquote {\bibinfo {title} {Structure factors in granular experiments with
  homogeneous fluidization},}\ }\href@noop {} {\bibfield  {journal} {\bibinfo
  {journal} {J. Chem. Phys.}\ }\textbf {\bibinfo {volume} {136}},\ \bibinfo
  {pages} {{014}{704}} (\bibinfo {year} {2012})}\BibitemShut {NoStop}%
\bibitem [{\citenamefont {Brey}\ \emph {et~al.}(1998)\citenamefont {Brey},
  \citenamefont {Dufty}, \citenamefont {Kim},\ and\ \citenamefont
  {Santos}}]{BDKS98}%
  \BibitemOpen
  \bibfield  {author} {\bibinfo {author} {\bibfnamefont {J.~J.}\ \bibnamefont
  {Brey}}, \bibinfo {author} {\bibfnamefont {J.~W.}\ \bibnamefont {Dufty}},
  \bibinfo {author} {\bibfnamefont {C.~S.}\ \bibnamefont {Kim}}, \ and\
  \bibinfo {author} {\bibfnamefont {A.}~\bibnamefont {Santos}},\ }\bibfield
  {title} {\enquote {\bibinfo {title} {Hydrodynamics for granular flows at low
  density},}\ }\href@noop {} {\bibfield  {journal} {\bibinfo  {journal} {Phys.
  Rev. E}\ }\textbf {\bibinfo {volume} {58}},\ \bibinfo {pages} {4638--4653}
  (\bibinfo {year} {1998})}\BibitemShut {NoStop}%
\bibitem [{\citenamefont {Santos}, \citenamefont {Garz\'o},\ and\ \citenamefont
  {Dufty}(2004)}]{SGD04}%
  \BibitemOpen
  \bibfield  {author} {\bibinfo {author} {\bibfnamefont {A.}~\bibnamefont
  {Santos}}, \bibinfo {author} {\bibfnamefont {V.}~\bibnamefont {Garz\'o}}, \
  and\ \bibinfo {author} {\bibfnamefont {J.~W.}\ \bibnamefont {Dufty}},\
  }\bibfield  {title} {\enquote {\bibinfo {title} {Inherent rheology of a
  granular fluid in uniform shear flow},}\ }\href@noop {} {\bibfield  {journal}
  {\bibinfo  {journal} {Phys. Rev. E}\ }\textbf {\bibinfo {volume} {69}},\
  \bibinfo {pages} {{061}{303}} (\bibinfo {year} {2004})}\BibitemShut {NoStop}%
\bibitem [{\citenamefont {Cordero}, \citenamefont {Risso},\ and\ \citenamefont
  {Soto}(2005)}]{CRS05}%
  \BibitemOpen
  \bibfield  {author} {\bibinfo {author} {\bibfnamefont {P.}~\bibnamefont
  {Cordero}}, \bibinfo {author} {\bibfnamefont {D.}~\bibnamefont {Risso}}, \
  and\ \bibinfo {author} {\bibfnamefont {R.}~\bibnamefont {Soto}},\ }\bibfield
  {title} {\enquote {\bibinfo {title} {Steady quasi-homogeneous granular gas
  state},}\ }\href@noop {} {\bibfield  {journal} {\bibinfo  {journal} {Physica
  A}\ }\textbf {\bibinfo {volume} {356}},\ \bibinfo {pages} {54--60} (\bibinfo
  {year} {2005})}\BibitemShut {NoStop}%
\bibitem [{\citenamefont {Olafsen}\ and\ \citenamefont {Urbach}(1998)}]{OU98}%
  \BibitemOpen
  \bibfield  {author} {\bibinfo {author} {\bibfnamefont {J.~S.}\ \bibnamefont
  {Olafsen}}\ and\ \bibinfo {author} {\bibfnamefont {J.~S.}\ \bibnamefont
  {Urbach}},\ }\bibfield  {title} {\enquote {\bibinfo {title} {Clustering,
  order, and collapse in a driven granular monolayer},}\ }\href@noop {}
  {\bibfield  {journal} {\bibinfo  {journal} {Phys. Rev. Lett.}\ }\textbf
  {\bibinfo {volume} {81}},\ \bibinfo {pages} {4369--4372} (\bibinfo {year}
  {1998})}\BibitemShut {NoStop}%
\bibitem [{\citenamefont {Prevost}\ \emph {et~al.}(2004)\citenamefont
  {Prevost}, \citenamefont {Melby}, \citenamefont {Egolf},\ and\ \citenamefont
  {Urbach}}]{PMEU04}%
  \BibitemOpen
  \bibfield  {author} {\bibinfo {author} {\bibfnamefont {A.}~\bibnamefont
  {Prevost}}, \bibinfo {author} {\bibfnamefont {P.}~\bibnamefont {Melby}},
  \bibinfo {author} {\bibfnamefont {D.~A.}\ \bibnamefont {Egolf}}, \ and\
  \bibinfo {author} {\bibfnamefont {J.~S.}\ \bibnamefont {Urbach}},\ }\bibfield
   {title} {\enquote {\bibinfo {title} {Nonequilibrium two-phase coexistence in
  a confined granular layer},}\ }\href@noop {} {\bibfield  {journal} {\bibinfo
  {journal} {Phys. Rev. E}\ }\textbf {\bibinfo {volume} {70}},\ \bibinfo
  {pages} {{050}{301}(R)} (\bibinfo {year} {2004})}\BibitemShut {NoStop}%
\bibitem [{\citenamefont {Melby}\ \emph {et~al.}(2005)\citenamefont {Melby},
  \citenamefont {Vega~Reyes}, \citenamefont {Prevost}, \citenamefont
  {Robertson}, \citenamefont {Kumar}, \citenamefont {Egolf},\ and\
  \citenamefont {Urbach}}]{MVPRKEU05}%
  \BibitemOpen
  \bibfield  {author} {\bibinfo {author} {\bibfnamefont {P.}~\bibnamefont
  {Melby}}, \bibinfo {author} {\bibfnamefont {F.}~\bibnamefont {Vega~Reyes}},
  \bibinfo {author} {\bibfnamefont {A.}~\bibnamefont {Prevost}}, \bibinfo
  {author} {\bibfnamefont {R.}~\bibnamefont {Robertson}}, \bibinfo {author}
  {\bibfnamefont {P.}~\bibnamefont {Kumar}}, \bibinfo {author} {\bibfnamefont
  {D.~A.}\ \bibnamefont {Egolf}}, \ and\ \bibinfo {author} {\bibfnamefont
  {J.~S.}\ \bibnamefont {Urbach}},\ }\bibfield  {title} {\enquote {\bibinfo
  {title} {The dynamics of thin vibrated granular layers},}\ }\href@noop {}
  {\bibfield  {journal} {\bibinfo  {journal} {J. Phys. C: Condens. Matter}\
  }\textbf {\bibinfo {volume} {17}},\ \bibinfo {pages} {S2689} (\bibinfo {year}
  {2005})}\BibitemShut {NoStop}%
\bibitem [{\citenamefont {Clerc}\ \emph {et~al.}(2008)\citenamefont {Clerc},
  \citenamefont {Cordero}, \citenamefont {Dunstan}, \citenamefont {Huff},
  \citenamefont {Mujica}, \citenamefont {Risso},\ and\ \citenamefont
  {Varas}}]{CCDHMRV08}%
  \BibitemOpen
  \bibfield  {author} {\bibinfo {author} {\bibfnamefont {M.~G.}\ \bibnamefont
  {Clerc}}, \bibinfo {author} {\bibfnamefont {P.}~\bibnamefont {Cordero}},
  \bibinfo {author} {\bibfnamefont {J.}~\bibnamefont {Dunstan}}, \bibinfo
  {author} {\bibfnamefont {K.}~\bibnamefont {Huff}}, \bibinfo {author}
  {\bibfnamefont {N.}~\bibnamefont {Mujica}}, \bibinfo {author} {\bibfnamefont
  {D.}~\bibnamefont {Risso}}, \ and\ \bibinfo {author} {\bibfnamefont
  {G.}~\bibnamefont {Varas}},\ }\bibfield  {title} {\enquote {\bibinfo {title}
  {Liquid-solid-like transition in quasi-one-dimensional driven granular
  media},}\ }\href@noop {} {\bibfield  {journal} {\bibinfo  {journal} {Nature
  Phys.}\ }\textbf {\bibinfo {volume} {4}},\ \bibinfo {pages} {249} (\bibinfo
  {year} {2008})}\BibitemShut {NoStop}%
\bibitem [{\citenamefont {Pacheco-V\'azquez}, \citenamefont
  {Caballero-Robledo},\ and\ \citenamefont {Ruiz-Su\'arez}(2009)}]{PCR09}%
  \BibitemOpen
  \bibfield  {author} {\bibinfo {author} {\bibfnamefont {F.}~\bibnamefont
  {Pacheco-V\'azquez}}, \bibinfo {author} {\bibfnamefont {G.~A.}\ \bibnamefont
  {Caballero-Robledo}}, \ and\ \bibinfo {author} {\bibfnamefont {J.~C.}\
  \bibnamefont {Ruiz-Su\'arez}},\ }\bibfield  {title} {\enquote {\bibinfo
  {title} {Superheating in granular matter},}\ }\href@noop {} {\bibfield
  {journal} {\bibinfo  {journal} {Phys. Rev. Lett.}\ }\textbf {\bibinfo
  {volume} {102}},\ \bibinfo {pages} {{170}{601}} (\bibinfo {year}
  {2009})}\BibitemShut {NoStop}%
\bibitem [{\citenamefont {Rivas}\ \emph {et~al.}(2011)\citenamefont {Rivas},
  \citenamefont {Ponce}, \citenamefont {Gallet}, \citenamefont {Risso},
  \citenamefont {Soto}, \citenamefont {Cordero},\ and\ \citenamefont
  {M\'ujica}}]{RPGRSCM11}%
  \BibitemOpen
  \bibfield  {author} {\bibinfo {author} {\bibfnamefont {N.}~\bibnamefont
  {Rivas}}, \bibinfo {author} {\bibfnamefont {S.}~\bibnamefont {Ponce}},
  \bibinfo {author} {\bibfnamefont {B.}~\bibnamefont {Gallet}}, \bibinfo
  {author} {\bibfnamefont {D.}~\bibnamefont {Risso}}, \bibinfo {author}
  {\bibfnamefont {R.}~\bibnamefont {Soto}}, \bibinfo {author} {\bibfnamefont
  {P.}~\bibnamefont {Cordero}}, \ and\ \bibinfo {author} {\bibfnamefont
  {N.}~\bibnamefont {M\'ujica}},\ }\bibfield  {title} {\enquote {\bibinfo
  {title} {Sudden chain energy transfer events in vibrated granular media},}\
  }\href@noop {} {\bibfield  {journal} {\bibinfo  {journal} {Phys. Rev. Lett.}\
  }\textbf {\bibinfo {volume} {106}},\ \bibinfo {pages} {088001} (\bibinfo
  {year} {2011})}\BibitemShut {NoStop}%
\bibitem [{\citenamefont {Castillo}, \citenamefont {M\'ujica},\ and\
  \citenamefont {Soto}(2012)}]{CMS12}%


  \BibitemOpen
  \bibfield  {author} {\bibinfo {author} {\bibfnamefont {G.}~\bibnamefont
  {Castillo}}, \bibinfo {author} {\bibfnamefont {N.}~\bibnamefont {M\'ujica}},
  \ and\ \bibinfo {author} {\bibfnamefont {R.}~\bibnamefont {Soto}},\
  }\bibfield  {title} {\enquote {\bibinfo {title} {Fluctuations and criticality
  of a granular solid-liquid-like phase transition},}\ }\href@noop {}
  {\bibfield  {journal} {\bibinfo  {journal} {Phys. Rev. Lett.}\ }\textbf
  {\bibinfo {volume} {109}},\ \bibinfo {pages} {095701} (\bibinfo {year}
  {2012})}\BibitemShut {NoStop}%
\bibitem [{\citenamefont {Mujica}\ and\ \citenamefont
  {Soto}(2016)}]{mujica2016dynamics}%
  \BibitemOpen
  \bibfield  {author} {\bibinfo {author} {\bibfnamefont {N.}~\bibnamefont
  {Mujica}}\ and\ \bibinfo {author} {\bibfnamefont {R.}~\bibnamefont {Soto}},\
  }\bibfield  {title} {\enquote {\bibinfo {title} {Dynamics of noncohesive
  confined granular media},}\ }in\ \href@noop {} {\emph {\bibinfo {booktitle}
  {Recent Advances in Fluid Dynamics with Environmental Applications}}}\
  (\bibinfo  {publisher} {Springer},\ \bibinfo {year} {2016})\ pp.\ \bibinfo
  {pages} {445--463}\BibitemShut {NoStop}%
  \bibitem [{\citenamefont {Brito}, \citenamefont {Risso},\ and\ \citenamefont
  {Soto}(2013)}]{BRS13}%
  \BibitemOpen
  \bibfield  {author} {\bibinfo {author} {\bibfnamefont {R.}~\bibnamefont
  {Brito}}, \bibinfo {author} {\bibfnamefont {D.}~\bibnamefont {Risso}}, \ and\
  \bibinfo {author} {\bibfnamefont {R.}~\bibnamefont {Soto}},\ }\bibfield
  {title} {\enquote {\bibinfo {title} {Hydrodynamic modes in a confined
  granular fluid},}\ }\href@noop {} {\bibfield  {journal} {\bibinfo  {journal}
  {Phys. Rev. E}\ }\textbf {\bibinfo {volume} {87}},\ \bibinfo {pages} {022209}
  (\bibinfo {year} {2013})}\BibitemShut {NoStop}%
\bibitem [{\citenamefont {Maynar}, \citenamefont {Garc\'{\i}a~de Soria},\ and\
  \citenamefont {Brey}(2019)}]{MGB19a}%
  \BibitemOpen
  \bibfield  {author} {\bibinfo {author} {\bibfnamefont {P.}~\bibnamefont
  {Maynar}}, \bibinfo {author} {\bibfnamefont {I.}~\bibnamefont {Garc\'{\i}a~de
  Soria}}, \ and\ \bibinfo {author} {\bibfnamefont {J.~J.}\ \bibnamefont
  {Brey}},\ }\bibfield  {title} {\enquote {\bibinfo {title} {Homogeneous
  dynamics in a vibrated granular monolayer},}\ }\href@noop {} {\bibfield
  {journal} {\bibinfo  {journal} {J. Stat. Mech.}\ }\textbf {\bibinfo {volume}
  {{093}{205}}} (\bibinfo {year} {2019})}\BibitemShut {NoStop}%
\bibitem [{\citenamefont {Risso}, \citenamefont {Soto},\ and\ \citenamefont
  {Guzm\'an}(2018)}]{RSG18}%
  \BibitemOpen
  \bibfield  {author} {\bibinfo {author} {\bibfnamefont {D.}~\bibnamefont
  {Risso}}, \bibinfo {author} {\bibfnamefont {R.}~\bibnamefont {Soto}}, \ and\
  \bibinfo {author} {\bibfnamefont {R.}~\bibnamefont {Guzm\'an}},\ }\bibfield
  {title} {\enquote {\bibinfo {title} {Effective two-dimensional model for
  granular matter with phase separation},}\ }\href@noop {} {\bibfield
  {journal} {\bibinfo  {journal} {Phys. Rev. E}\ }\textbf {\bibinfo {volume}
  {98}},\ \bibinfo {pages} {{022}{901}} (\bibinfo {year} {2018})}\BibitemShut
  {NoStop}%
\bibitem [{\citenamefont {Brey}\ \emph {et~al.}(2013)\citenamefont {Brey},
  \citenamefont {Garc\'ia~de {S}oria}, \citenamefont {Maynar},\ and\
  \citenamefont {Buz\'on}}]{BGMB13}%
  \BibitemOpen
  \bibfield  {author} {\bibinfo {author} {\bibfnamefont {J.~J.}\ \bibnamefont
  {Brey}}, \bibinfo {author} {\bibfnamefont {M.~I.}\ \bibnamefont {Garc\'ia~de
  {S}oria}}, \bibinfo {author} {\bibfnamefont {P.}~\bibnamefont {Maynar}}, \
  and\ \bibinfo {author} {\bibfnamefont {V.}~\bibnamefont {Buz\'on}},\
  }\bibfield  {title} {\enquote {\bibinfo {title} {Homogeneous steady state of
  a confined granular gas},}\ }\href@noop {} {\bibfield  {journal} {\bibinfo
  {journal} {Phys. Rev. E}\ }\textbf {\bibinfo {volume} {88}},\ \bibinfo
  {pages} {062205} (\bibinfo {year} {2013})}\BibitemShut {NoStop}%
\bibitem [{\citenamefont {Brey}\ \emph {et~al.}(2014)\citenamefont {Brey},
  \citenamefont {Maynar}, \citenamefont {Garc\'ia~de {S}oria},\ and\
  \citenamefont {Buz\'on}}]{BMGB14}%
  \BibitemOpen
  \bibfield  {author} {\bibinfo {author} {\bibfnamefont {J.~J.}\ \bibnamefont
  {Brey}}, \bibinfo {author} {\bibfnamefont {P.}~\bibnamefont {Maynar}},
  \bibinfo {author} {\bibfnamefont {M.~I.}\ \bibnamefont {Garc\'ia~de
  {S}oria}}, \ and\ \bibinfo {author} {\bibfnamefont {V.}~\bibnamefont
  {Buz\'on}},\ }\bibfield  {title} {\enquote {\bibinfo {title} {Homogeneous
  hydrodynamics of a collisional model of confined granular gases},}\
  }\href@noop {} {\bibfield  {journal} {\bibinfo  {journal} {Phys. Rev. E}\
  }\textbf {\bibinfo {volume} {89}},\ \bibinfo {pages} {052209} (\bibinfo
  {year} {2014})}\BibitemShut {NoStop}%
\bibitem [{\citenamefont {Brey}\ \emph {et~al.}(2015)\citenamefont {Brey},
  \citenamefont {Buz\'on}, \citenamefont {Maynar},\ and\ \citenamefont
  {Garc\'{\i}a~de Soria}}]{BBMG15}%
  \BibitemOpen
  \bibfield  {author} {\bibinfo {author} {\bibfnamefont {J.~J.}\ \bibnamefont
  {Brey}}, \bibinfo {author} {\bibfnamefont {V.}~\bibnamefont {Buz\'on}},
  \bibinfo {author} {\bibfnamefont {P.}~\bibnamefont {Maynar}}, \ and\ \bibinfo
  {author} {\bibfnamefont {M.}~\bibnamefont {Garc\'{\i}a~de Soria}},\
  }\bibfield  {title} {\enquote {\bibinfo {title} {Hydrodynamics for a model of
  a confined quasi-two-dimensional granular gas},}\ }\href@noop {} {\bibfield
  {journal} {\bibinfo  {journal} {Phys. Rev. E}\ }\textbf {\bibinfo {volume}
  {91}},\ \bibinfo {pages} {052201} (\bibinfo {year} {2015})}\BibitemShut
  {NoStop}%
\bibitem [{\citenamefont {Brey}\ \emph {et~al.}(2016)\citenamefont {Brey},
  \citenamefont {Buz\'on}, \citenamefont {Garc\'ia~de Soria},\ and\
  \citenamefont {Maynar}}]{BBGM16}%
  \BibitemOpen
  \bibfield  {author} {\bibinfo {author} {\bibfnamefont {J.~J.}\ \bibnamefont
  {Brey}}, \bibinfo {author} {\bibfnamefont {V.}~\bibnamefont {Buz\'on}},
  \bibinfo {author} {\bibfnamefont {M.~I.}\ \bibnamefont {Garc\'ia~de Soria}},
  \ and\ \bibinfo {author} {\bibfnamefont {P.}~\bibnamefont {Maynar}},\
  }\bibfield  {title} {\enquote {\bibinfo {title} {Stability analysis of the
  homogeneous hydrodynamics of a model for a confined granular gas},}\
  }\href@noop {} {\bibfield  {journal} {\bibinfo  {journal} {Phys. Rev. E}\
  }\textbf {\bibinfo {volume} {93}},\ \bibinfo {pages} {{062}{907}} (\bibinfo
  {year} {2016})}\BibitemShut {NoStop}%
\bibitem [{\citenamefont {Soto}, \citenamefont {Risso},\ and\ \citenamefont
  {Brito}(2014)}]{SRB14}%
  \BibitemOpen
  \bibfield  {author} {\bibinfo {author} {\bibfnamefont {R.}~\bibnamefont
  {Soto}}, \bibinfo {author} {\bibfnamefont {D.}~\bibnamefont {Risso}}, \ and\
  \bibinfo {author} {\bibfnamefont {R.}~\bibnamefont {Brito}},\ }\bibfield
  {title} {\enquote {\bibinfo {title} {Shear viscosity of a model for confined
  granular media},}\ }\href@noop {} {\bibfield  {journal} {\bibinfo  {journal}
  {Phys. Rev. E}\ }\textbf {\bibinfo {volume} {90}},\ \bibinfo {pages} {062204}
  (\bibinfo {year} {2014})}\BibitemShut {NoStop}%
  \bibitem{GBS18}V. Garz\'o, R. Brito, and R. Soto, ``Enskog kinetic theory for a model of a
  confined quasi-two-dimensional granular fluid,'' Phys. Rev. E \textbf{98}, 052904 (2018)
  [Erratum] Phys. Rev. E \textbf{102}, 05901 (2020).
  \bibitem [{\citenamefont {Brito}, \citenamefont {Soto},\ and\ \citenamefont
  {Garz\'o}(2020)}]{BSG20}%
  \BibitemOpen
  \bibfield  {author} {\bibinfo {author} {\bibfnamefont {R.}~\bibnamefont
  {Brito}}, \bibinfo {author} {\bibfnamefont {R.}~\bibnamefont {Soto}}, \ and\
  \bibinfo {author} {\bibfnamefont {V.}~\bibnamefont {Garz\'o}},\ }\bibfield
  {title} {\enquote {\bibinfo {title} {Energy nonequipartition in a collisional
  model of a confined quasi-two-dimensional granular mixture},}\ }\href@noop {}
  {\bibfield  {journal} {\bibinfo  {journal} {arXiv:2009.02957}\ } (\bibinfo
  {year} {2020})}\BibitemShut {NoStop}%
\bibitem [{\citenamefont {Garz\'o}\ and\ \citenamefont {Dufty}(2002)}]{GD02}%
  \BibitemOpen
  \bibfield  {author} {\bibinfo {author} {\bibfnamefont {V.}~\bibnamefont
  {Garz\'o}}\ and\ \bibinfo {author} {\bibfnamefont {J.~W.}\ \bibnamefont
  {Dufty}},\ }\bibfield  {title} {\enquote {\bibinfo {title} {Hydrodynamics for
  a granular binary mixture at low density},}\ }\href@noop {} {\bibfield
  {journal} {\bibinfo  {journal} {Phys. Fluids.}\ }\textbf {\bibinfo {volume}
  {14}},\ \bibinfo {pages} {1476--1490} (\bibinfo {year} {2002})}\BibitemShut
  {NoStop}%
\bibitem [{\citenamefont {Garz\'o}, \citenamefont {Montanero},\ and\
  \citenamefont {Dufty}(2006)}]{GMD06}%
  \BibitemOpen
  \bibfield  {author} {\bibinfo {author} {\bibfnamefont {V.}~\bibnamefont
  {Garz\'o}}, \bibinfo {author} {\bibfnamefont {J.~M.}\ \bibnamefont
  {Montanero}}, \ and\ \bibinfo {author} {\bibfnamefont {J.~W.}\ \bibnamefont
  {Dufty}},\ }\bibfield  {title} {\enquote {\bibinfo {title} {Mass and heat
  fluxes for a binary granular mixture at low density},}\ }\href@noop {}
  {\bibfield  {journal} {\bibinfo  {journal} {Phys. Fluids}\ }\textbf {\bibinfo
  {volume} {18}},\ \bibinfo {pages} {{083}{305}} (\bibinfo {year}
  {2006})}\BibitemShut {NoStop}%
\bibitem [{\citenamefont {Serero}\ \emph {et~al.}(2006)\citenamefont {Serero},
  \citenamefont {Goldhirsch}, \citenamefont {Noskowicz},\ and\ \citenamefont
  {Tan}}]{SGNT06}%
  \BibitemOpen
  \bibfield  {author} {\bibinfo {author} {\bibfnamefont {D.}~\bibnamefont
  {Serero}}, \bibinfo {author} {\bibfnamefont {I.}~\bibnamefont {Goldhirsch}},
  \bibinfo {author} {\bibfnamefont {S.~H.}\ \bibnamefont {Noskowicz}}, \ and\
  \bibinfo {author} {\bibfnamefont {M.~L.}\ \bibnamefont {Tan}},\ }\bibfield
  {title} {\enquote {\bibinfo {title} {Hydrodynamics of granular gases and
  granular gas mixtures},}\ }\href@noop {} {\bibfield  {journal} {\bibinfo
  {journal} {J. Fluid Mech.}\ }\textbf {\bibinfo {volume} {554}},\ \bibinfo
  {pages} {237--258} (\bibinfo {year} {2006})}\BibitemShut {NoStop}%
\bibitem [{\citenamefont {Garz\'o}\ and\ \citenamefont
  {Montanero}(2007)}]{GM07}%
  \BibitemOpen
  \bibfield  {author} {\bibinfo {author} {\bibfnamefont {V.}~\bibnamefont
  {Garz\'o}}\ and\ \bibinfo {author} {\bibfnamefont {J.~M.}\ \bibnamefont
  {Montanero}},\ }\bibfield  {title} {\enquote {\bibinfo {title}
  {\textsc{N}avier--\textsc{S}tokes transport coefficients of $d$-dimensional
  granular binary mixtures at low-density},}\ }\href@noop {} {\bibfield
  {journal} {\bibinfo  {journal} {J. Stat. Phys.}\ }\textbf {\bibinfo {volume}
  {129}},\ \bibinfo {pages} {27--58} (\bibinfo {year} {2007})}\BibitemShut
  {NoStop}%
\bibitem [{\citenamefont {Serero}\ \emph {et~al.}(2009)\citenamefont {Serero},
  \citenamefont {Noskowicz}, \citenamefont {Tan},\ and\ \citenamefont
  {Goldhirsch}}]{SNTG09}%
  \BibitemOpen
  \bibfield  {author} {\bibinfo {author} {\bibfnamefont {D.}~\bibnamefont
  {Serero}}, \bibinfo {author} {\bibfnamefont {S.~H.}\ \bibnamefont
  {Noskowicz}}, \bibinfo {author} {\bibfnamefont {M.~L.}\ \bibnamefont {Tan}},
  \ and\ \bibinfo {author} {\bibfnamefont {I.}~\bibnamefont {Goldhirsch}},\
  }\bibfield  {title} {\enquote {\bibinfo {title} {Binary granular gas
  mixtures: {T}heory, layering effects and some open questions},}\ }\href@noop
  {} {\bibfield  {journal} {\bibinfo  {journal} {Eur. Phys. J. Special Topics}\
  }\textbf {\bibinfo {volume} {179}},\ \bibinfo {pages} {221--247} (\bibinfo
  {year} {2009})}\BibitemShut {NoStop}%
\bibitem [{\citenamefont {Garz\'o}, \citenamefont {Murray},\ and\ \citenamefont
  {Vega~Reyes}(2013)}]{GMV13a}%
  \BibitemOpen
  \bibfield  {author} {\bibinfo {author} {\bibfnamefont {V.}~\bibnamefont
  {Garz\'o}}, \bibinfo {author} {\bibfnamefont {J.~A.}\ \bibnamefont {Murray}},
  \ and\ \bibinfo {author} {\bibfnamefont {F.}~\bibnamefont {Vega~Reyes}},\
  }\bibfield  {title} {\enquote {\bibinfo {title} {Diffusion transport
  coefficients for granular binary mixtures at low density: {T}hermal diffusion
  segregation},}\ }\href@noop {} {\bibfield  {journal} {\bibinfo  {journal}
  {Phys. Fluids}\ }\textbf {\bibinfo {volume} {25}},\ \bibinfo {pages}
  {{043}{302}} (\bibinfo {year} {2013})}\BibitemShut {NoStop}%
\bibitem [{\citenamefont {Chapman}\ and\ \citenamefont {Cowling}(1970)}]{CC70}%
  \BibitemOpen
  \bibfield  {author} {\bibinfo {author} {\bibfnamefont {S.}~\bibnamefont
  {Chapman}}\ and\ \bibinfo {author} {\bibfnamefont {T.~G.}\ \bibnamefont
  {Cowling}},\ }\href@noop {} {\emph {\bibinfo {title} {The Mathematical Theory
  of Nonuniform Gases}}}\ (\bibinfo  {publisher} {Cambridge University Press,
  Cambridge},\ \bibinfo {year} {1970})\BibitemShut {NoStop}%
\bibitem [{\citenamefont {Karkheck}\ and\ \citenamefont {Stell}(1979)}]{KS79b}%
  \BibitemOpen
  \bibfield  {author} {\bibinfo {author} {\bibfnamefont {J.}~\bibnamefont
  {Karkheck}}\ and\ \bibinfo {author} {\bibfnamefont {G.}~\bibnamefont
  {Stell}},\ }\bibfield  {title} {\enquote {\bibinfo {title} {Bulk viscosity of
  fluid mixtures},}\ }\href@noop {} {\bibfield  {journal} {\bibinfo  {journal}
  {J. Chem. Phys.}\ }\textbf {\bibinfo {volume} {71}},\ \bibinfo {pages}
  {3636--3639} (\bibinfo {year} {1979})}\BibitemShut {NoStop}%
\bibitem [{\citenamefont {G\'omez~Gonz\'alez}\ and\ \citenamefont
  {Garz\'o}(2019)}]{GGG19b}%
  \BibitemOpen
  \bibfield  {author} {\bibinfo {author} {\bibfnamefont {R.}~\bibnamefont
  {G\'omez~Gonz\'alez}}\ and\ \bibinfo {author} {\bibfnamefont
  {V.}~\bibnamefont {Garz\'o}},\ }\bibfield  {title} {\enquote {\bibinfo
  {title} {Influence of the first-order contributions to the partial
  temperatures on transport properties in polydisperse dense granular
  mixtures},}\ }\href@noop {} {\bibfield  {journal} {\bibinfo  {journal} {Phys.
  Rev. E}\ }\textbf {\bibinfo {volume} {100}},\ \bibinfo {pages} {032904}
  (\bibinfo {year} {2019})}\BibitemShut {NoStop}%
\bibitem [{\citenamefont {Lutsko}(2004)}]{L04bis}%
  \BibitemOpen
  \bibfield  {author} {\bibinfo {author} {\bibfnamefont {J.~F.}\ \bibnamefont
  {Lutsko}},\ }\bibfield  {title} {\enquote {\bibinfo {title} {Kinetic theory
  and hydrodynamics of dense, reacting fluids far from equilibrium},}\
  }\href@noop {} {\bibfield  {journal} {\bibinfo  {journal} {J. Chem. Phys.}\
  }\textbf {\bibinfo {volume} {120}},\ \bibinfo {pages} {6325} (\bibinfo {year}
  {2004})}\BibitemShut {NoStop}%
\bibitem [{\citenamefont {Ferziger}\ and\ \citenamefont {Kaper}(1972)}]{FK72}%
  \BibitemOpen
  \bibfield  {author} {\bibinfo {author} {\bibfnamefont {J.~H.}\ \bibnamefont
  {Ferziger}}\ and\ \bibinfo {author} {\bibfnamefont {G.~H.}\ \bibnamefont
  {Kaper}},\ }\href@noop {} {\emph {\bibinfo {title} {Mathematical Theory of
  Transport Processes in Gases}}}\ (\bibinfo  {publisher} {North-Holland,
  Amsterdam},\ \bibinfo {year} {1972})\BibitemShut {NoStop}%
\bibitem [{\citenamefont {Garz\'o}\ and\ \citenamefont {Santos}(2003)}]{GS03}%
  \BibitemOpen
  \bibfield  {author} {\bibinfo {author} {\bibfnamefont {V.}~\bibnamefont
  {Garz\'o}}\ and\ \bibinfo {author} {\bibfnamefont {A.}~\bibnamefont
  {Santos}},\ }\href@noop {} {\emph {\bibinfo {title} {Kinetic Theory of Gases
  in Shear Flows. Nonlinear Transport}}}\ (\bibinfo  {publisher} {Kluwer
  Academic Publishers, Dordrecht},\ \bibinfo {year} {2003})\BibitemShut
  {NoStop}%
 \bibitem{S16}R. Soto, \emph{Kinetic Theory and Transport Phenomena}, Vol. 25 (Oxford University Press, Oxford, 2016).

\bibitem [{\citenamefont {Garz\'o}\ and\ \citenamefont {Dufty}(1999)}]{GD99b}%
  \BibitemOpen
  \bibfield  {author} {\bibinfo {author} {\bibfnamefont {V.}~\bibnamefont
  {Garz\'o}}\ and\ \bibinfo {author} {\bibfnamefont {J.~W.}\ \bibnamefont
  {Dufty}},\ }\bibfield  {title} {\enquote {\bibinfo {title} {Homogeneous
  cooling state for a granular mixture},}\ }\href@noop {} {\bibfield  {journal}
  {\bibinfo  {journal} {Phys. Rev. E}\ }\textbf {\bibinfo {volume} {60}},\
  \bibinfo {pages} {5706--5713} (\bibinfo {year} {1999})}\BibitemShut {NoStop}%
\bibitem [{\citenamefont {Brey}, \citenamefont {Ruiz-Montero},\ and\
  \citenamefont {Cubero}(1999)}]{BRC99}%
  \BibitemOpen
  \bibfield  {author} {\bibinfo {author} {\bibfnamefont {J.~J.}\ \bibnamefont
  {Brey}}, \bibinfo {author} {\bibfnamefont {M.~J.}\ \bibnamefont
  {Ruiz-Montero}}, \ and\ \bibinfo {author} {\bibfnamefont {D.}~\bibnamefont
  {Cubero}},\ }\bibfield  {title} {\enquote {\bibinfo {title} {On the validity
  of linear hydrodynamics for low-density granular flows described by the
  {B}oltzmann equation},}\ }\href@noop {} {\bibfield  {journal} {\bibinfo
  {journal} {Europhys. Lett.}\ }\textbf {\bibinfo {volume} {48}},\ \bibinfo
  {pages} {359--364} (\bibinfo {year} {1999})}\BibitemShut {NoStop}%
\bibitem [{\citenamefont {Brey}\ \emph {et~al.}(2000)\citenamefont {Brey},
  \citenamefont {Ruiz-Montero}, \citenamefont {Cubero},\ and\ \citenamefont
  {Garc\'{\i}a-Rojo}}]{BRCG00}%
  \BibitemOpen
  \bibfield  {author} {\bibinfo {author} {\bibfnamefont {J.~J.}\ \bibnamefont
  {Brey}}, \bibinfo {author} {\bibfnamefont {M.~J.}\ \bibnamefont
  {Ruiz-Montero}}, \bibinfo {author} {\bibfnamefont {D.}~\bibnamefont
  {Cubero}}, \ and\ \bibinfo {author} {\bibfnamefont {R.}~\bibnamefont
  {Garc\'{\i}a-Rojo}},\ }\bibfield  {title} {\enquote {\bibinfo {title}
  {Self-diffusion in freely evolving granular gases},}\ }\href@noop {}
  {\bibfield  {journal} {\bibinfo  {journal} {Phys. Fluids.}\ }\textbf
  {\bibinfo {volume} {12}},\ \bibinfo {pages} {876--883} (\bibinfo {year}
  {2000})}\BibitemShut {NoStop}%
\bibitem [{\citenamefont {Brey}, \citenamefont {Ruiz-Montero},\ and\
  \citenamefont {Moreno}(2001)}]{BRM01}%
  \BibitemOpen
  \bibfield  {author} {\bibinfo {author} {\bibfnamefont {J.~J.}\ \bibnamefont
  {Brey}}, \bibinfo {author} {\bibfnamefont {M.~J.}\ \bibnamefont
  {Ruiz-Montero}}, \ and\ \bibinfo {author} {\bibfnamefont {F.}~\bibnamefont
  {Moreno}},\ }\bibfield  {title} {\enquote {\bibinfo {title} {Hydrodynamics of
  an open vibrated granular system},}\ }\href@noop {} {\bibfield  {journal}
  {\bibinfo  {journal} {Phys. Rev. E}\ }\textbf {\bibinfo {volume} {63}},\
  \bibinfo {pages} {{061}{305}} (\bibinfo {year} {2001})}\BibitemShut {NoStop}%
\bibitem [{\citenamefont {Dahl}\ \emph {et~al.}(2002)\citenamefont {Dahl},
  \citenamefont {Hrenya}, \citenamefont {Garz\'o},\ and\ \citenamefont
  {Dufty}}]{DHGD02}%
  \BibitemOpen
  \bibfield  {author} {\bibinfo {author} {\bibfnamefont {S.~R.}\ \bibnamefont
  {Dahl}}, \bibinfo {author} {\bibfnamefont {C.~M.}\ \bibnamefont {Hrenya}},
  \bibinfo {author} {\bibfnamefont {V.}~\bibnamefont {Garz\'o}}, \ and\
  \bibinfo {author} {\bibfnamefont {J.~W.}\ \bibnamefont {Dufty}},\ }\bibfield
  {title} {\enquote {\bibinfo {title} {Kinetic temperatures for a granular
  mixture},}\ }\href@noop {} {\bibfield  {journal} {\bibinfo  {journal} {Phys.
  Rev. E}\ }\textbf {\bibinfo {volume} {66}},\ \bibinfo {pages} {{041}{301}}
  (\bibinfo {year} {2002})}\BibitemShut {NoStop}%
\bibitem [{\citenamefont {Lutsko}, \citenamefont {Brey},\ and\ \citenamefont
  {Dufty}(2002)}]{LBD02}%
  \BibitemOpen
  \bibfield  {author} {\bibinfo {author} {\bibfnamefont {J.~F.}\ \bibnamefont
  {Lutsko}}, \bibinfo {author} {\bibfnamefont {J.~J.}\ \bibnamefont {Brey}}, \
  and\ \bibinfo {author} {\bibfnamefont {J.~W.}\ \bibnamefont {Dufty}},\
  }\bibfield  {title} {\enquote {\bibinfo {title} {Diffusion in a granular
  fluid. \textsc{II}. \textsc{S}imulation},}\ }\href@noop {} {\bibfield
  {journal} {\bibinfo  {journal} {Phys. Rev. E}\ }\textbf {\bibinfo {volume}
  {65}},\ \bibinfo {pages} {{051}{304}} (\bibinfo {year} {2002})}\BibitemShut
  {NoStop}%
\bibitem [{\citenamefont {Montanero}\ and\ \citenamefont
  {Garz\'o}(2002)}]{MG02}%
  \BibitemOpen
  \bibfield  {author} {\bibinfo {author} {\bibfnamefont {J.~M.}\ \bibnamefont
  {Montanero}}\ and\ \bibinfo {author} {\bibfnamefont {V.}~\bibnamefont
  {Garz\'o}},\ }\bibfield  {title} {\enquote {\bibinfo {title} {Monte {C}arlo
  simulation of the homogeneous cooling state for a granular mixture},}\
  }\href@noop {} {\bibfield  {journal} {\bibinfo  {journal} {Granular Matter}\
  }\textbf {\bibinfo {volume} {4}},\ \bibinfo {pages} {17--24} (\bibinfo {year}
  {2002})}\BibitemShut {NoStop}%
\bibitem [{\citenamefont {Montanero}\ and\ \citenamefont
  {Garz\'o}(2003)}]{MG03}%
  \BibitemOpen
  \bibfield  {author} {\bibinfo {author} {\bibfnamefont {J.~M.}\ \bibnamefont
  {Montanero}}\ and\ \bibinfo {author} {\bibfnamefont {V.}~\bibnamefont
  {Garz\'o}},\ }\bibfield  {title} {\enquote {\bibinfo {title} {Shear viscosity
  for a heated granular binary mixture at low density},}\ }\href@noop {}
  {\bibfield  {journal} {\bibinfo  {journal} {Phys. Rev. E}\ }\textbf {\bibinfo
  {volume} {67}},\ \bibinfo {pages} {{021}{308}} (\bibinfo {year}
  {2003})}\BibitemShut {NoStop}%
\bibitem [{\citenamefont {Garz\'o}\ and\ \citenamefont
  {Montanero}(2003)}]{GM03}%
  \BibitemOpen
  \bibfield  {author} {\bibinfo {author} {\bibfnamefont {V.}~\bibnamefont
  {Garz\'o}}\ and\ \bibinfo {author} {\bibfnamefont {J.~M.}\ \bibnamefont
  {Montanero}},\ }\bibfield  {title} {\enquote {\bibinfo {title} {Shear
  viscosity for a moderately dense granular binary mixture},}\ }\href@noop {}
  {\bibfield  {journal} {\bibinfo  {journal} {Phys. Rev. E}\ }\textbf {\bibinfo
  {volume} {68}},\ \bibinfo {pages} {{041}{302}} (\bibinfo {year}
  {2003})}\BibitemShut {NoStop}%
\bibitem [{\citenamefont {Brey}, \citenamefont {Ruiz-Montero},\ and\
  \citenamefont {Moreno}(2005)}]{BRM05}%
  \BibitemOpen
  \bibfield  {author} {\bibinfo {author} {\bibfnamefont {J.~J.}\ \bibnamefont
  {Brey}}, \bibinfo {author} {\bibfnamefont {M.~J.}\ \bibnamefont
  {Ruiz-Montero}}, \ and\ \bibinfo {author} {\bibfnamefont {F.}~\bibnamefont
  {Moreno}},\ }\bibfield  {title} {\enquote {\bibinfo {title} {Energy partition
  and segregation for an intruder in a vibrated granular system under
  gravity},}\ }\href@noop {} {\bibfield  {journal} {\bibinfo  {journal} {Phys.
  Rev. Lett.}\ }\textbf {\bibinfo {volume} {95}},\ \bibinfo {pages}
  {{098}{001}} (\bibinfo {year} {2005})}\BibitemShut {NoStop}%
\bibitem [{\citenamefont {Lois}, \citenamefont {Lema\^itre},\ and\
  \citenamefont {Carlson}(2007)}]{LLC07}%
  \BibitemOpen
  \bibfield  {author} {\bibinfo {author} {\bibfnamefont {G.}~\bibnamefont
  {Lois}}, \bibinfo {author} {\bibfnamefont {A.}~\bibnamefont {Lema\^itre}}, \
  and\ \bibinfo {author} {\bibfnamefont {J.~M.}\ \bibnamefont {Carlson}},\
  }\bibfield  {title} {\enquote {\bibinfo {title} {Spatial force correlations
  in granular shear flow. {II}. \textsc{T}heoretical implications},}\
  }\href@noop {} {\bibfield  {journal} {\bibinfo  {journal} {Phys. Rev. E}\
  }\textbf {\bibinfo {volume} {76}},\ \bibinfo {pages} {{021}{303}} (\bibinfo
  {year} {2007})}\BibitemShut {NoStop}%
\bibitem [{\citenamefont {Mitrano}\ \emph {et~al.}(2011)\citenamefont
  {Mitrano}, \citenamefont {Dhal}, \citenamefont {Cromer}, \citenamefont
  {Pacella},\ and\ \citenamefont {Hrenya}}]{MDCPH11}%
  \BibitemOpen
  \bibfield  {author} {\bibinfo {author} {\bibfnamefont {P.~P.}\ \bibnamefont
  {Mitrano}}, \bibinfo {author} {\bibfnamefont {S.~R.}\ \bibnamefont {Dhal}},
  \bibinfo {author} {\bibfnamefont {D.~J.}\ \bibnamefont {Cromer}}, \bibinfo
  {author} {\bibfnamefont {M.~S.}\ \bibnamefont {Pacella}}, \ and\ \bibinfo
  {author} {\bibfnamefont {C.~M.}\ \bibnamefont {Hrenya}},\ }\bibfield  {title}
  {\enquote {\bibinfo {title} {Instabilities in the homogeneous cooling of a
  granular gas: a quantitative assessment of kinetic-theory predictions},}\
  }\href@noop {} {\bibfield  {journal} {\bibinfo  {journal} {Phys. Fluids}\
  }\textbf {\bibinfo {volume} {23}},\ \bibinfo {pages} {{093}{303}} (\bibinfo
  {year} {2011})}\BibitemShut {NoStop}%
\bibitem [{\citenamefont {Brey}\ and\ \citenamefont
  {Ruiz-Montero}(2013)}]{BR13}%
  \BibitemOpen
  \bibfield  {author} {\bibinfo {author} {\bibfnamefont {J.~J.}\ \bibnamefont
  {Brey}}\ and\ \bibinfo {author} {\bibfnamefont {M.~J.}\ \bibnamefont
  {Ruiz-Montero}},\ }\bibfield  {title} {\enquote {\bibinfo {title} {Shearing
  instability of a dilute granular mixture},}\ }\href@noop {} {\bibfield
  {journal} {\bibinfo  {journal} {Phys. Rev. E}\ }\textbf {\bibinfo {volume}
  {87}},\ \bibinfo {pages} {{022}{210}} (\bibinfo {year} {2013})}\BibitemShut
  {NoStop}%
\bibitem [{\citenamefont {Mitrano}, \citenamefont {Garz\'o},\ and\
  \citenamefont {Hrenya}(2014)}]{MGH14}%
  \BibitemOpen
  \bibfield  {author} {\bibinfo {author} {\bibfnamefont {P.~P.}\ \bibnamefont
  {Mitrano}}, \bibinfo {author} {\bibfnamefont {V.}~\bibnamefont {Garz\'o}}, \
  and\ \bibinfo {author} {\bibfnamefont {C.~M.}\ \bibnamefont {Hrenya}},\
  }\bibfield  {title} {\enquote {\bibinfo {title} {Instabilities in granular
  binary mixtures at moderate densities},}\ }\href@noop {} {\bibfield
  {journal} {\bibinfo  {journal} {Phys. Rev. E}\ }\textbf {\bibinfo {volume}
  {89}},\ \bibinfo {pages} {{020}{201}(R)} (\bibinfo {year}
  {2014})}\BibitemShut {NoStop}%
\bibitem{note1}In standard textbooks like Ref.\ 44, a factor $\epsilon ^{-1}$ is usually assigned to the Boltzmann
  collision operator $J_{ij}[f_i,f_j]$ so that the operators $\partial _t$ and
  $\nabla $ are formally of order $\epsilon ^0$. In this scheme, the operator
  $\partial _t^{(0)}$ is the same as our operator $\partial _t^{(1)}$ of Eqs.\
  (3.8)--(3.10) The results derived from both formulations are of course
  completely equivalent for elastic collisions.
\bibitem{note2} As discussed in Ref.\ 39, strictly speaking the thermal conductivity in a mixture is generally measured
in the absence of diffusion (i.e., when $\protect \mathbf {j}_1=\protect
\mathbf {0}$). To identify this coefficient, we have to express the heat flow
$\protect \mathbf {J}_q=\protect \mathbf
{q}-(d+2)/2)T((m_2-m_1)/m_1m_2)\protect \mathbf {j}_1$ in terms of $\protect
\mathbf {j}_1$, $T$, and $p$. The corresponding coefficient of $\nabla T$
defines the thermal conductivity coefficient.
\bibitem{KCLH87}J. M. Kincaid, E. G. D. Cohen, and M. L\'opez de Haro, ``The Enskog theory for multicomponent mixtures. IV. Thermal diffusion''J. Chem. Phys. \textbf{86}, 963--975 (1987).
  \bibitem [{\citenamefont {Jenkins}\ and\ \citenamefont {Yoon}(2002)}]{JY02}%
  \BibitemOpen
  \bibfield  {author} {\bibinfo {author} {\bibfnamefont {J.~T.}\ \bibnamefont
  {Jenkins}}\ and\ \bibinfo {author} {\bibfnamefont {D.~K.}\ \bibnamefont
  {Yoon}},\ }\bibfield  {title} {\enquote {\bibinfo {title} {Segregation in
  binary mixtures under gravity},}\ }\href@noop {} {\bibfield  {journal}
  {\bibinfo  {journal} {Phys. Rev. Lett.}\ }\textbf {\bibinfo {volume} {88}},\
  \bibinfo {pages} {{194}{301}} (\bibinfo {year} {2002})}\BibitemShut {NoStop}%
\bibitem [{\citenamefont {Brey}, \citenamefont {Ruiz-Montero},\ and\
  \citenamefont {Moreno}(2006)}]{BRM06}%
  \BibitemOpen
  \bibfield  {author} {\bibinfo {author} {\bibfnamefont {J.~J.}\ \bibnamefont
  {Brey}}, \bibinfo {author} {\bibfnamefont {M.~J.}\ \bibnamefont
  {Ruiz-Montero}}, \ and\ \bibinfo {author} {\bibfnamefont {F.}~\bibnamefont
  {Moreno}},\ }\bibfield  {title} {\enquote {\bibinfo {title} {Hydrodynamic
  profiles for an impurity in an open vibrated granular gas},}\ }\href@noop {}
  {\bibfield  {journal} {\bibinfo  {journal} {Phys. Rev. E}\ }\textbf {\bibinfo
  {volume} {73}},\ \bibinfo {pages} {{031}{301}} (\bibinfo {year}
  {2006})}\BibitemShut {NoStop}%
\bibitem [{\citenamefont {Garz\'o}(2006)}]{G06}%
  \BibitemOpen
  \bibfield  {author} {\bibinfo {author} {\bibfnamefont {V.}~\bibnamefont
  {Garz\'o}},\ }\bibfield  {title} {\enquote {\bibinfo {title} {Segregation in
  granular binary mixtures: {T}hermal diffusion},}\ }\href@noop {} {\bibfield
  {journal} {\bibinfo  {journal} {Europhys. Lett.}\ }\textbf {\bibinfo {volume}
  {75}},\ \bibinfo {pages} {521--527} (\bibinfo {year} {2006})}\BibitemShut
  {NoStop}%
\bibitem [{\citenamefont {Garz\'o}(2008)}]{G08a}%
  \BibitemOpen
  \bibfield  {author} {\bibinfo {author} {\bibfnamefont {V.}~\bibnamefont
  {Garz\'o}},\ }\bibfield  {title} {\enquote {\bibinfo {title} {Brazil-nut
  effect versus reverse {B}razil-nut effect in a moderately granular dense
  gas},}\ }\href@noop {} {\bibfield  {journal} {\bibinfo  {journal} {Phys. Rev.
  E}\ }\textbf {\bibinfo {volume} {78}},\ \bibinfo {pages} {{020}{301} (R)}
  (\bibinfo {year} {2008})}\BibitemShut {NoStop}%
\bibitem [{\citenamefont {Brito}\ \emph {et~al.}(2008)\citenamefont {Brito},
  \citenamefont {Enr\'{\i}quez}, \citenamefont {Godoy},\ and\ \citenamefont
  {Soto}}]{BEGS08}%
  \BibitemOpen
  \bibfield  {author} {\bibinfo {author} {\bibfnamefont {R.}~\bibnamefont
  {Brito}}, \bibinfo {author} {\bibfnamefont {H.}~\bibnamefont
  {Enr\'{\i}quez}}, \bibinfo {author} {\bibfnamefont {S.}~\bibnamefont
  {Godoy}}, \ and\ \bibinfo {author} {\bibfnamefont {R.}~\bibnamefont {Soto}},\
  }\bibfield  {title} {\enquote {\bibinfo {title} {Segregation induced by
  inelasticity in a vibrofluidized granular mixture},}\ }\href@noop {}
  {\bibfield  {journal} {\bibinfo  {journal} {Phys. Rev. E}\ }\textbf {\bibinfo
  {volume} {77}},\ \bibinfo {pages} {{061}{301}} (\bibinfo {year}
  {2008})}\BibitemShut {NoStop}%
\bibitem [{\citenamefont {Garz\'o}(2009)}]{G09}%
  \BibitemOpen
  \bibfield  {author} {\bibinfo {author} {\bibfnamefont {V.}~\bibnamefont
  {Garz\'o}},\ }\bibfield  {title} {\enquote {\bibinfo {title} {Segregation by
  thermal diffusion in moderately dense granular mixtures},}\ }\href@noop {}
  {\bibfield  {journal} {\bibinfo  {journal} {Eur. Phys. J. E}\ }\textbf
  {\bibinfo {volume} {29}},\ \bibinfo {pages} {261--274} (\bibinfo {year}
  {2009})}\BibitemShut {NoStop}%
\bibitem [{\citenamefont {Garz\'o}(2011)}]{G11}%
  \BibitemOpen
  \bibfield  {author} {\bibinfo {author} {\bibfnamefont {V.}~\bibnamefont
  {Garz\'o}},\ }\bibfield  {title} {\enquote {\bibinfo {title} {Thermal
  diffusion segregation in granular binary mixtures described by the {E}nskog
  equation},}\ }\href@noop {} {\bibfield  {journal} {\bibinfo  {journal} {New
  J. Phys.}\ }\textbf {\bibinfo {volume} {13}},\ \bibinfo {pages} {{055}{020}}
  (\bibinfo {year} {2011})}\BibitemShut {NoStop}%
\bibitem [{\citenamefont {de~Groot}\ and\ \citenamefont {Mazur}(1984)}]{GM84}%
  \BibitemOpen
  \bibfield  {author} {\bibinfo {author} {\bibfnamefont {S.~R.}\ \bibnamefont
  {de~Groot}}\ and\ \bibinfo {author} {\bibfnamefont {P.}~\bibnamefont
  {Mazur}},\ }\href@noop {} {\emph {\bibinfo {title} {Nonequilibrium
  Thermodynamics}}}\ (\bibinfo  {publisher} {Dover, New York},\ \bibinfo {year}
  {1984})\BibitemShut {NoStop}%
\bibitem [{\citenamefont {Soto}, \citenamefont {Mareschal},\ and\ \citenamefont
  {Risso}(1999)}]{SMR99}%
  \BibitemOpen
  \bibfield  {author} {\bibinfo {author} {\bibfnamefont {R.}~\bibnamefont
  {Soto}}, \bibinfo {author} {\bibfnamefont {M.}~\bibnamefont {Mareschal}}, \
  and\ \bibinfo {author} {\bibfnamefont {D.}~\bibnamefont {Risso}},\ }\bibfield
   {title} {\enquote {\bibinfo {title} {Departure from Fourier's law for
  fluidized granular media},}\ }\href@noop {} {\bibfield  {journal} {\bibinfo
  {journal} {Phys. Rev. Lett.}\ }\textbf {\bibinfo {volume} {83}},\ \bibinfo
  {pages} {5003} (\bibinfo {year} {1999})}\BibitemShut {NoStop}%
\bibitem [{\citenamefont {Khalil}\ and\ \citenamefont {Garz\'o}(2013)}]{KG13}%
  \BibitemOpen
  \bibfield  {author} {\bibinfo {author} {\bibfnamefont {N.}~\bibnamefont
  {Khalil}}\ and\ \bibinfo {author} {\bibfnamefont {V.}~\bibnamefont
  {Garz\'o}},\ }\bibfield  {title} {\enquote {\bibinfo {title} {Transport
  coefficients for driven granular mixtures at low-density},}\ }\href@noop {}
  {\bibfield  {journal} {\bibinfo  {journal} {Phys. Rev. E}\ }\textbf {\bibinfo
  {volume} {88}},\ \bibinfo {pages} {052201} (\bibinfo {year}
  {2013})}\BibitemShut {NoStop}%
\bibitem [{\citenamefont {Khalil}\ and\ \citenamefont {Garz\'o}(2018)}]{KG18}%
  \BibitemOpen
  \bibfield  {author} {\bibinfo {author} {\bibfnamefont {N.}~\bibnamefont
  {Khalil}}\ and\ \bibinfo {author} {\bibfnamefont {V.}~\bibnamefont
  {Garz\'o}},\ }\bibfield  {title} {\enquote {\bibinfo {title} {Heat flux of
  driven granular mixtures at low density: {S}tability analysis of the
  homogeneous steady state},}\ }\href@noop {} {\bibfield  {journal} {\bibinfo
  {journal} {Phys. Rev. E}\ }\textbf {\bibinfo {volume} {97}},\ \bibinfo
  {pages} {022902} (\bibinfo {year} {2018})}\BibitemShut {NoStop}%
\bibitem [{\citenamefont {G\'omez~Gonz\'alez}, \citenamefont {Khalil},\ and\
  \citenamefont {Garz\'o}(2020)}]{GGKG20}%
  \BibitemOpen
  \bibfield  {author} {\bibinfo {author} {\bibfnamefont {R.}~\bibnamefont
  {G\'omez~Gonz\'alez}}, \bibinfo {author} {\bibfnamefont {N.}~\bibnamefont
  {Khalil}}, \ and\ \bibinfo {author} {\bibfnamefont {V.}~\bibnamefont
  {Garz\'o}},\ }\bibfield  {title} {\enquote {\bibinfo {title} {Enskog kinetic
  theory for multicomponent granular suspensions},}\ }\href@noop {} {\bibfield
  {journal} {\bibinfo  {journal} {Phys. Rev. E}\ }\textbf {\bibinfo {volume}
  {101}},\ \bibinfo {pages} {012904} (\bibinfo {year} {2020})}\BibitemShut
  {NoStop}%
\bibitem [{\citenamefont {Brey}\ \emph {et~al.}(2017)\citenamefont {Brey},
  \citenamefont {Buz\'on}, \citenamefont {Maynar},\ and\ \citenamefont
  {Garc\'ia~de Soria}}]{BBGM17}%
  \BibitemOpen
  \bibfield  {author} {\bibinfo {author} {\bibfnamefont {J.~J.}\ \bibnamefont
  {Brey}}, \bibinfo {author} {\bibfnamefont {V.}~\bibnamefont {Buz\'on}},
  \bibinfo {author} {\bibfnamefont {P.}~\bibnamefont {Maynar}}, \ and\ \bibinfo
  {author} {\bibfnamefont {M.~I.}\ \bibnamefont {Garc\'ia~de Soria}},\
  }\bibfield  {title} {\enquote {\bibinfo {title} {Kinetic theory of a confined
  quasi-two-dimensional gas of hard spheres},}\ }\href@noop {} {\bibfield
  {journal} {\bibinfo  {journal} {Entropy}\ }\textbf {\bibinfo {volume} {19}},\
  \bibinfo {pages} {68} (\bibinfo {year} {2017})}\BibitemShut {NoStop}%
\bibitem [{\citenamefont {Brey}, \citenamefont {Maynar},\ and\ \citenamefont
  {Garc\'ia~de Soria}(2016)}]{BMG16}%
  \BibitemOpen
  \bibfield  {author} {\bibinfo {author} {\bibfnamefont {J.~J.}\ \bibnamefont
  {Brey}}, \bibinfo {author} {\bibfnamefont {P.}~\bibnamefont {Maynar}}, \ and\
  \bibinfo {author} {\bibfnamefont {M.~I.}\ \bibnamefont {Garc\'ia~de Soria}},\
  }\bibfield  {title} {\enquote {\bibinfo {title} {Kinetic equation and
  non-equilibrium entropy for a quasi-two-dimensional gas},}\ }\href@noop {}
  {\bibfield  {journal} {\bibinfo  {journal} {Phys. Rev. E}\ }\textbf {\bibinfo
  {volume} {94}},\ \bibinfo {pages} {{040}{103} (R)} (\bibinfo {year}
  {2016})}\BibitemShut {NoStop}%
\bibitem [{\citenamefont {Brey}, \citenamefont {Garc\'ia~de Soria},\ and\
  \citenamefont {Maynar}(2019)}]{BGM19}%
  \BibitemOpen
  \bibfield  {author} {\bibinfo {author} {\bibfnamefont {J.~J.}\ \bibnamefont
  {Brey}}, \bibinfo {author} {\bibfnamefont {M.~I.}\ \bibnamefont {Garc\'ia~de
  Soria}}, \ and\ \bibinfo {author} {\bibfnamefont {P.}~\bibnamefont
  {Maynar}},\ }\bibfield  {title} {\enquote {\bibinfo {title} {Inhomogeneous
  cooling state of a strongly confined granular gas at low density},}\
  }\href@noop {} {\bibfield  {journal} {\bibinfo  {journal} {Phys. Rev. E}\
  }\textbf {\bibinfo {volume} {100}},\ \bibinfo {pages} {{052}{901}} (\bibinfo
  {year} {2019})}\BibitemShut {NoStop}%
\bibitem [{\citenamefont {Maynar}, \citenamefont {Garc\'ia~de Soria},\ and\
  \citenamefont {Brey}(2019)}]{MGB19}%
  \BibitemOpen
  \bibfield  {author} {\bibinfo {author} {\bibfnamefont {P.}~\bibnamefont
  {Maynar}}, \bibinfo {author} {\bibfnamefont {M.~I.}\ \bibnamefont
  {Garc\'ia~de Soria}}, \ and\ \bibinfo {author} {\bibfnamefont {J.~J.}\
  \bibnamefont {Brey}},\ }\bibfield  {title} {\enquote {\bibinfo {title}
  {Understanding an instability in vibrated granular monolayers},}\ }\href@noop
  {} {\bibfield  {journal} {\bibinfo  {journal} {Phys. Rev. E}\ }\textbf
  {\bibinfo {volume} {99}},\ \bibinfo {pages} {{032}{903}} (\bibinfo {year}
  {2019})}\BibitemShut {NoStop}%
\bibitem [{\citenamefont {Garz\'o}\ and\ \citenamefont
  {Vega~Reyes}(2009)}]{GV09}%
  \BibitemOpen
  \bibfield  {author} {\bibinfo {author} {\bibfnamefont {V.}~\bibnamefont
  {Garz\'o}}\ and\ \bibinfo {author} {\bibfnamefont {F.}~\bibnamefont
  {Vega~Reyes}},\ }\bibfield  {title} {\enquote {\bibinfo {title} {Mass
  transport of impurities in a moderately dense granular gas},}\ }\href@noop {}
  {\bibfield  {journal} {\bibinfo  {journal} {Phys. Rev. E}\ }\textbf {\bibinfo
  {volume} {79}},\ \bibinfo {pages} {{041}{303}} (\bibinfo {year}
  {2009})}\BibitemShut {NoStop}%
\bibitem [{\citenamefont {Garz\'o}\ and\ \citenamefont
  {Vega~Reyes}(2012)}]{GV12}%
  \BibitemOpen
  \bibfield  {author} {\bibinfo {author} {\bibfnamefont {V.}~\bibnamefont
  {Garz\'o}}\ and\ \bibinfo {author} {\bibfnamefont {F.}~\bibnamefont
  {Vega~Reyes}},\ }\bibfield  {title} {\enquote {\bibinfo {title} {Segregation
  of an intruder in a heated granular gas},}\ }\href@noop {} {\bibfield
  {journal} {\bibinfo  {journal} {Phys. Rev. E}\ }\textbf {\bibinfo {volume}
  {85}},\ \bibinfo {pages} {{021}{308}} (\bibinfo {year} {2012})}\BibitemShut
  {NoStop}%
\end{thebibliography}
\end{document}